\documentclass[12pt,a4paper]{article}
\usepackage[applemac]{inputenc}
\usepackage[T1]{fontenc}
\usepackage{amsmath,amsfonts,amssymb}
\usepackage[breaklinks=true]{hyperref}
\usepackage{mathtools}
\usepackage{graphicx,epsfig,float}
\usepackage{color,xcolor}
\usepackage{textcomp}
\usepackage{indentfirst}
\usepackage{multirow}
\usepackage{multicol}
\usepackage{enumerate}
\usepackage[font=small]{caption}
\usepackage{adjustbox}
\usepackage{amsthm}
\usepackage{graphics}
\usepackage{pstricks,pstricks-add,pst-math,pst-xkey}
\usepackage{bm}
\usepackage{nccmath}
\usepackage{url}
\usepackage{bbm}
\usepackage{setspace}
\usepackage[left=1in,right=1in,bottom=1.3in,top=1.3in]{geometry}
\usepackage{parskip}
\usepackage{breakcites}
\usepackage{algorithm}
\usepackage{algpseudocode}
\usepackage{tikz}
\usetikzlibrary{automata, positioning}
\usetikzlibrary{matrix,arrows,decorations.pathmorphing,positioning,graphs,calc}
\usepackage{natbib}
\usepackage{subcaption}
\usepackage{subfiles}
\usepackage[title]{appendix}
\usepackage{titling}

\newcommand{\Lcal}{{\mathcal{L}}}
\theoremstyle{definition}

\theoremstyle{plain}

\theoremstyle{definition}

\providecommand{\keywords}[1]{{\small{\it Keywords:} #1}}

\begin{document}
\title{Joint modelling of the body and tail of bivariate data}
\author{L. M. Andr\'e$^{1*}$, J. L. Wadsworth$^{2}$, A. O'Hagan$^{3,4}$\\
\small $^{1}$ STOR-i Centre for Doctoral Training, Lancaster University, LA1 4YR, United Kingdom \\
\small $^{2}$ Department of Mathematics and Statistics, Lancaster University, LA1 4YF, United Kingdom \\
\small $^{3}$ School of Mathematics and Statistics, University College Dublin, Ireland \\
\small $^{4}$ Insight Centre for Data Analytics, University College Dublin, Ireland \\
\small $^*$ Correspondence to: \href{mailto:l.andre@lancaster.ac.uk}{l.andre@lancaster.ac.uk} }
\date{August 25, 2023}

\maketitle
\pagenumbering{arabic}

\begin{abstract}
    \noindent In situations where both extreme and non-extreme data are of interest, mo\-delling the whole data set accurately is important. In a univariate framework, modelling the bulk and tail of a distribution has been extensively studied before. However, when more than one variable is of concern, models that aim specifically at capturing both regions correctly are scarce in the literature. A dependence model that blends two copulas with different characteristics over the whole range of the data support is proposed. One copula is tailored to the bulk and the other to the tail, with a dynamic weighting function employed to transition smoothly between them. Tail dependence properties are investigated numerically and simulation is used to confirm that the blended model is sufficiently flexible to capture a wide variety of structures. The model is applied to study the dependence between temperature and ozone concentration at two sites in the UK and compared with a single copula fit. The proposed model provides a better, more flexi\-ble, fit to the data, and is also capable of captu\-ring complex dependence structures.
\end{abstract}


\newpage

\section{Introduction} \label{section:introduction}

\subsection{Motivation} \label{subsection:motivation}

\noindent When dealing with environmental phenomena such as high temperatures, wind speeds or air pollution, or with financial applications such as insurance losses, interest often lies in modelling the extreme observations, which are typically scarce. For such cases, a model with focus on the tail of the distribution is required as common statistical models that may be used to fit the entire data set lead to poor estimates of the extremes. To overcome this issue, models based on extreme value theory (EVT) can be applied; these aim to quantify the behaviour of a process at extremely large (or small) values of a series. Typically, the generalised extreme value (GEV) distribution is fitted to block maxima, often annual maxima, or the generalised Pareto distribution (GPD) is fitted to data exceeding a high threshold. The former can be seen as a wasteful approach if there are more data on extremes available, while the latter usually requires a subjective choice of threshold, which inevitably leads to uncertainty, with different choices leading to different results; see \citet{Coles2001}.

\noindent However, in some cases, interest not only lies in modelling the extreme observations accurately but also fitting the non-extremes well, meaning a flexible model over the whole support of the distribution is required. For instance, the concentration of pollutants in the air may be so high that harmful levels are actually in the body of the data set. Thus, from a public health pers\-pective, we care not only about the probability of exceeding extreme, and potentially more dangerous, pollutant levels but also about the probability of exceeding harmful yet locally moderate levels. Fitting a model to both the bulk (i.e., the non-extreme observations) and tail (i.e., the extreme obser\-vations) of a data set has been dealt with in the univariate framework but little work has been done in extending to a multivariate setting. In this work, we outline an approach that offers dependence models for the bulk and tail, while ensuring a smooth transition between the two.

\subsection{Background}\label{subsection:background}

\noindent In the univariate setting, several models have been proposed to join one distribution for the bulk to a GPD for the tail. \citet{ScarrotMacDonald2012} review several of these approaches, hereafter referred to as extreme value mixture models, or EVMMs. These models aim to account for the uncertainty in the choice of threshold, by implicitly or explicitly estimating it. With EVMMs, care is needed so that the bulk and tail are not excessively influenced by each other, though they cannot be fully disjoint since they share information. Parametric EVMMs entail fitting a specified distribution to the bulk and a GPD to the tail, while semi-parametric models fit a GPD to the tail with a more flexible model in the bulk. \citet{BehrensAll2004} propose a parametric model, which exhibits discontinuity at the threshold; \citet{CarreauBengio2009a} avoid this by forcing continuity up to the first derivative of the density function. On the other hand, \citet{FrigessiAll2002} fit two distributions to the whole data, giving more weight to the bulk at low ranges in the support and to the GPD in the upper tail by means of a dynamic weighting function $p(x;\theta) \in (0,1].$ The density of their model is defined as
\begin{equation*}\label{eq:frigessi}
    h(x;\theta,\bm\beta,\bm\alpha)=\frac{[1-p(x;\theta)]f(x;\bm \beta)+p(x;\theta)g(x;\bm \alpha)}{K(\theta,\bm \beta,\bm \alpha)}, 
\end{equation*}
where $g(x;\bm \alpha)$ is the density of the GPD with vector of parameters $\bm\alpha$,  $f(x;\bm \beta)$ is a density with a lighter tail and vector of parameters $\bm\beta$, $K(\theta,\bm \beta,\bm \alpha)$ is a normalising constant and $p(x;\theta)$ is increasing in $x$ for all $\theta.$ Because $p(x;\theta)$ depends on $x,$ it favours the GPD in the upper tail whilst the lower tail is controlled by $f(x;\bm \beta).$ However, careful choice of the weighting function is needed since some functions, such as the unit step function, may lead to a discontinuity in the transition between the two distributions; see \citet{FrigessiAll2002} for details. More recently, methods introduced by \citet{NaveauAll2016} and \citet{Stein2021} aim to model the lower and upper tails of the data with GPDs, while ensuring a smooth transition between the regions. The former achieve this by constructing a model relying on compositions of functions, where one is a cumulative distribution function (CDF) of a GPD, and the other is a CDF that satisfies certain constraints to ensure both tails follow a generalised Pareto-type distribution. The model proposed by \cite{Stein2021} also assumes a composition of functions, where one is a monotone-increasing function that controls both the lower and upper tails, and the other is a Student t CDF. Finally, \citet{KrockEtAll2021} extend the latter approach to incorporate non-stationarity. The methods proposed by \citet{FrigessiAll2002}, \citet{NaveauAll2016}, \citet{Stein2021} and \citet{KrockEtAll2021} avoid the choice of threshold.

\noindent In a semi-parametric framework, \citet{CabrasCastellanos2010} approximate the bulk distribution by an equi-spaced binning of the data followed by a Poisson log-link gene\-ra\-lised linear model fit to the counts with a polynomial smoother for the mean parameter. \citet{NascimentoAll2011} define the bulk distribution as a weighted mixture of gamma densities, extending the method proposed by \citet{BehrensAll2004}, while \citet{HuangAll2019} estimate the log-density by first transforming the data and then applying a cubic spline to the histogram. \citet{TencaliecAll2020} propose a method based on the extension of the GPD proposed by \citet{NaveauAll2016}. Finally, \citet{TancrediAll2006} and \citet{MacDonaldAll2011} propose non-parametric fits to the data. In the former, the bulk model is fitted via a mixture of uniform distributions whereas in the latter a kernel density estimator is used instead.

\noindent When we move to the multivariate setting, there is an extra difficulty; not only is it important to model the margins of the data correctly, but the dependence between the variables is also of interest since the behaviour of one variable can influence the behaviour and value of another. It is common practice to measure this relationship using correlation coeffi\-cients, such as Pearson's linear correlation or Kendall's concordance \citep{Kendallstau}. However, these only give information about the association between variables as a whole. An alternative is to use copulas, which fully capture the dependence between two or more variables. According to Sklar's Theorem \citep{Sklar1959}, the multivariate distribution function, $F,$ of the random vector $(X_1,\ldots,X_d)$ can be written as the composition of a copula, $C,$ and the marginal distributions of each $X_i,$ $F_{X_i}(X_i),$ $i=1, \ldots, d,$ $d\geq 2,$ as follows
\begin{equation*}
    F(x_1,\ldots,x_d)=C\left(F_{X_1}(x_1),\ldots,F_{X_d}(x_d)\right).
\end{equation*}
\noindent If the variables are continuous, then the copula $C$ is unique. One advantage of copulas is that they are able to describe the dependence structure of two or more variables in a way that does not depend on the margins. Where it exists, the copula density $c\left(F_{X_1}(x_1),\ldots,F_{X_d}(x_d)\right)$ can be obtained by taking the $d^{\text{th}}$ order derivative with respect to the variables $F_{X_1}(x_1),\ldots,F_{X_d}(x_d).$

\noindent There is a large literature on dependence modelling for extremes, which usually involves defining a multivariate threshold above which an asymptotically-motivated copula is assu\-med to hold. However, models specifically aimed at capturing the behaviour of extremes as well as the body of the data, while permitting a likelihood-based approach to inference, are scarce in the literature. Both defining and performing inference on such models can be challenging compared to univariate models.  

\noindent Methods for constructing more flexible copula families have been increasing in recent years, especially in financial applications. For instance, \citet{DurrlemanEtAll2000}, \citet{Morillas2005}, \citet{KlementEtAll2005} and \citet{DuranteEtAll2010} propose transforming known copulas, especially from the Archimedean family, by means of bijections on $[0,1].$ In particular, the methods proposed by \citet{DurrlemanEtAll2000} and \citet{DuranteEtAll2010} allow for a more accurate fit of the dependence structure. Given a bijection $\gamma:[0,1]\rightarrow [0,1],$ the copula $C$ is transformed into a new copula $C_\gamma$ in the following way $C_\gamma(x,y)=\gamma^{-1}(C(\gamma(x),\gamma(y))).$ Moreover, depending on specific conditions imposed on $\gamma,$ the dependence structure of $C_\gamma$ contrasts with that of $C$ in different ways. Specifically, in the method proposed by \citet{DurrlemanEtAll2000}, changes in the overall dependence measures of $C$, such as Kendall's $\tau,$ are possible while $C$ and $C_\gamma$ share the same extremal behaviour. On the other hand, \citet{DuranteEtAll2010} study how the dependence in the extremes changes from $C$ to $C_{\gamma^{-1}},$ while the fit in the body remains the same between the two.

\noindent Other possibilities for building new copula families rely on piecewise cons\-tructions or convex combinations. For the former, by constructing box co\-pu\-las (i.e., copulas nested in each other), \citet{Hummel2009} is able to control and modify the dependence in the tail. For the latter, \citet{BacigalEtAll2010} propose new construction techniques through additive generators of binary Archimedean copulas, whereas \citet{ShamiriEtAll2011} construct a Clayton-Gumbel copula, where, by means of a standard mixture model, two individual copulas are joined into one. This model allows for asymmetry in the data while being able to capture strong dependence in both tails.

\noindent Methods based on transformation of copulas or convex combinations allow for different, more flexible, dependence structures beyond the usual copulas. However, their main focus lies in providing a way of constructing new copula families, rather than offering an accurate representation of the bulk and tail regions simultaneously.

\noindent Alternatively, patchwork copulas can offer a way to capture dependence structures that are not well suited to standard copulas. These allow for different copula models to be fitted to several regions of $[0,1]^2$ based on their characteristics; see for example \citet{PfeiferRagulina2021}. Particular cases of patchwork copulas include those based on ordinal sums \citep{AlsinaEtAll2006}; gluing copulas, where two or more copulas are scaled back to boxes in a region of the unit square and glued together along some hyperplane \citep{MesiarEtAll2008,SiburgStoimenov2008}; and copulas based on rectangular constructions, where it is possible to have a copula in the body and another in the upper tail by defining two rectangles (disjoint up to their boundaries) over the diagonal, for example; see \citet{DuranteEtAll2009} for more details. A generalised method to construct patchwork copulas that include the above mentioned cases is given in \citet{DuranteEtAll2013}. Given a copula $C,$ a patchwork copula derived from it features the same probability mass distribution as $C,$ excluding a $d$-dimensional box ($\subseteq [0,1]$) in which the probability mass is distributed differently. These models can be used to modify the extremal behaviour of a copula in two or more corners of $[0,1]^d,$ and allow strong positive tail dependence to be induced if the application requires it. In this way, patchwork copulas aim to overcome the issue of misrepresentation of the extremes, when considering the whole data set. However, the transition \-between the non-extreme and the extreme regions is not smooth and therefore may be unsuitable in many real applications. 

\begin{figure}[H]
    \centering
    \includegraphics[width=0.55\textwidth]{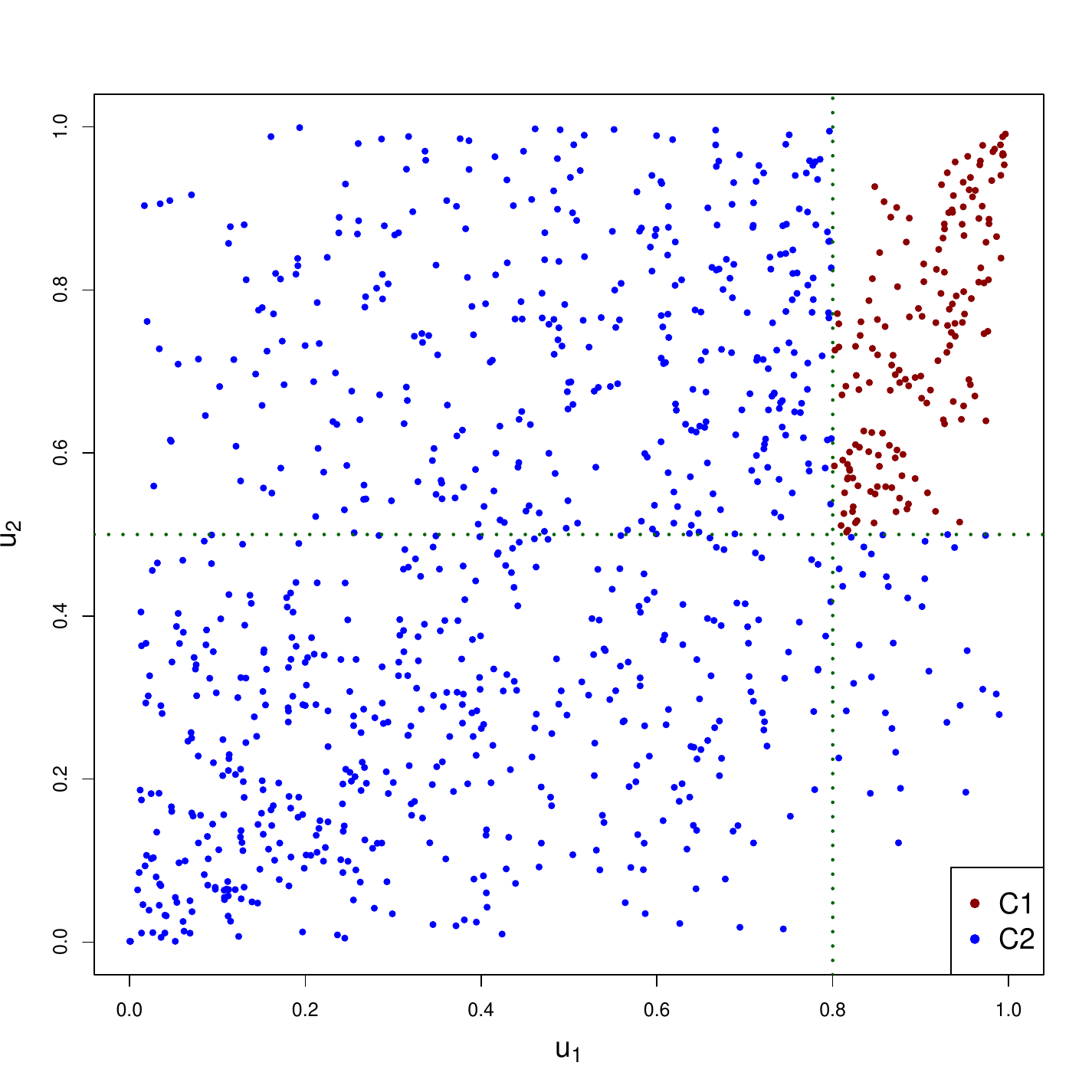}
    \caption{Example of $\bm Q$ simulated according to equation \eqref{eq:q} with a Gumbel copula with para\-me\-ter $\alpha=2$ selected for the upper tail copula $C_1$ and Gaussian copula with parameter $\rho=0.6$ selected for the body copula $C_2$ of the model proposed by \citet{AulbachAll2012a}. For illustration purposes, the vector of thresholds was chosen to be ${\bm t}=(0.8,0.5).$}
    \label{fig:aulbach}
\end{figure}

\noindent \citet{AulbachAll2012a, AulbachAll2012b} suggest an extension to the multivariate setting of the model proposed by \citet{BehrensAll2004}. They define a novel co\-pula model by joining two $d$-dimensional ($d\geq 2$) copulas, one for the upper tail and the other for the body, in a manner that produces a new copula. Specifically, the authors assume two independent random vectors, each of which follow an arbitrary copula, that is ${\bm V}=(V_1,\ldots,V_d)\sim C_1$ and ${\bm Y}=(Y_1,\ldots,Y_d)\sim C_2$. It is also required that the copulas are defined in $[-1,0]^d,$ which is not a problem since, if $\bm U$ follows a copula $C: [0,1]^d\rightarrow[0,1]$, then $\widetilde{\bm U}=\bm U-1$ follows a copula $\widetilde{C}$ with shifted support i.e., $\widetilde{C}: [-1,0]^d\rightarrow [0,1].$ Then, by an appropriate choice of threshold vector ${\bm t}=(t_1,\ldots,t_d),$ they construct a random vector $\bm Q,$ whose $i^{\text{th}}$ element is given by
\begin{equation}\label{eq:q}
   Q_i\coloneqq Y_i\mathbbm{1}_{Y_i\leq t_i}-t_iV_i\mathbbm{1}_{Y_i> t_i}, \qquad i=1,\ldots, d.
\end{equation}
The authors prove that $\bm Q$ also follows a copula with support on $[-1,0]^d,$ which coincides with $C_1$ on the region $(t_1,0]\times\ldots \times (t_d,0]$ and with $C_2$ on the region $[-1,t_1]\times\ldots \times [-1,t_d].$ An exact representation of the method is presented in \citet{AulbachAll2012b}.
However, the model not only requires a choice of cut-off values $t_i,\, i=1,\ldots,d,$ to define the regions to fit each copula but, as with patchwork copulas, the transition between the two copulas may not be smooth. Figure \ref{fig:aulbach} displays an example of a data set simulated according to equation \eqref{eq:q}; the discontinuity at the threshold is evident. Moreover, this method does not offer a convenient formulation of the likelihood, which results in difficulties for inference. 
 
\noindent More recently, \citet{PfeiferEtAll2017} and \citet{PfeiferEtAll2019} propose infinite discrete and continuous partition-of unity copulas, respectively; these are flexible in higher dimensions and can be applied when there is asymmetry in the data. Similar to patchwork copulas, these copulas allow for implementing positive dependence in the tails; the density of the proposed model is approximated by an infinite mixture of functions, and careful choice of these functions can modify the tail behaviour if required.

\noindent A different type of approach was taken by \citet{HuOHagan2021}, who consider averaging different copula families that have been fitted to the whole distribution, in order to obtain a more robust estimate of the tail dependence of the data set. However, the use of BIC in the calculation of the weights assigned to each copula places the focus on the body and not on the tail of the data.

\noindent In a spatial context, \citet{Graler2014} proposes capturing the dependence of skewed spatial random fields (that display extreme events) by considering convex combinations of bivariate copulas in the construction of a spatial copula. In this way, between each location, a different dependence model is obtained. More recently, \citet{KrupskiiAll2016} and \citet{ZhangAll2021} each propose models fitted to both the body and tail of a distribution. The former outlines a copula model based on the assumption that there exists a common factor which affects the joint dependence of all the observations of the underlying process, and which is able to model both tail dependence and asymmetry. Numerical integration over this factor variable leads to a likelihood that can be fitted to all data. The latter propose using the generalised hyperbolic copula, which is flexible due to having a relatively large number of parameters. For both of these models, the authors show that there is reasonable flexibility for capturing both body and tail, yet a primary motivation for fitting to all data is the desire to avoid the computational difficulty involved in using censored likelihoods for extremes. 

\subsection{Extremal dependence properties} \label{subsection:depproperties}

\noindent When the focus lies on extreme values, studying the extremal dependence \-between the variables is of interest. Two variables are said to be asympto\-tically dependent (AD) if joint extremes occur at a similar frequency to marginal extremes, or asymptotically independent (AI) otherwise. This dependence can be quantified through the measure $\chi=\displaystyle{\lim_{r\rightarrow 1}\,\chi(r)}\in [0,1],$ where the limit exists, with
\begin{equation}\label{eq:chir}
    \chi(r)=P[F_Y(Y)>r \mid F_X(X)>r]=\frac{1-2r+C(r,r)}{1-r} , \qquad r\in (0,1),
\end{equation}
 where $C$ is the copula of $(X,Y);$ see \citet{Joe1997} or \citet{ColesAll1999}. The random variables $X$ and $Y$ are asymptotically independent if $\chi=0,$ whereas if $\chi>0$ they are asymptotically dependent.

\noindent A complementary measure to $\chi$ is the residual tail dependence coefficient $\eta\in (0,1]$ proposed by \citet{LedfordTawn1996}. For a function $\Lcal$ that is slowly-varying at zero, they assume that the joint tail can be written as
\begin{equation}\label{eq:eta}
    P[F_Y(Y)>r \mid F_X(X)>r]\sim \Lcal(1-r)(1-r)^{\frac{1}{\eta}-1} \quad \text{as }r\rightarrow 1.
\end{equation}
The variables are asymptotically dependent if $\eta=1$ and $\Lcal(1-r)\not\rightarrow 0$ as $r\rightarrow 1,$ and asymptotically independent otherwise. Additionally, if $\eta \in \left(0,1\slash 2\right),$ the variables show negative extremal association; positive extremal association if $\eta \in \left(1\slash 2,1\right]$ and they exhibit near extremal independence if $\eta=1\slash 2$.

\noindent Similarly to $\chi(r)$, for a particular value of $r\in (0,1),$ $\eta(r)$ can be obtained as
\begin{equation}\label{eq:etar}
    \eta(r)=\frac{\log\left(P[F_X(X)>r]\right)}{\log\left(P[F_X(X)>r,F_Y(Y)>r]\right)},
\end{equation}
with $\eta=\displaystyle{\lim_{r\rightarrow 1}\eta(r)}.$

\noindent This paper is organised as follows: in Section \ref{section:model} we present our proposed model and its properties. Inference for the model is studied in Section \ref{section:inference}, complemented by a simulation study to demonstrate performance in correctly specified and misspecified scenarios. We then apply our methodology to ozone and temperature data in the UK in Section \ref{section:casestudy} and conclude with a discussion in Section \ref{section:conclusion}.

\section{Weighted copula model} \label{section:model}

\subsection{Model definition} \label{subsection:modeldef}

\noindent Our interest lies in accurately modelling both the bulk and the tail of the whole distribution. From existing literature in the dependence context, \citet{Hummel2009}, \citet{AulbachAll2012a,AulbachAll2012b}, \citet{DuranteEtAll2013} and \citet{PfeiferEtAll2017,PfeiferEtAll2019} are concerned with representing both regions correctly. However, our model differs from these approaches in that we aim for a smooth transition between the two regions and allow for likelihood-based inference. To do so, we propose a mixture model where we fit two copulas to the whole range of the support and blend them by means of a dynamic weighting function $\pi;$ in this way, data can be allowed to favour the ``best" copula for each region, avoiding the subjective choice of thresholds often present in EVT applications. This approach can be seen as an extension to the multivaria\-te framework of the model proposed by \citet{FrigessiAll2002} mentioned in Section \ref{subsection:background}. 

\noindent Although our ideas could theoretically be applied in higher dimensions, we restrict ourselves to the bivariate setting for computational simplicity. Let $c_t$ and $c_b$ be copula densities re\-pre\-senting the tail and the body, with vectors of parameters $\bm\alpha$ and $\bm \beta,$ respectively. For $(u^*,v^*) \in [0,1]^2,$ we define a new density $c^*$ by
\begin{equation}\label{eq:model}
    c^*(u^*,v^*;\bm \gamma)=\frac{\pi(u^*,v^*;\theta)c_t(u^*,v^*;\bm \alpha)+[1-\pi(u^*,v^*;\theta)]c_b(u^*,v^*;\bm \beta)}{K(\bm\gamma)},
\end{equation}
where $\bm\gamma=(\theta,\bm\alpha,\bm\beta)$ is the vector of model parameters and  
\begin{equation*}
    K(\bm\gamma)=\int_{0}^{1}\int_{0}^{1}\left[\pi(u^*,v^*;\theta)c_t(u^*,v^*;\bm \alpha)+(1-\pi(u^*,v^*;\theta))c_b(u^*,v^*;\bm \beta)\right]\text{d}u^*\,\text{d}v^*
\end{equation*}
is a normalising constant. The weighting function $\pi$ depends on the data, and is spe\-ci\-fied such that, for small values of $u^*$ and $v^*,$ more weight is given to $c_b$ and, for larger values, more weight is given to $c_t.$ Thus, for a fixed value of the parameter $\theta,$ the function $\pi:(0,1)^2\rightarrow (0,1)$ should be increasing in $u^*$ and $v^*.$ We note that having a dynamic weighting function is a modelling choice, but without this equation \eqref{eq:model} simply represents a standard mixture model. Moreover, $\pi$ is not required to be monotonic and can be defined based on the application, which might make more sense outside of the extreme value context.

\noindent A direct consequence of $\pi(u^*,v^*;\theta)$ depending on the data is that the margins of the density $c^*$ are non-uniform; this leads to complications for inference. That is, we cannot fit $c^*$ directly to the data as it is not a copula density. We overcome these issues by fitting the copula of the density in equation \eqref{eq:model}, which requires numerical integration to calculate. The first stage is to obtain the true margins of $(U^*,V^*)\sim c^*$ as
\begin{equation*}\label{eq:margins}
    F_{U^*}(u^*)=P[U^*\leq u^*]=\int_{0}^{u^*}\int_{0}^{1}c^*(u,v)\text{d}v\,\text{d}u,
\end{equation*}
and similarly for $F_{V^*},$ and then the corresponding inverse functions, $F^{-1}_{U^*}$ and $F^{-1}_{V^*}$ so that we can transform the margins to Uniform$(0,1)$ via the probability integral transform. The resulting copula is thus represented as
\begin{equation} \label{eq:copmodel}
    c(u,v;\bm \gamma)=\frac{c^*\left(F^{-1}_{U^*}(u),F^{-1}_{V^*}(v);\bm \gamma\right)}{f_{U^*}\left(F^{-1}_{U^*}(u)\right)f_{V^*}\left(F^{-1}_{V^*}(v)\right)},
\end{equation}
where $f_{U^*}$ and $f_{V^*}$ are the marginal probability density functions of $c^*$ and $\bm \gamma=(\theta,\bm\alpha,\bm\beta)$ is the vector of model parameters, common to the density in equation \eqref{eq:model}. Note that each of $f_{U^*},$ $f_{V^*},$ $F_{U^*}$ and $F_{V^*}$ depends on $\bm \gamma,$ but this is suppressed in the notation for readability.

\subsection{Simulation} \label{subsection:simulation}

\noindent It is important to be able to sample from the proposed model so that it can be validated. To do so, we first note that we can rewrite the density \eqref{eq:model} as a standard mixture of two densities
\begin{equation*}
    c^*(u^*,v^*;\bm \gamma)=\frac{K_t}{K}f_t(u^*,v^*;\theta,\bm\alpha)+\left(1-\frac{K_t}{K}\right)f_b(u^*,v^*;\theta,\bm\beta),
\end{equation*}
where $K=K(\bm\gamma)$ and
\begin{align*}
    f_t(u^*,v^*;\theta,\bm\alpha)&=\frac{\pi(u^*,v^*;\theta)c_t(u^*,v^*;\bm\alpha)}{K_t}, \\
    f_b(u^*,v^*;\theta,\bm\beta)&=\frac{[1-\pi(u^*,v^*;\theta)]c_b(u^*,v^*;\bm\beta)}{K_b}, \\
    K_t&=\int_{0}^{1}\int_{0}^{1}\pi(u^*,v^*;\theta)c_t(u^*,v^*;\bm \alpha)\text{d}u^*\,\text{d}v^*, \\
    K_b&=\int_{0}^{1}\int_{0}^{1}[1-\pi(u^*,v^*;\theta)]c_b(u^*,v^*;\bm \beta)\text{d}u^*\,\text{d}v^*.
\end{align*}
Note that $K=K_t+K_b$. Thus, to simulate from $c^*(u^*,v^*;\bm \gamma)$ we need to be able to sample from the two densities $f_t(u^*,v^*;\theta,\bm\alpha)$ and $f_b(u^*,v^*;\theta,\bm\beta),$ which are non-standard as they depend on the weighting function $\pi(u^*,v^*;\theta)$ as well as the copula densities. However, as we can sample from the densities $c_t(u^*,v^*;\bm\alpha)$ and $c_b(u^*,v^*;\bm\beta),$ we can use a rejection sampling scheme to simulate from the required densities $f_t$ and $f_b.$

\noindent Note that, since the weighting function $\pi(u^*,v^*;\theta)$ is in $(0,1),$ it is the case that
\begin{equation*}
    \sup_{(u^*,v^*)\in (0,1)^2} \frac{f_t(u^*,v^*;\bm\alpha)}{c_t(u^*,v^*;\bm\alpha)}=\sup_{(u^*,v^*)\in (0,1)^2} \frac{\pi(u^*,v^*;\theta)c_t(u^*,v^*;\bm\alpha)}{K_t c_t(u^*,v^*;\bm\alpha)} = \frac{\pi(u^*,v^*;\theta)}{K_t}\leq \frac{1}{K_t}.
\end{equation*}
Similarly, the ratio $f_b\slash c_b$ is bounded by $1\slash K_b.$ The rejection algorithm for sampling from $c^*$ via $f_t$ and $f_b$ is then as follows:
\begin{enumerate}
    \item Simulate $n$ draws from $c_t(u^*,v^*;\bm\alpha)$ and keep each with probability  $$\frac{f_t(u^*,v^*;\theta,\bm\alpha)}{(1\slash K_t) c_t(u^*,v^*;\bm\alpha)}=\frac{K_t\pi(u^*,v^*;\theta)c_t(u^*,v^*;\bm\alpha)}{K_t c_t(u^*,v^*;\bm\alpha)}=\pi(u^*,v^*;\theta).$$
    The expected number of returned draws from $f_t$ is $nK_t.$
    \item Simulate $n$ draws from $c_b(u^*,v^*;\bm\beta)$ and keep each with probability $$\frac{f_b(u^*,v^*;\theta,\bm\beta)}{(1\slash K_b) c_b(u^*,v^*;\bm\beta)}=\frac{K_b[1-\pi(u^*,v^*;\theta)]c_b(u^*,v^*;\bm\beta)}{K_b c_b(u^*,v^*;\bm\beta)}=1-\pi(u^*,v^*;\theta).$$
    The expected number of returned draws from $f_b$ is $nK_b$.
\end{enumerate}

\noindent The total expected number of draws from both distributions together is $n(K_t+K_b)=nK;$ these are in proportions $K_t\slash K$ and $K_b\slash K=1-K_t\slash K,$ and consequently we have a random sample from density $c^*.$ To get a fixed sample size $n',$ we simply take sufficiently large $n$ and keep $n'$ draws at random.

\begin{figure}[H]
    \centering
    \includegraphics[width=\textwidth]{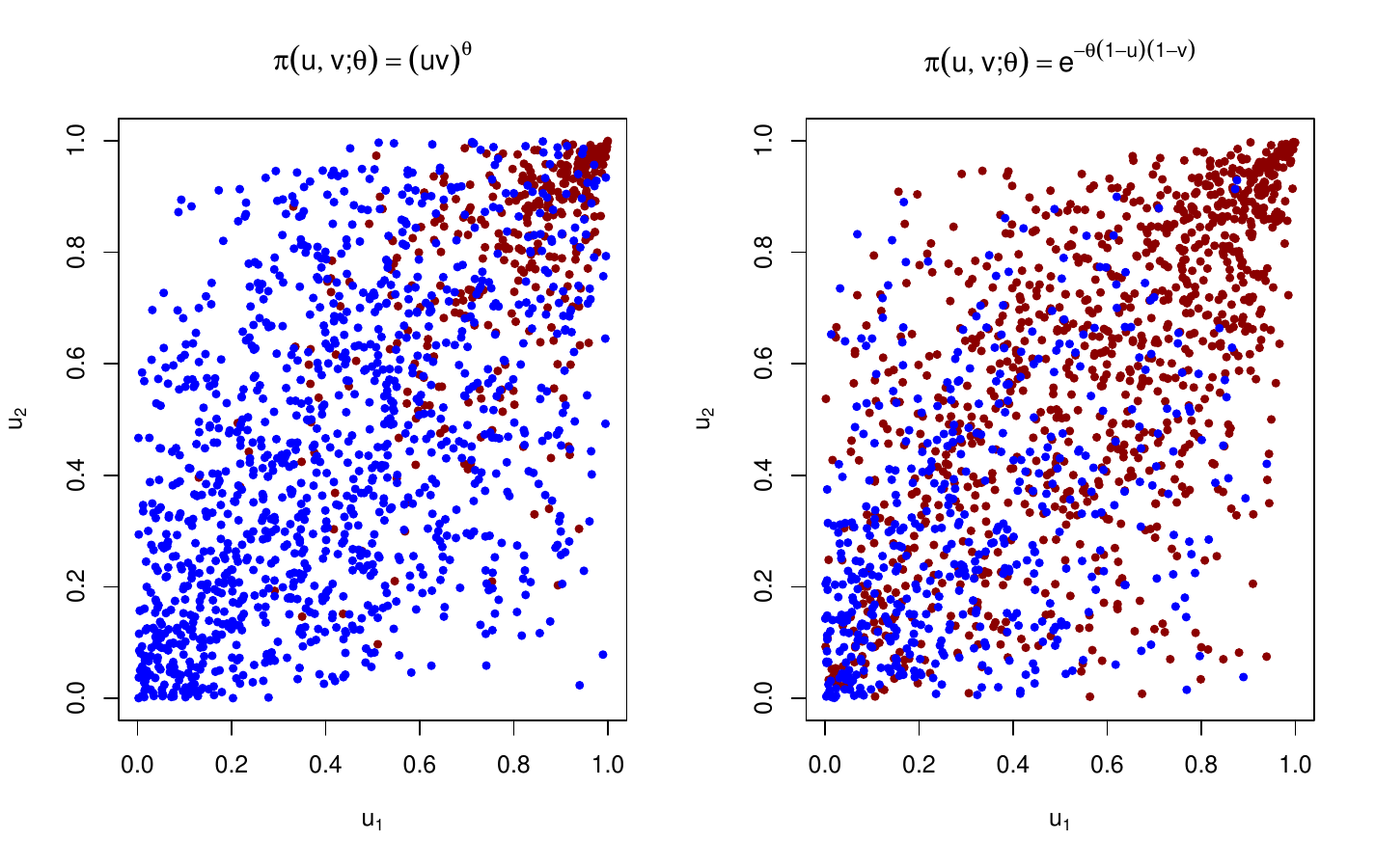}
    \caption{Example of data points from two weighted copula models simulated according to the sampling procedure detailed in Section \ref{subsection:simulation}. In both cases, a Gumbel copula with parameter $\alpha=2$ is taken as $c_t$ and a Gaussian copula with parameter $\rho=0.6$ as $c_b.$ Two weighting functions are used with $\theta=1.5$ in both: $\pi(u^*,v^*;\theta)=(u^*v^*)^\theta$ (left) and $\pi(u^*,v^*;\theta)=\exp\{-\theta(1-u^*)(1-v^*)\}$ (right). Points in blue originate from $c_b$ and points in red originate from $c_t.$}
    \label{fig:simmodel}
\end{figure}

\noindent Figure \ref{fig:simmodel} illustrates two examples of random samples from our weighted co\-pula model with different weighting functions. In each case we take a Gumbel copula with $\alpha=2$ as $c_t$ and a Gaussian copula with $\rho=0.6$ as $c_b,$ which are the same components as the example in Figure \ref{fig:aulbach}. See \ref{section:appendix} for a directory of copula models and their parameterisations. Contrary to the \cite{AulbachAll2012a} approach, we see that there is no cut-off between the two regions, with a smooth transition from data points mainly derived from $c_b$ in the bottom left to those mainly derived from $c_t$ in the top right. The influence of the choice of weighting function is also visible; for the same value of $\theta,$ a preference for $c_t$ over $c_b$ is shown in the right plot.

\subsection{Extremal dependence properties} \label{subsubsection:chietacstar}

\noindent We are interested in understanding the extremal dependence properties of the proposed model and, to do so, we compute the dependence measures $\chi$ and $\eta$ mentioned in Section \ref{subsection:depproperties}. However, since they are defined in terms of the joint survival function of $\left(F_X(X),F_Y(Y)\right),$ which we do not have, and the integral of the density in equation \eqref{eq:model} is intractable, $\chi$ and $\eta$ are mainly obtained numerically. We have, however, derived these measures for one particular case with two different weighting functions; these are presented in the Supplementary Material. For a set of bivariate copulas, \citet{Heffernan2000} and \citet{Joe2014} study these dependence measures; a selection of which are summarised in Table \ref{tab:chietacopulas}.

\begin{table}[H]
\caption{$\chi$ and $\eta$ for a selection of copulas; $\rho$ is the parameter of the Gaussian copula, and $\alpha$ the parameter of the Gumbel and H\"usler-Reiss copulas.}
    \centering
    \begin{tabular}{lcc}
      Copula  & $\chi$ & $\eta$ \\
      \hline
      Gaussian  & 0 & $(1+\rho)\slash 2$ \\
      Frank  & 0 & $1\slash 2$\\
      Gumbel  & $2-2^{1\slash \alpha}$ & 1 \\
      H\"usler-Reiss  & $2-2\Phi(1\slash \alpha)$ & 1 \\
    \end{tabular}
    \label{tab:chietacopulas}
\end{table}

\begin{figure}[H]
\vspace{-1.75cm}
    \centering
    \begin{subfigure}[b]{0.9\textwidth}
        \includegraphics[width=\textwidth]{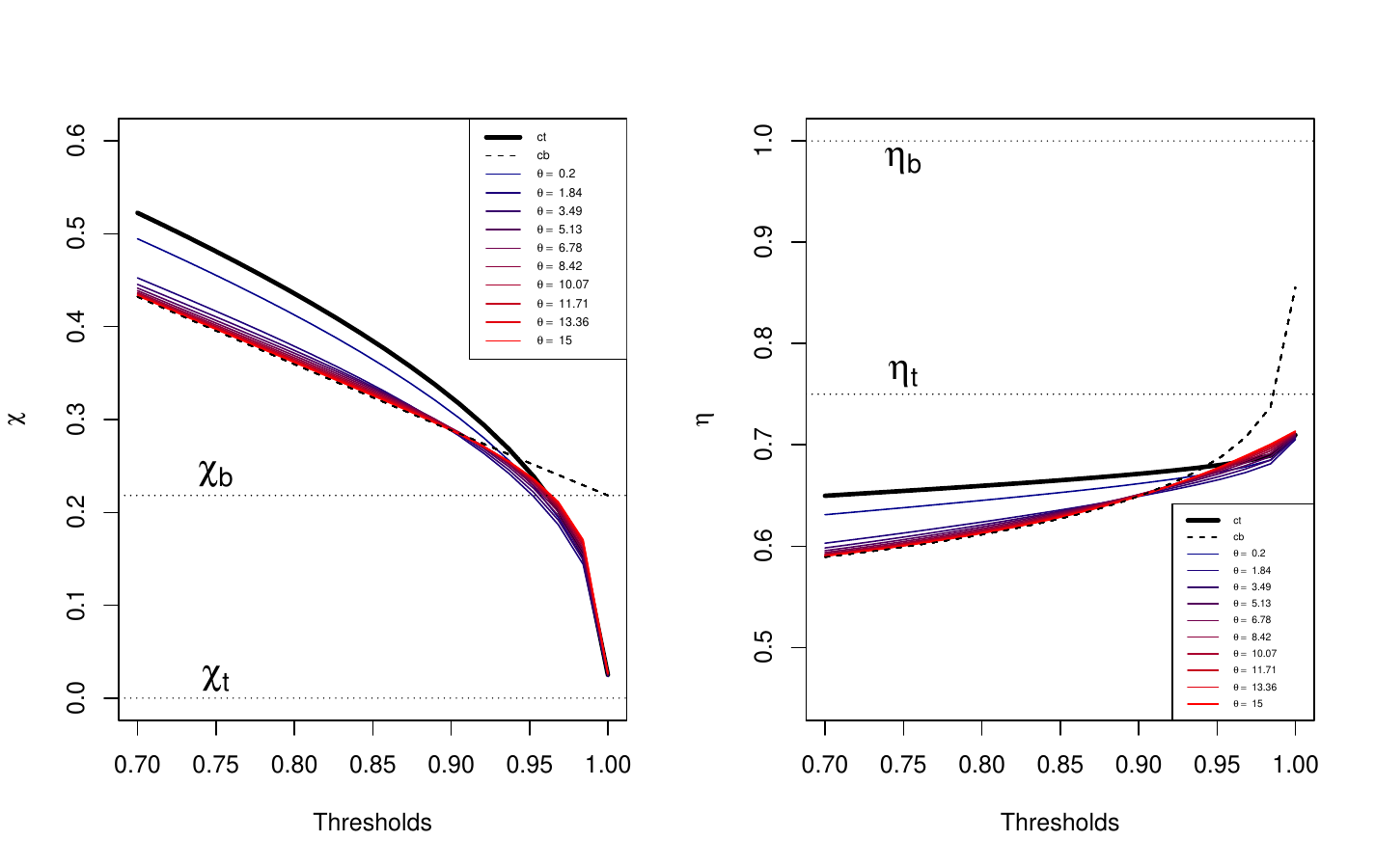}
        \caption{$\chi(r)$ and $\eta(r)$ with weighting function $\pi(u^*,v^*;\theta)=(u^*v^*)^{\theta}.$}
    \end{subfigure}
    \begin{subfigure}[b]{0.9\textwidth}
        \includegraphics[width=\textwidth]{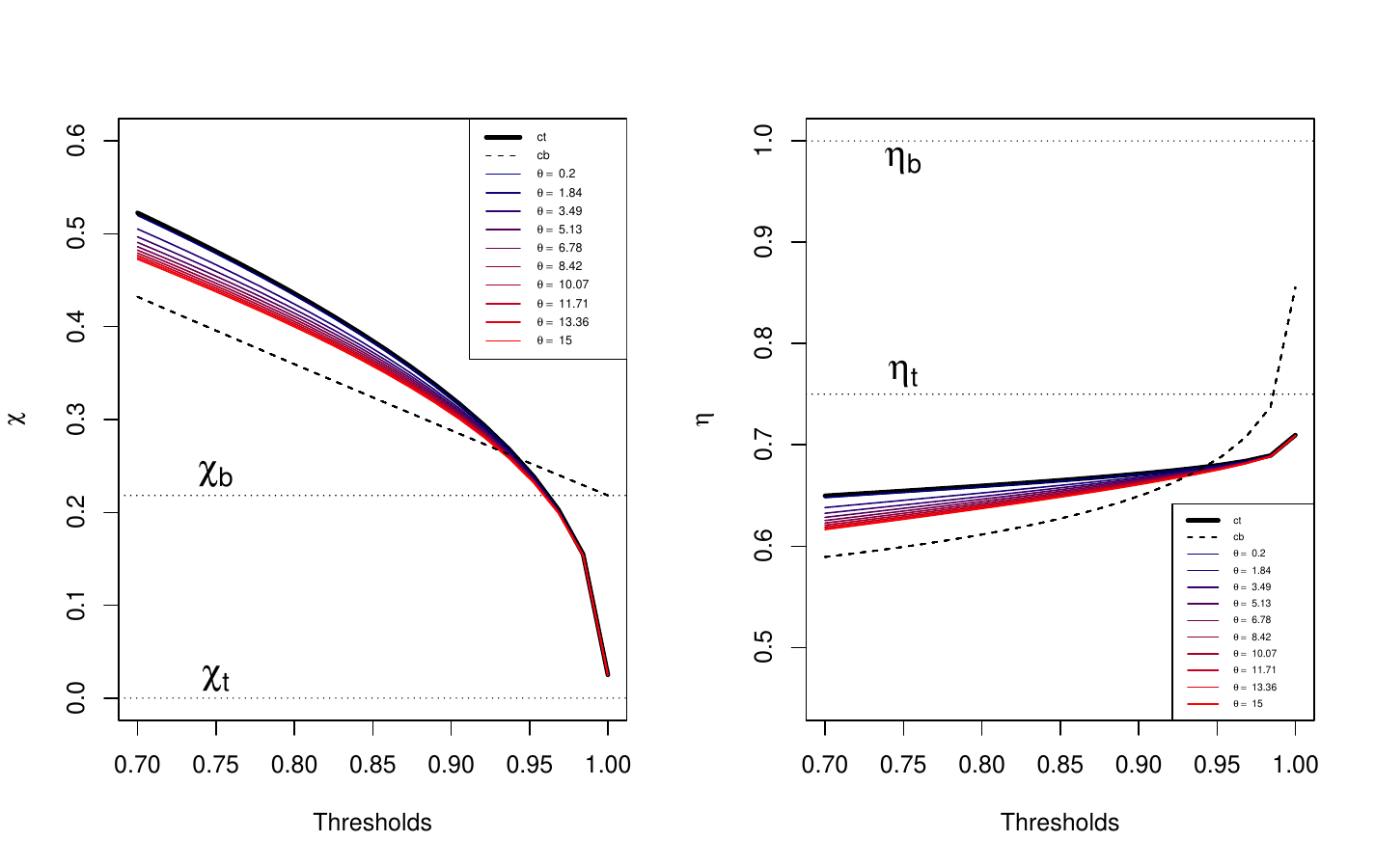}
        \caption{$\chi(r)$ and $\eta(r)$ with weighting function $\pi(u^*,v^*;\theta)=\exp\{-\theta(1-u^*)(1-v^*)\}.$}
    \end{subfigure}
    \caption{$\chi(r)$ and $\eta(r)$ for different thresholds $r\in[0.7,1)$ for the proposed model with both $\pi(u^*,v^*;\theta)$ when $c_b$ is Gumbel (AD) and $c_t$ is Gaussian (AI). The coloured lines represent the 10 different models depending on different values of $\theta;$ the thick black lines represent the single copula models - Gumbel (dashed) and Gaussian (solid). The theore\-tical values for the Gumbel and Gaussian copulas based on Table \ref{tab:depprop} are represented by the horizontal dashed lines.}
    \label{fig:depprop}
\end{figure}

\noindent We consider mixtures of these four copulas to study the dependence pro\-perties of our model. In addition, we study the influence of the weighting function $\pi(u^*,v^*;\theta)$ and its parameter $\theta$. Thus, we consider two functions, $\pi(u^*,v^*;\theta)=(u^*v^*)^\theta$ and $\pi(u^*,v^*;\theta)=\exp\{-\theta(1-u^*)(1-v^*)\},$ each with $\theta\in[0.2,15].$ The dependence measures $\chi(r)$ and $\eta(r)$ were computed for 10 different threshold values $r$ ranging from 0.7 to 0.9998779, which is $1-(2\times\text{Machine Epsilon})^{0.25}$ in R, according to equations \eqref{eq:chir} and \eqref{eq:etar}. For small $\theta$, the weighting functions are closer to 1 at lower levels $u^*$ and $v^*,$ meaning that the tail copula dominates over a larger region, and vice versa for large $\theta.$ In general, we expect that, in the limit $r\rightarrow 1$ and with a weighting function that goes to 1 with $u^*$ and $v^*,$ the dependence properties of our model are dominated by those from the copula tailored to the tail, with similarities to the body copula for large $\theta$ and smaller $r.$ Table \ref{tab:depprop} shows the theoretical values for $\chi$ and $\eta$ for each of the copulas used in the four weighted copula models, and Figure \ref{fig:depprop} shows the outcomes of our numerical investigations for Case 3. The remaining results are shown in the Supplementary Material. For use in Table \ref{tab:depprop} and beyond, we let $\eta_t$ and $\chi_t$ represent $\eta$ and $\chi$ for the tail copula, and similarly $\eta_b$ and $\chi_b$ for the body copula.

\begin{table}[H]
\caption{Theoretical values for $\chi$ and $\eta$ for each of the copulas considered in the weighted copula models studied based on Table \ref{tab:chietacopulas}. AD denotes ``asymptotically dependent"; AI denotes ``asymptotically independent".}
    \centering
    \begin{tabular}{cllllcccc}
    Case & 
    \multicolumn{2}{c}{Body Copula $c_b$} & \multicolumn{2}{c}{Tail Copula $c_t$} & $\chi_{t}$ & $\chi_{b}$ & $\eta_{t}$ & $\eta_{b}$\\
\hline
1 & Frank (AI) & $\alpha=2$ & Gaussian (AI) & $\rho=0.6$ & 0 & 0 & 0.8 & 0.5 \\
2 & Frank (AI) & $\alpha=1$ & Gumbel (AD) & $\alpha=3$ &  0.74 & 0 & 1 & 0.5 \\
3 & Gumbel (AD) & $\alpha=1.2$ & Gaussian (AI) & $\rho=0.5$ & 0.22 & 0 & 1 & 0.75 \\
4 & Gumbel (AD) & $\alpha=2$ & H\"usler-Reiss (AD) & $\alpha=2$ & 0.62 & 0.59 & 1 & 1 \\
\hline
\end{tabular}
    \label{tab:depprop}
\end{table}

\noindent We can see from Figure \ref{fig:depprop} that, in the limit $r\rightarrow 1,$ $\chi(r)$ and $\eta(r)$ of the weighted copula model tend towards $\chi_t$ and $\eta_t$ for both weighting functions. However, the results in the Supplementary Material suggest that this does not hold true for each of the combinations we consider. Depending on the weighting function, our investigations suggest that $c_b$ has an influence on the extremal dependence properties of the model in some cases. In parti\-cular, if $c_t$ is an asymptotically dependent copula and the weighting function is $\pi(u^*,v^*;\theta)=(u^*v^*)^\theta,$ we observe that the limiting value of $\chi$ for the weighted copula model is dominated by $\chi_t$ with an influence from $\chi_b.$ For an asymptotically independent tail copula and/or the weighting function $\pi(u^*,v^*;\theta)=\exp\{-\theta(1-u^*)(1-v^*)\},$ our investigations suggest that the limiting extremal dependence properties of the model are those from $c_t.$ Moreover, the influence of the parameter $\theta$ differs since $\pi(u^*,v^*;\theta)=(u^*v^*)^\theta$ grows more slowly than $\pi(u^*,v^*;\theta)=\exp\{-\theta(1-u^*)(1-v^*)\}$ as $u^*,\,v^*\rightarrow 1.$ When $\theta$ is larger, $\chi(r)$ and $\eta(r)$ are closer to $\chi_b(r)$ and $\eta_b(r),$ particularly for smaller $r,$ where $\chi_b(r)$ and $\eta_b(r)$ are the sub-asymptotic extremal dependence measures $\chi(r)$ and $\eta(r)$ for $c_b.$

\noindent We note that this investigation suggests that there are some interesting subtleties in the tail dependence of models constructed in this way, and does not provide general conclusions. As shown theoretically for some of the considered cases, the weighted copula model has some intriguing features, such as the influence that the body copula might have when the tail component is asymptotically dependent for a given weighting function, which are worth investigating further. However, for specific cases, similar numerical or theoretical investigations can be carried out for any copulas and weighting functions of interest.
 
\section{Inference} \label{section:inference}

\subsection{Parameter estimation} \label{subsection:simulation1}

\noindent In order to estimate $\bm\gamma$, we maximise the log-likelihood function of model \eqref{eq:copmodel}, 
\begin{equation}\label{eq:loglike}
    \ell(\bm\gamma)=\sum_{i=1}^{n}\log c(u_i,v_i;\bm\gamma), \qquad u_i,v_i\in [0,1]^2,\, i=1,\ldots,n,
\end{equation} 
assuming $n$ independent observations from the copula. Because $F^{-1}_{U^*}$ and $F^{-1}_{V^*}$ are computationally expensive to obtain by a root finding algorithm, these are approximated using a smooth spline, following \citet{ZhangEtAll2020}. We found the spline approximation produces results with a similar degree of precision to the root finding algorithm, while reducing the computational time considerably. 

\noindent We conduct a simulation study to verify that inference on the proposed model produces reasonable estimates for the vector of model parameters $\bm \gamma,$ and their inherent uncertainty. To do so, we consider two examples with different sample sizes: 500 and 1000 data points. Data are sampled from density \eqref{eq:model} via the sampling procedure outlined in Section \ref{subsection:simulation}.

\noindent For the first case, we take $c_b$ to be the Clayton copula density with $\alpha=1,$ and $c_t$ to be the Gumbel copula density with $\alpha=2.$ For the second example, $c_b$ is taken as the Joe copula density with $\alpha=2$ and $c_t$ is the Gaussian copula density with $\rho=0.6.$ The parameter of the weighting function is set to be $\theta=0.8$ in the first example and $\theta=1$ in the second case. Each data set is simulated 100 times.

\begin{figure}[H]
    \centering
    \begin{subfigure}[b]{0.45\textwidth}
        \includegraphics[width=\textwidth]{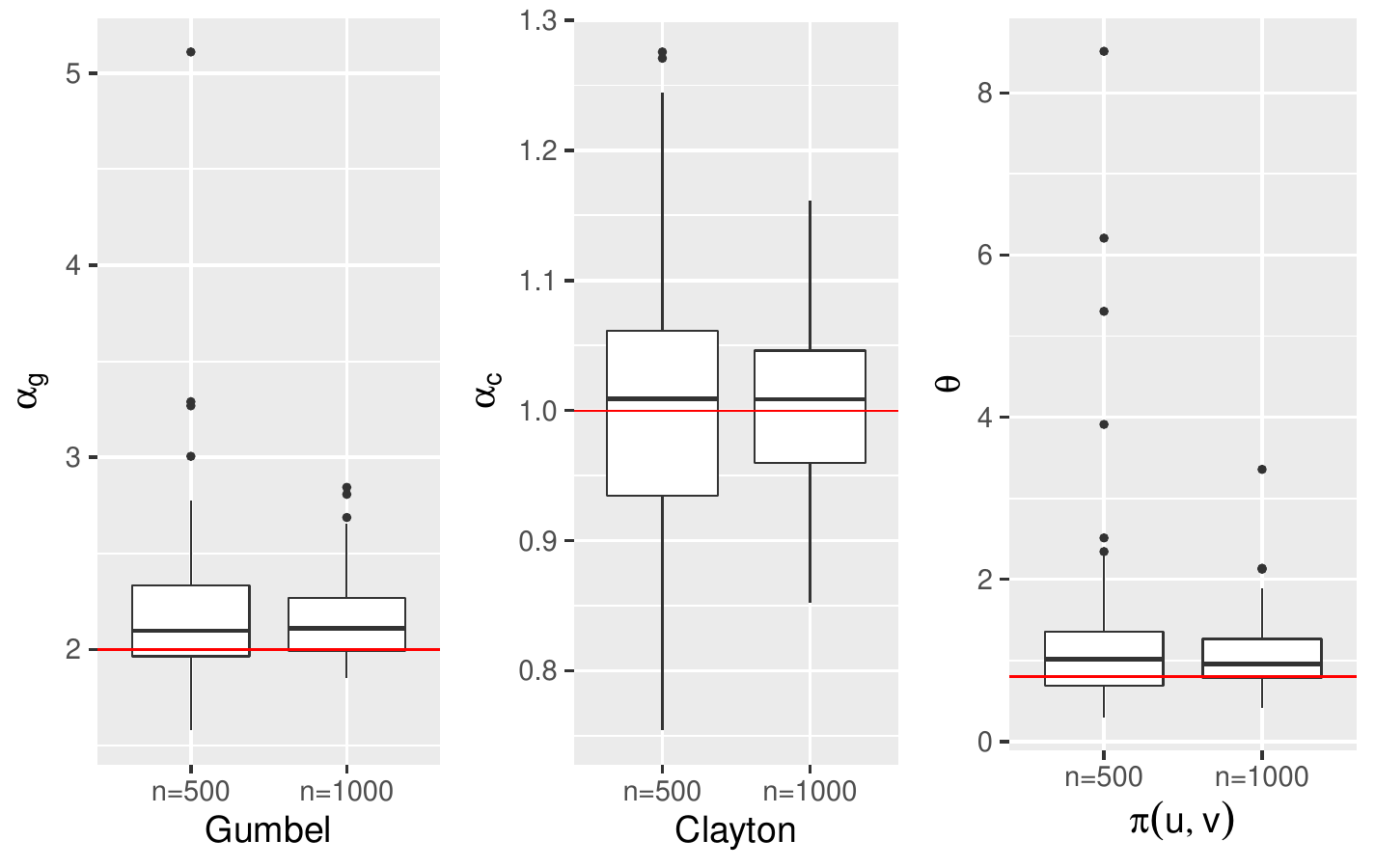}
        \caption{Parameter estimates of $c_t$ (left plot), $c_b$ (middle plot), and the weighting parameter $\theta$ (right plot) for $n=500$ and $n=1000$. The true values for the parameters are shown in red.}
        \label{subfig:partAmod2}
    \end{subfigure}
    \hfill
    \begin{subfigure}[b]{0.45\textwidth}
        \includegraphics[width=\textwidth]{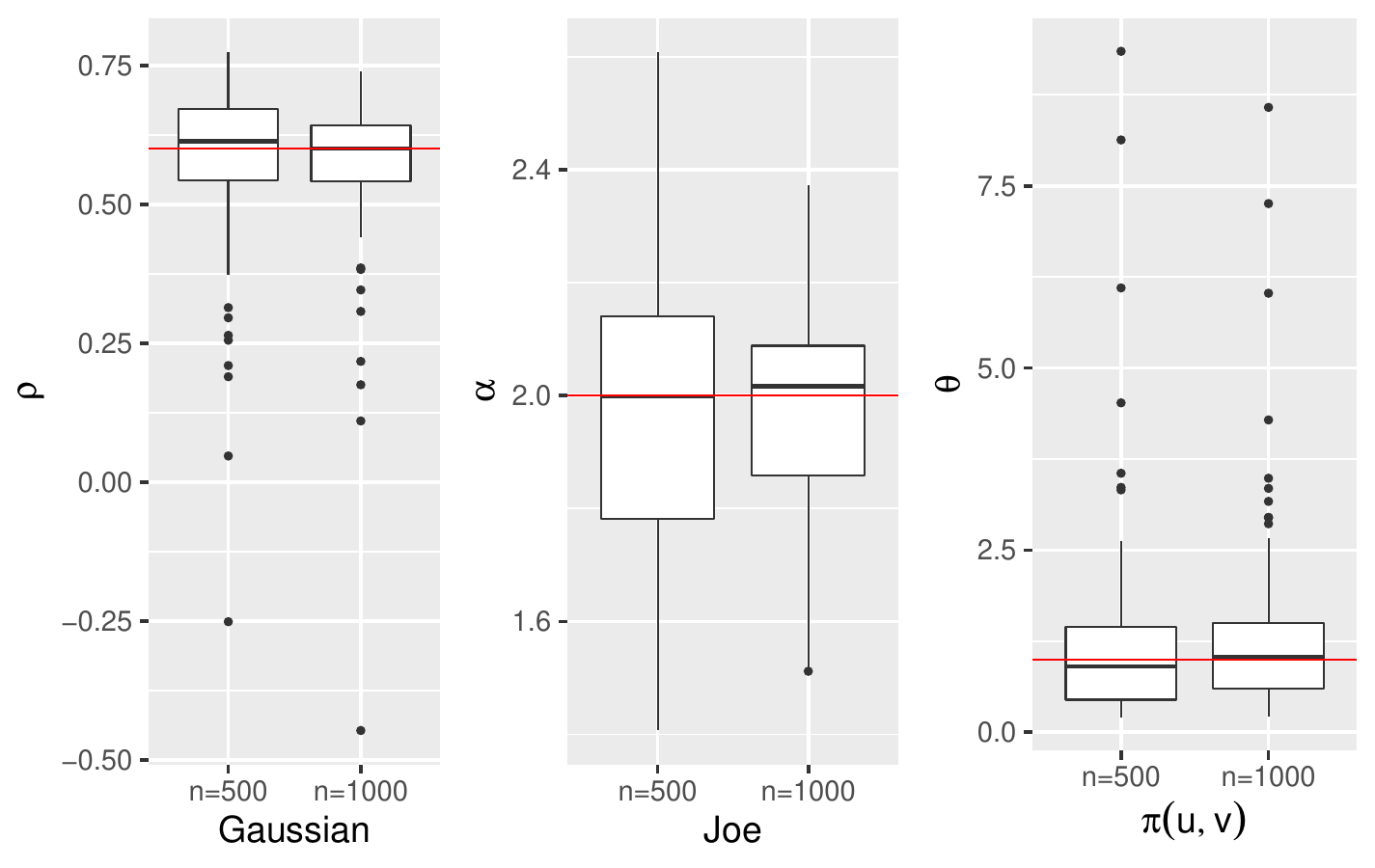}
        \caption{Parameter estimates of $c_t$ (left plot), $c_b$ (middle plot), and the weighting parameter $\theta$ (right plot) for $n=500$ and $n=1000$. The true values for the parameters are shown in red.}
        \label{subfig:partAmod1}
    \end{subfigure}
    \caption{Estimation variability obtained by simulating each case 100 times.}
    \label{fig:boxplotpartA}
\end{figure}

\noindent Figure \ref{fig:boxplotpartA} displays the results of the simulation study. For each parameter, the left boxplot shows the spread of estimates when $n=500,$ and the right boxplot displays this for $n=1000.$ We observe that estimation seems generally unbiased and uncertainty reduces when the sample size increases. 

\begin{figure}[H]
    \centering
    \begin{subfigure}[b]{0.45\textwidth}
        \includegraphics[width=\textwidth]{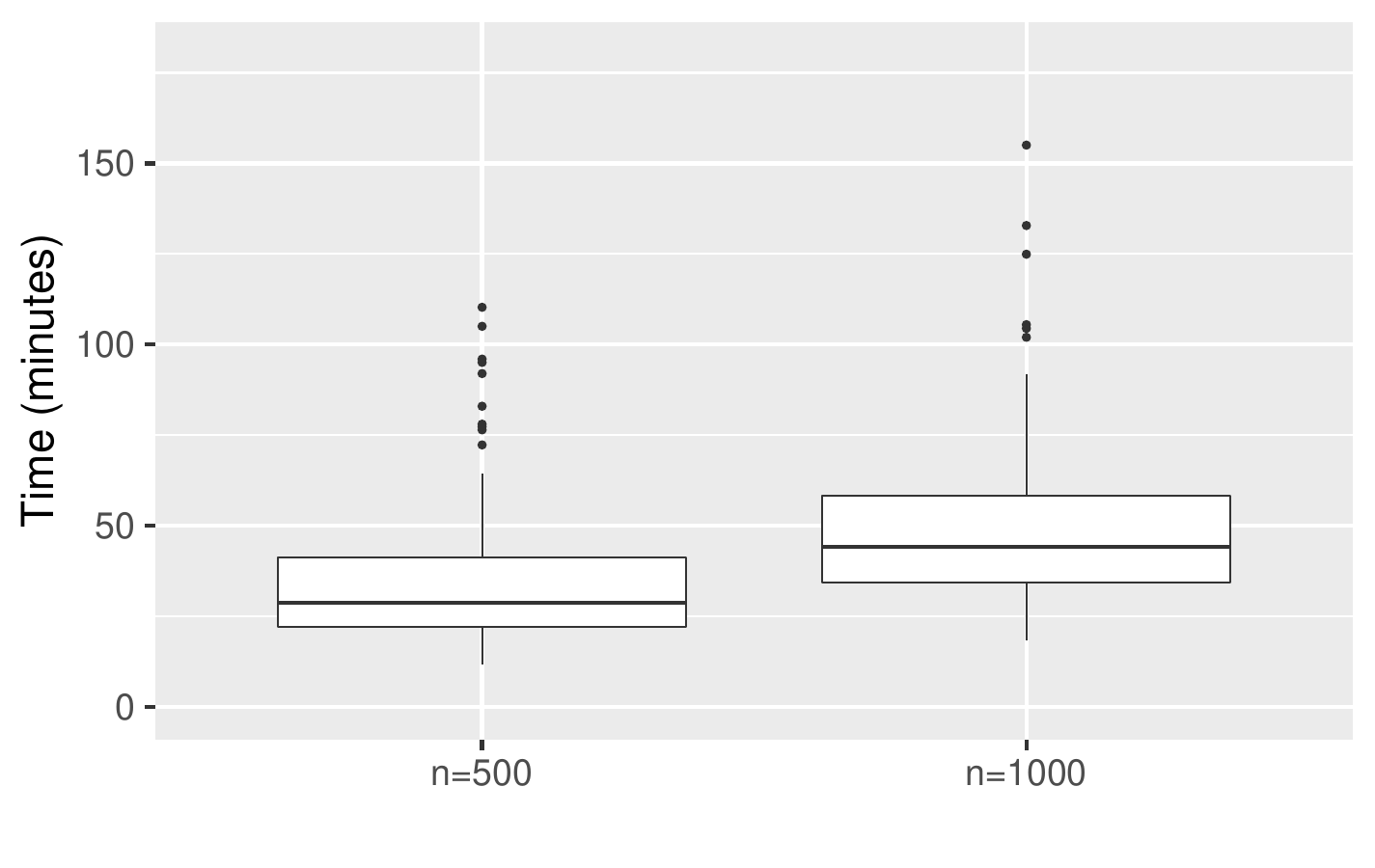}
        \caption{Case when $c_t$ is Gumbel and $c_b$ is Clayton for $n=500$ (left) and $n=1000$ (right).}
        \label{subfig:partAmod2time}
    \end{subfigure}
    \hfill
    \begin{subfigure}[b]{0.45\textwidth}
        \includegraphics[width=\textwidth]{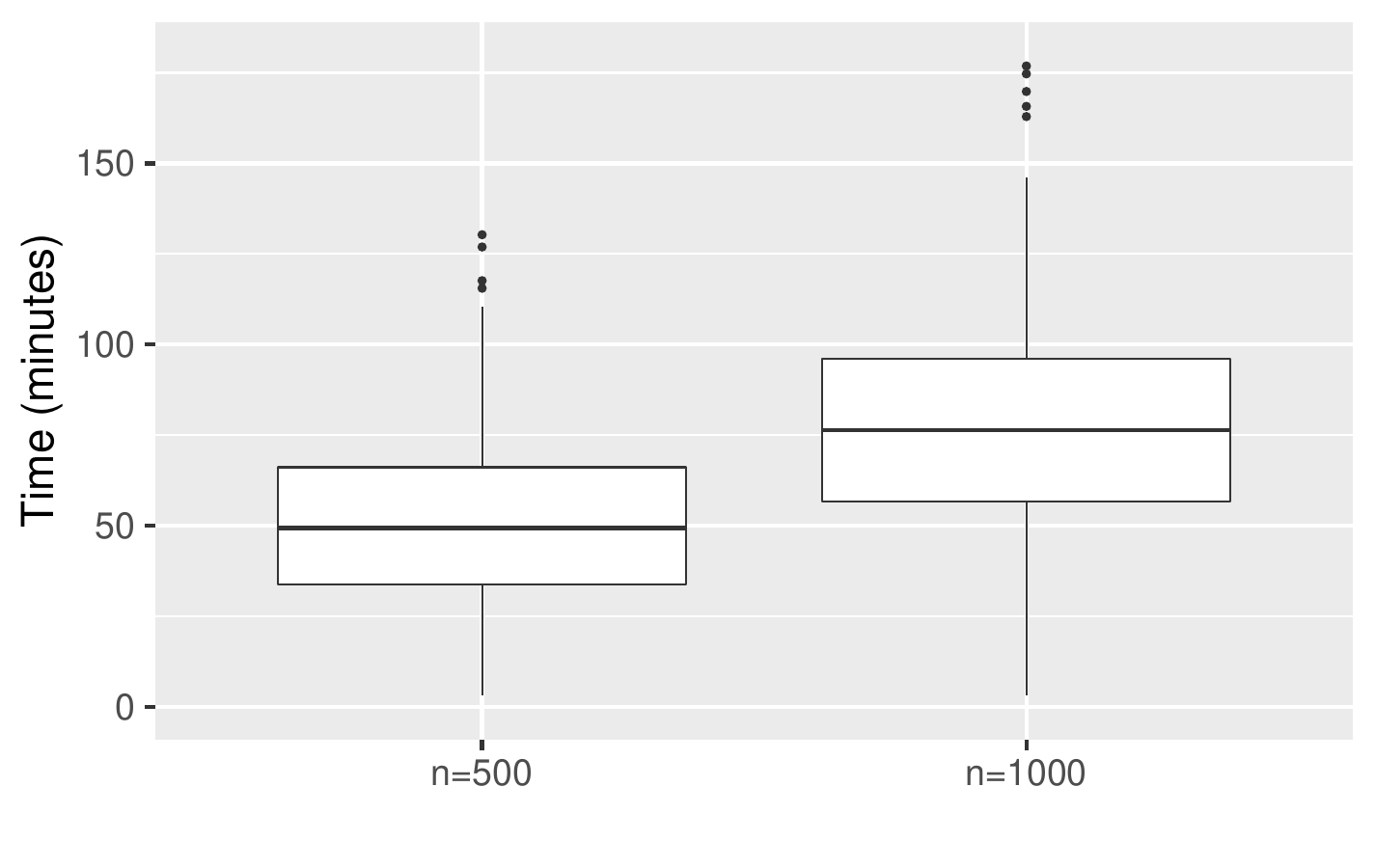}
        \caption{Case when $c_t$ is Gaussian and $c_b$ is Joe for $n=500$ (left) and $n=1000$ (right).}
        \label{subfig:partAmod1time}
    \end{subfigure}
    \caption{Time (minutes) taken to optimise the log-likelihood \eqref{eq:loglike} for each simulation.}
    \label{fig:timepartA}
\end{figure}

\noindent Because the copula density \eqref{eq:copmodel} relies on numerical integration to obtain $F,$ $f$ and $F^{-1}$, it is important to assess the computational effort required to perform inference. Figure~\ref{fig:timepartA} displays the time taken to optimise the likelihoods on an internal computing node running CentOS Linux, with an Intel CPU running at 500GB of RAM. We can see that, for each of the models, the time taken increases with the sample size, which is to be expected. It also varies with the chosen copulas; for example, to evaluate the likelihood with $n=500$ data points, the first model took around 30 minutes while the second took around 50 minutes.

\subsection{Model misspecification}\label{subsection:simulation2}

\noindent In addition to checking if inference on the model produces reasonable estimates for $\bm\gamma$, we study the ability of the model to capture a misspecified dependence structure. We consider two situations: the case where the underlying data set comes from a single copula and we fit our model with this copula as one of the components; and the case where the fitted model does not contain the true copula. In the first case, we investigate whether the estimate of the parameter of the weighting function $\theta$ agrees with the true data. Since $\pi(u^*,v^*;\theta)$ is increasing in $(0,1),$ we expect $\hat\theta$ to be large (small) when the true copula is tailored to the body (tail) of the distribution. In the second case, we investigate whether our model still produces reliable estimates of various dependence summaries even though the true dependence structure cannot be captured.

\begin{figure}[H]
    \centering
    \begin{subfigure}[b]{0.45\textwidth}
        \includegraphics[width=\textwidth]{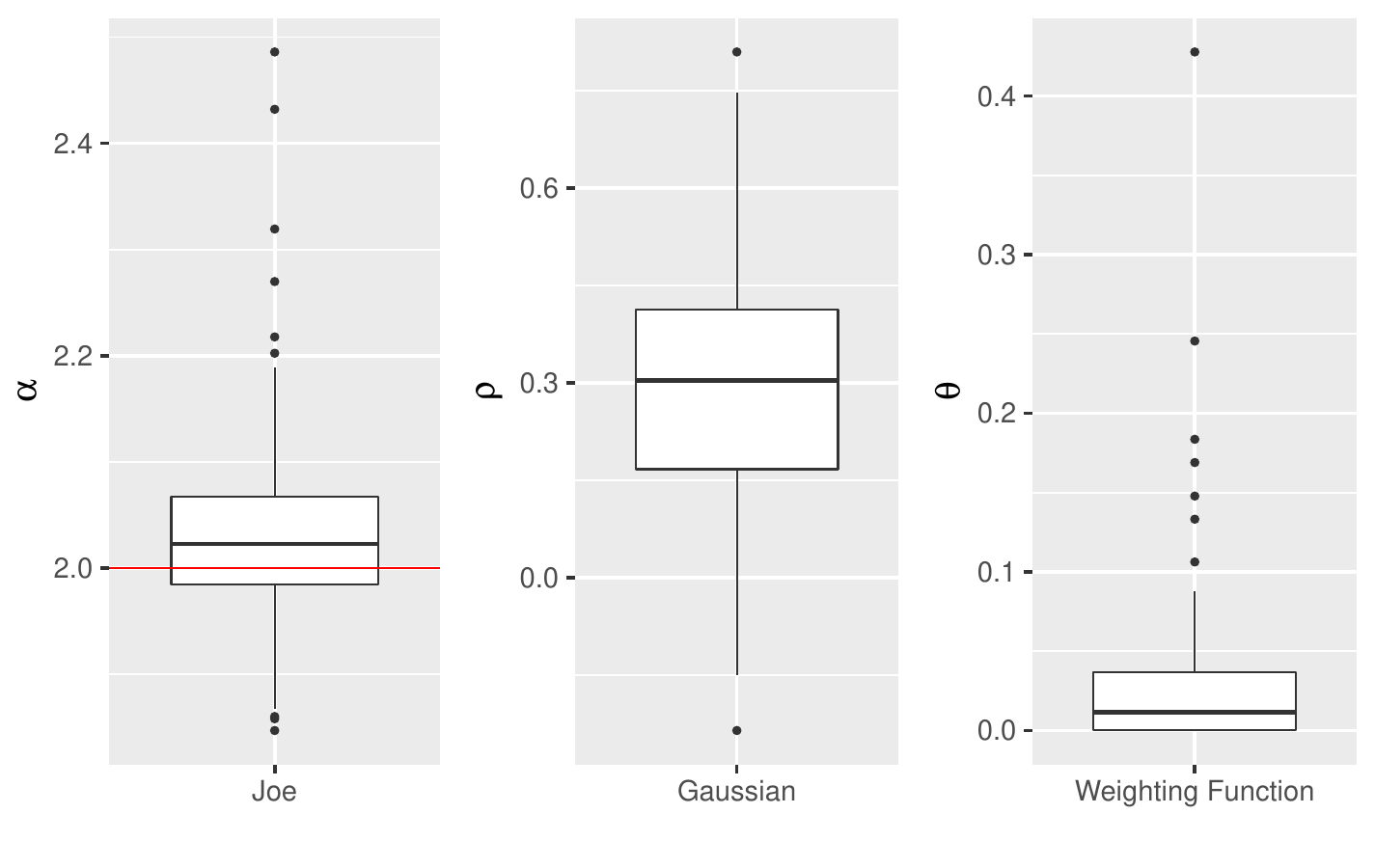}
        \caption{Parameter estimates when $c_t$ is taken as the true copula (left) and $c_b$ is taken as the Gaussian copula (middle). The true value for the parameter is shown in red. Estimates of $\theta$ are shown in the right boxplot.}
        \label{subfig:partBmod1}
    \end{subfigure}
    \hfill
    \begin{subfigure}[b]{0.45\textwidth}
        \includegraphics[width=\textwidth]{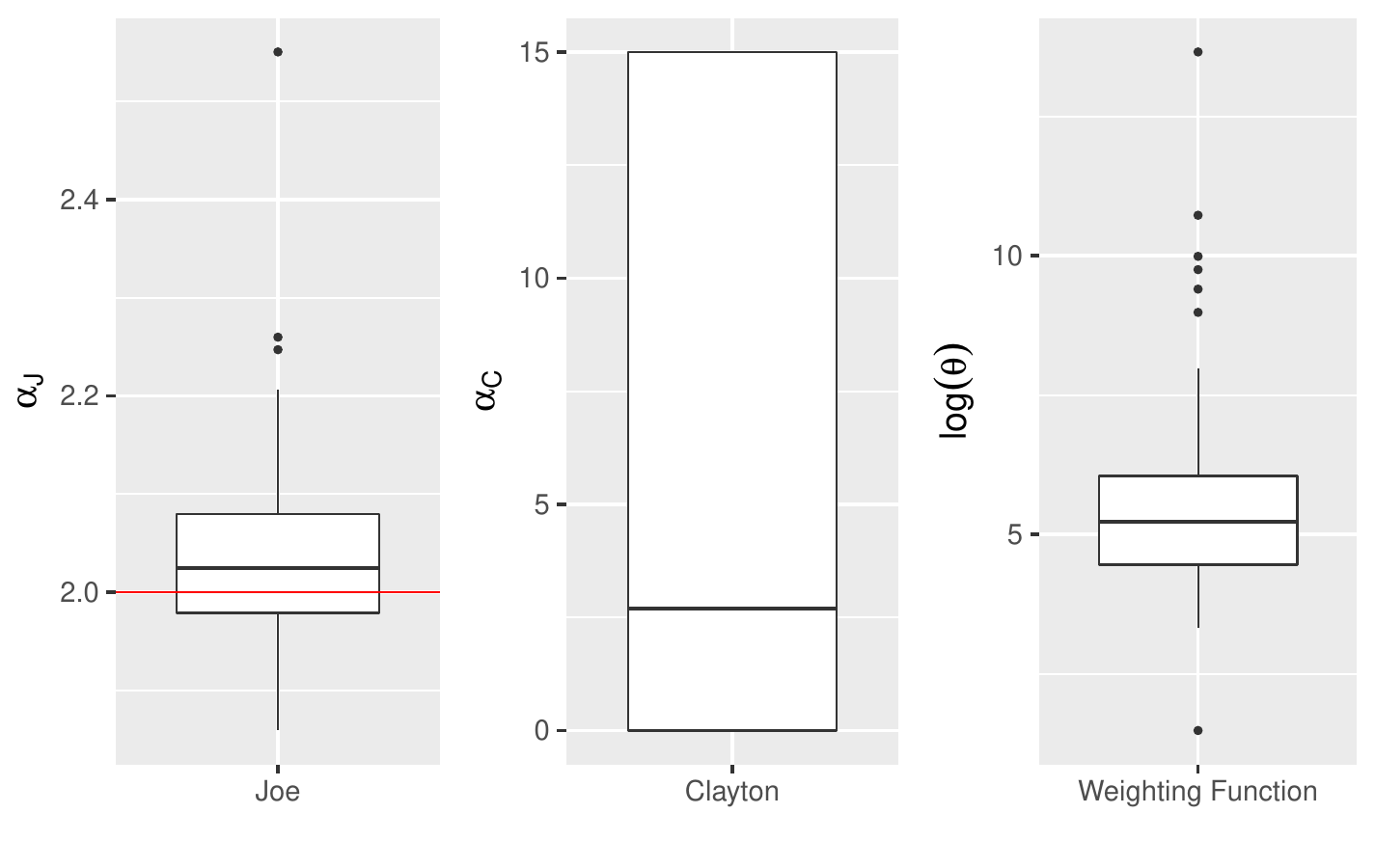}
        \caption{Parameter estimates when $c_b$ is taken as the true copula (left) and $c_t$ is taken as the Clayton copula (middle). The true value for the parameter is shown in red. Estimates of $\log(\theta)$ are shown in the right boxplot.}
        \label{subfig:partBmod2}
    \end{subfigure}
    \caption{Estimation variability obtained by simulating each case 100 times.}
    \label{fig:boxplotpartB}
\end{figure}

\noindent For the first case, we generate 1000 data points from a Joe copula with $\alpha=2$ and fit two weighted copula models: one with the true copula as $c_t$ and a Gaussian copula as $c_b,$ and the other with the true copula as $c_b$ and a Clayton copula as $c_t.$ As before, 100 simulations for each case were performed and the results are shown in the boxplots in Figure \ref{fig:boxplotpartB}.

\begin{figure}[H]
    \centering
    \begin{subfigure}[b]{0.45\textwidth}
        \includegraphics[width=\textwidth]{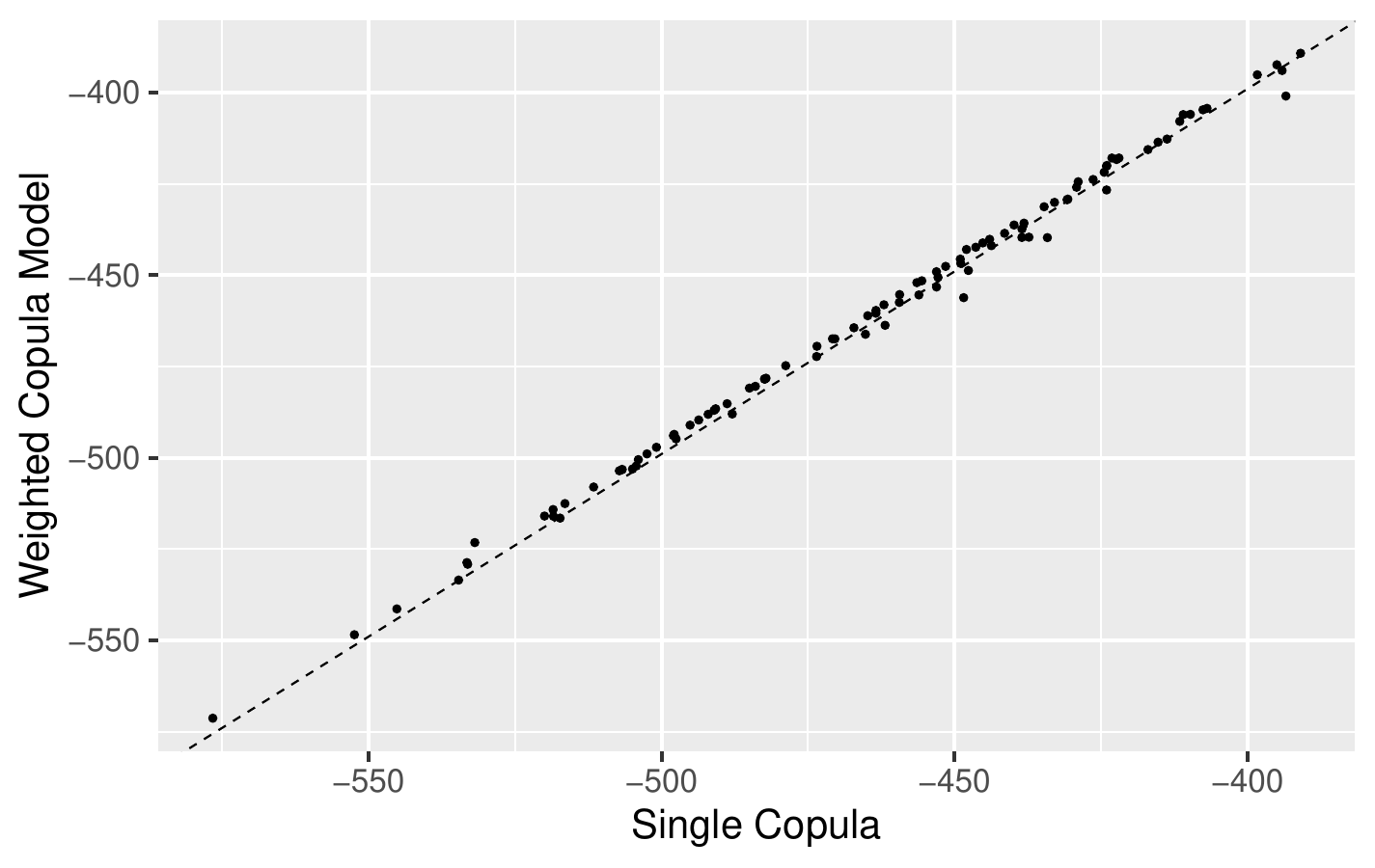}
        \caption{Case with the true copula as $c_t$ and a Gaussian copula as $c_b$.}
        \label{subfig:partBmod1aic}
    \end{subfigure}
    \hfill
    \begin{subfigure}[b]{0.45\textwidth}
        \includegraphics[width=\textwidth]{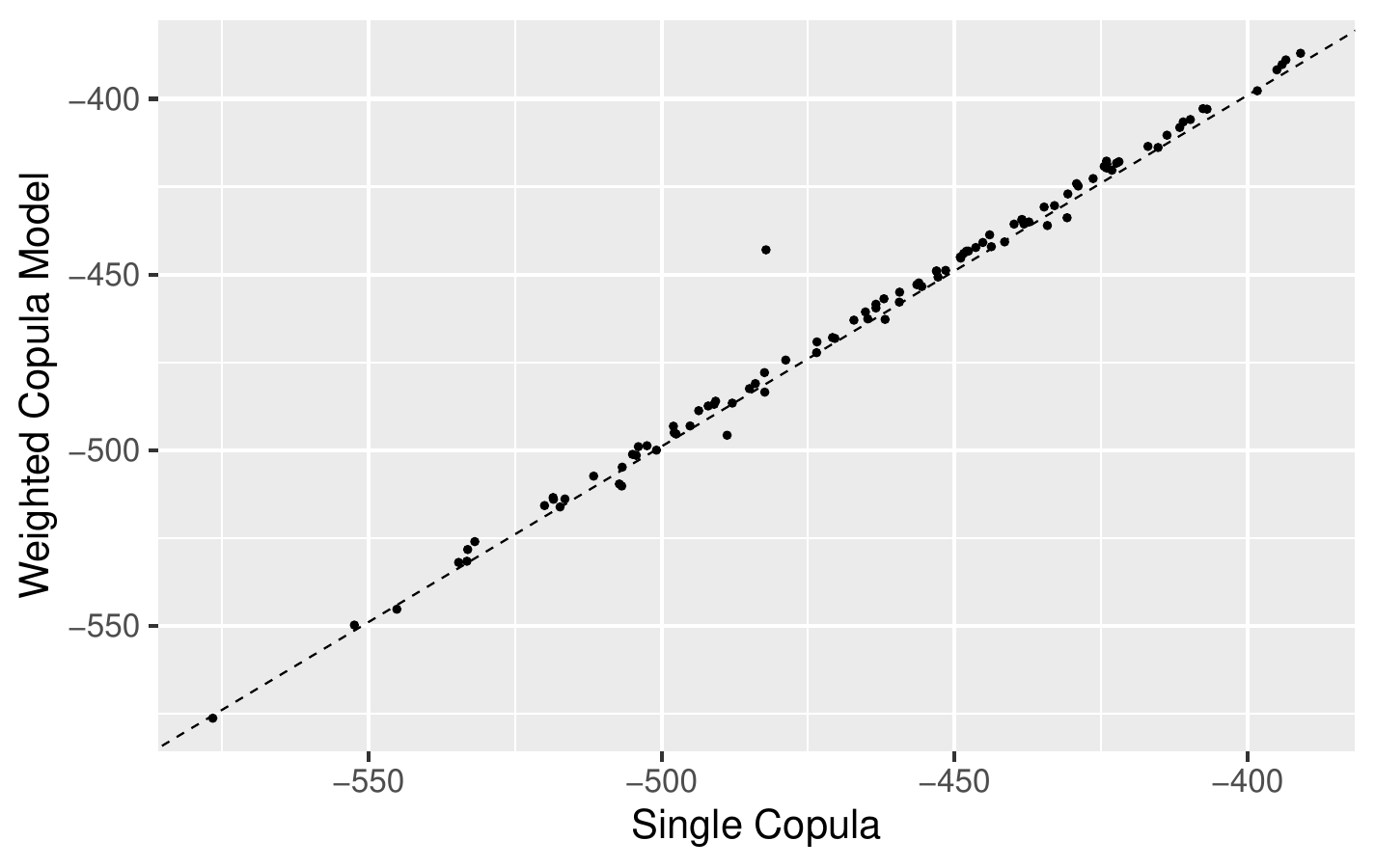}
        \caption{Case with true copula as $c_b$ and a Clayton copula as $c_t$.}
        \label{subfig:partBmod2aic}
    \end{subfigure}
    \caption{Comparison between the AIC of the true model and the fitted model.}
    \label{fig:aicpartB}
\end{figure}

\noindent We observe that, when the Joe copula is taken as $c_t,$ the estimates for $\theta$ are all less than 1, and when it is taken as $c_b,$ these are considerably larger (here we use the logarithm of $\theta$ for ease of visualisation). Looking at the estimates for the parameter of the true copula, although they show some bias, they are fairly close to the true values, represented by the red lines. Finally, the estimates for the parameters of the misspecified copula show larger variability, which is to be expected as most of the weight is on the true copula. Figure \ref{fig:aicpartB} shows a comparison between the AIC of the true and weighted copula models, respectively. In the majority of cases (89$\%$ for the first and 92$\%$ for the second), the true model outperforms the weighted copula model in terms of AIC, as expected.

\begin{figure}[H]
    \centering
    \begin{subfigure}[b]{0.45\textwidth}
        \includegraphics[width=\textwidth]{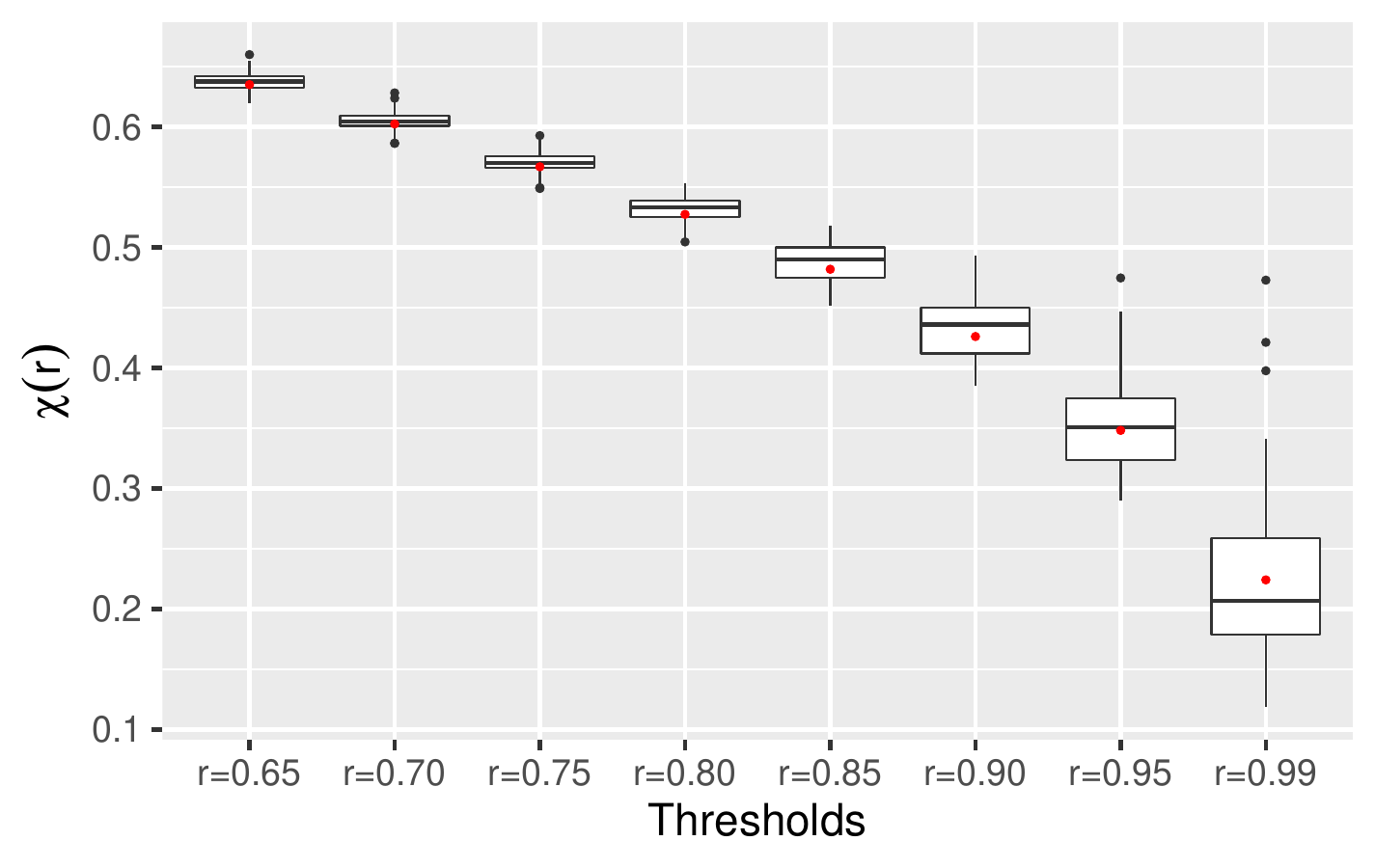}
        \label{subfig:partBmod1chi}
    \end{subfigure}
    \hfill
    \begin{subfigure}[b]{0.45\textwidth}
        \includegraphics[width=\textwidth]{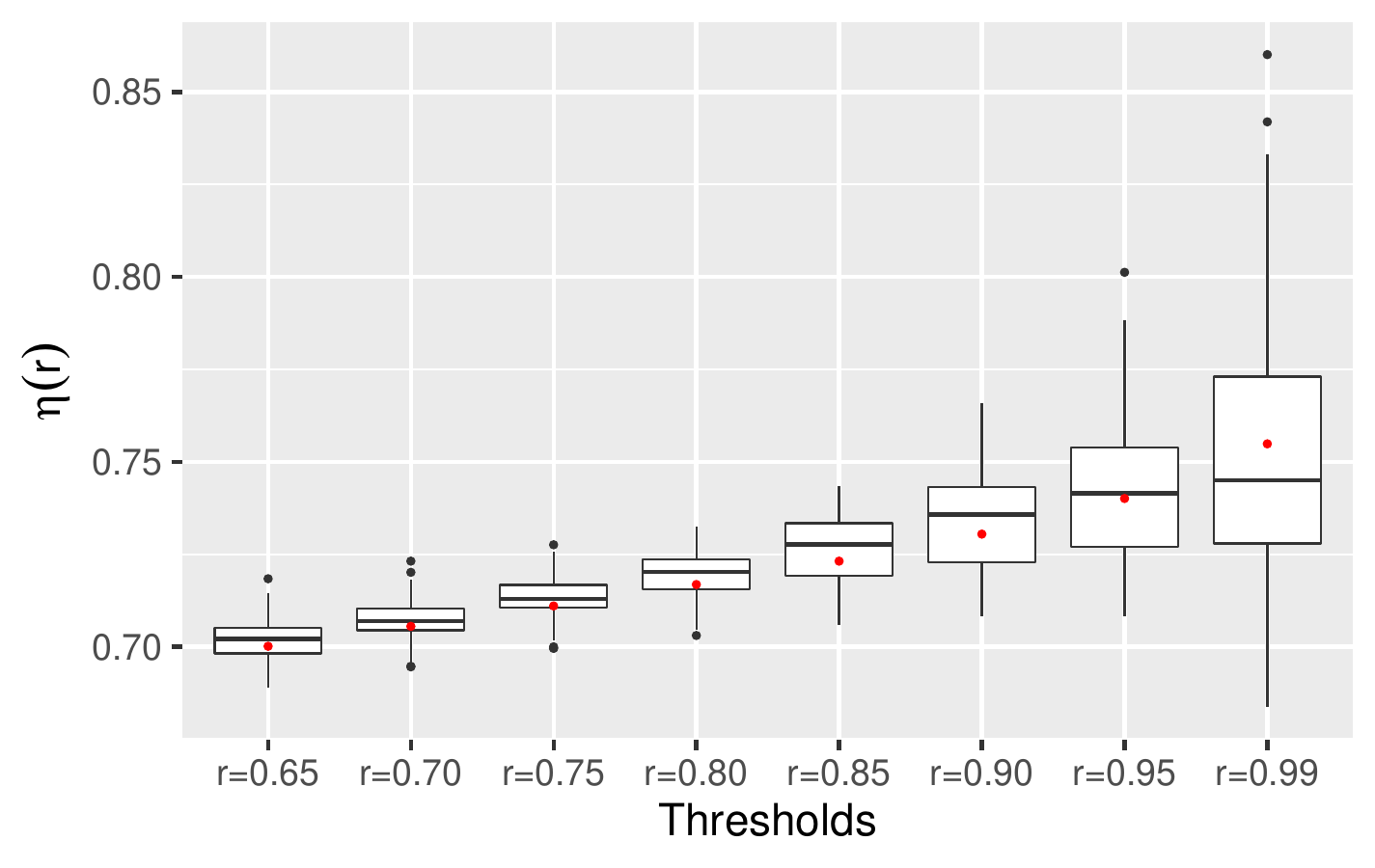}
        \label{subfig:partBmod1eta}
    \end{subfigure}
    \begin{subfigure}[b]{0.45\textwidth}
        \includegraphics[width=\textwidth]{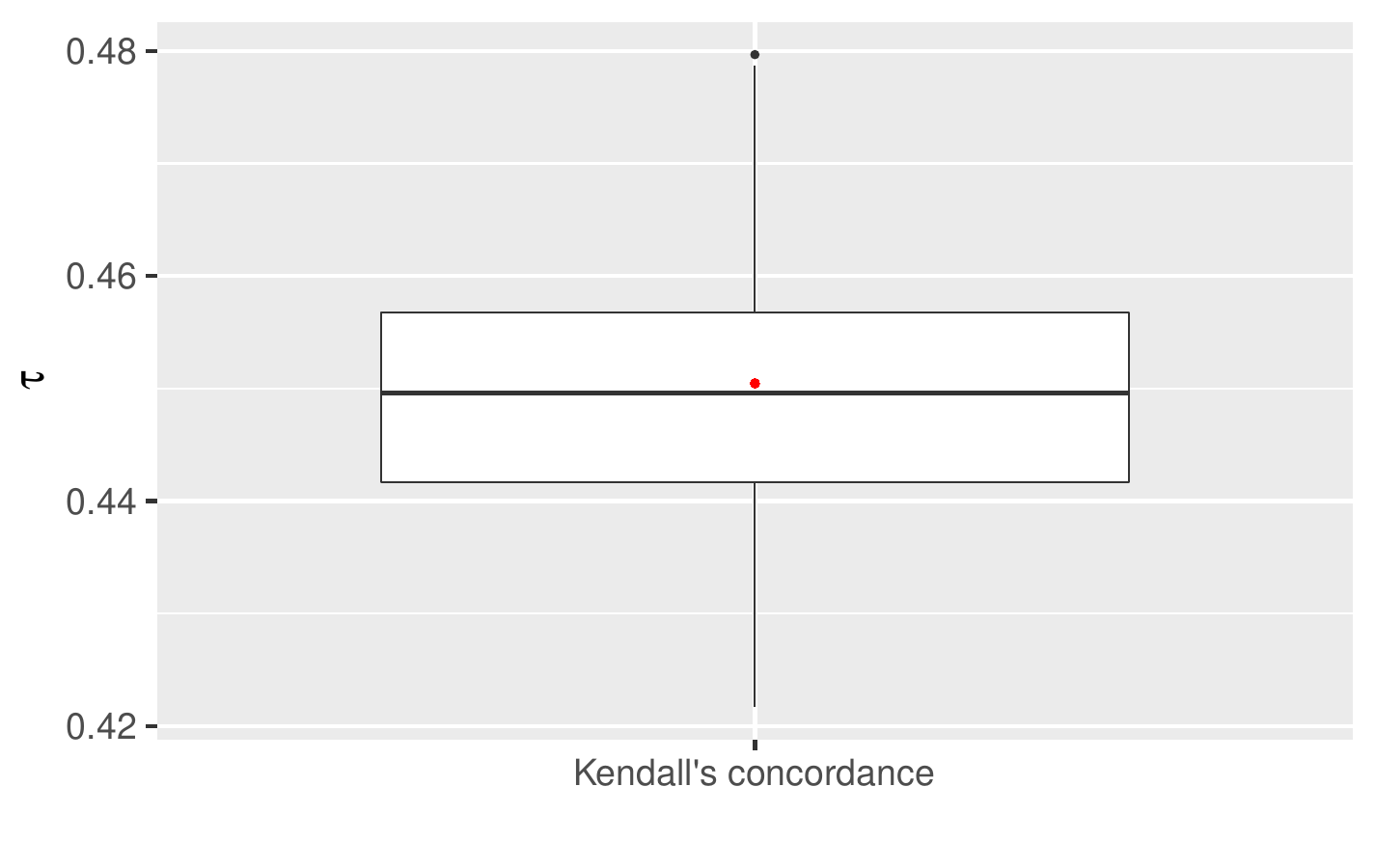}
        \label{subfig:partBmod1tau}
    \end{subfigure}
    \caption{Model and theoretical (in red) $\chi(r)$ (top left) and $\eta(r)$ (top right) at levels \linebreak$r\in \{0.65,0.7,0.75,0.8,0.85,0.9,0.95,0.99\},$ and Kendall's $\tau$ (bottom) for the selected model when the true model is Gaussian with $\rho=0.65$.}
    \label{fig:partBstats}
\end{figure}

\begin{figure}[H]
    \centering
    \begin{subfigure}[b]{0.45\textwidth}
        \includegraphics[width=\textwidth]{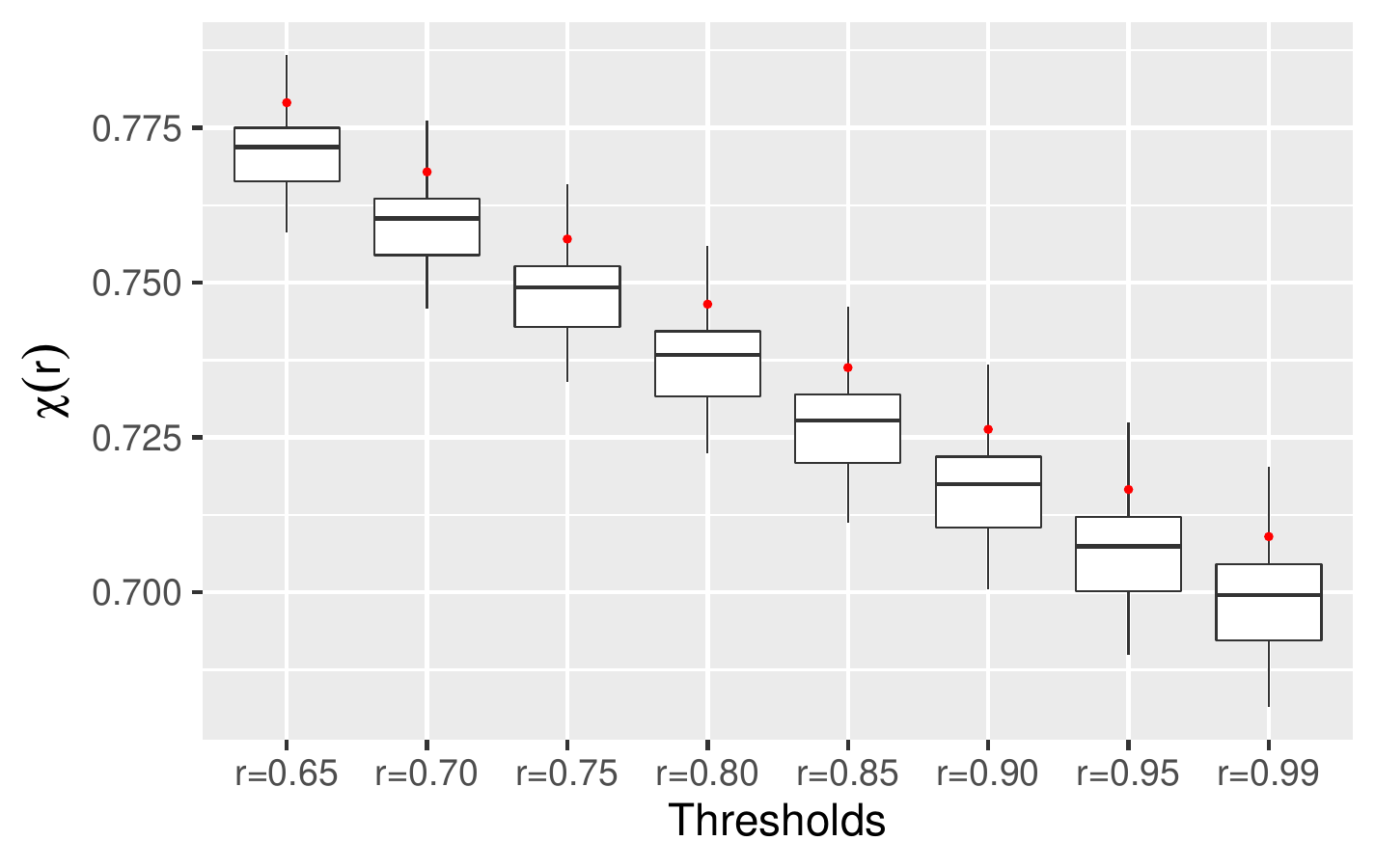}
        \label{subfig:partBmod3chi}
    \end{subfigure}
    \hfill
    \begin{subfigure}[b]{0.45\textwidth}
        \includegraphics[width=\textwidth]{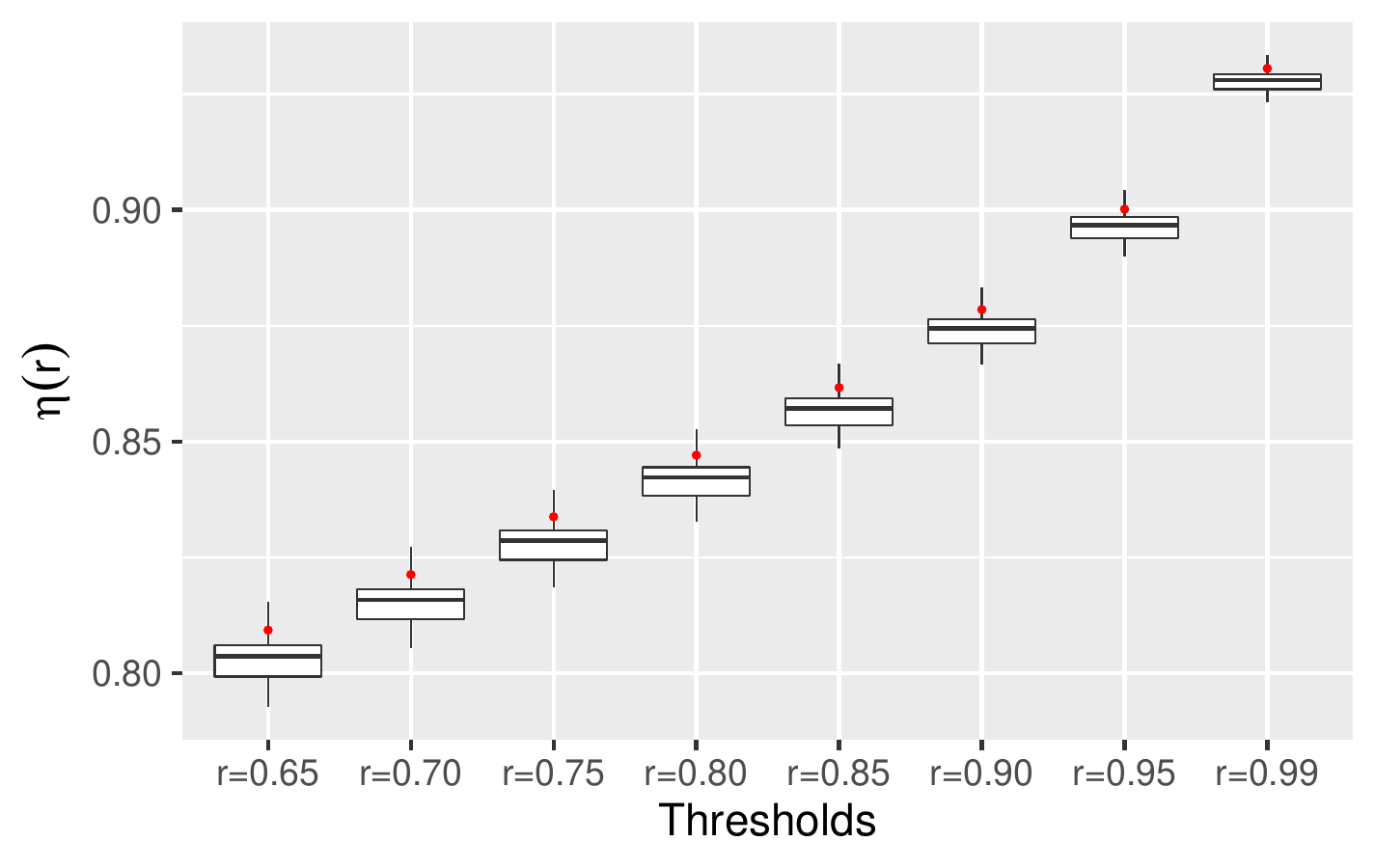}
        \label{subfig:partBmod3eta}
    \end{subfigure}
    \begin{subfigure}[b]{0.45\textwidth}
        \includegraphics[width=\textwidth]{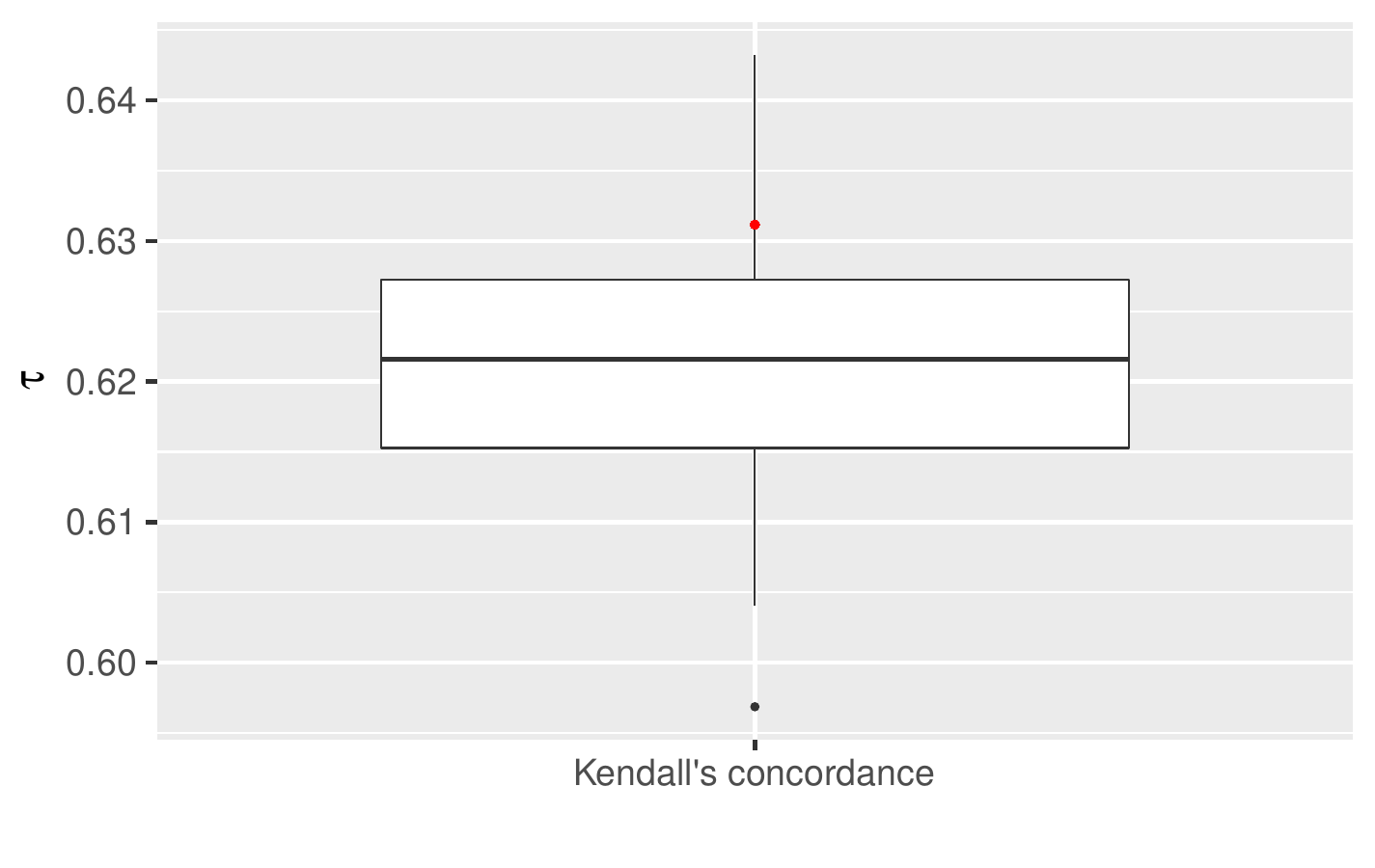}
        \label{subfig:partBmod3tau}
    \end{subfigure}
    \caption{Model and theoretical (in red) $\chi(r)$ (top left) and $\eta(r)$ (top right) at levels \linebreak$r\in \{0.65,0.7,0.75,0.8,0.85,0.9,0.95,0.99\},$ and Kendall's $\tau$ (bottom) for the selected model when the true model is Galambos with $\alpha=2$.}
    \label{fig:partBstats1}
\end{figure}

\noindent For our second experiment, to evaluate the outcome of not being able to capture the true dependence structure, we simulate 1000 data points from a Gaussian copula with $\rho=0.65$ and from a Galambos copula with $\alpha=2.$ For both cases, we generate $50$ repetitions of the data set and fit a variety of weighted copula models, selecting the best model based on the average AIC values. In order to assess if the selected weighted copula model is flexible enough to capture the dependence of the true data sets, we compute three measures of dependence: Kendall's $\tau,$ and $\chi(r)$ and $\eta(r)$ from equations \eqref{eq:chir} and \eqref{eq:etar}, respectively, at several thresholds $r\in(0,1).$ We show how the model performs by comparing with the theoretical values of the underlying models; the results are shown in Figures \ref{fig:partBstats} and \ref{fig:partBstats1}.

\noindent Figure \ref{fig:partBstats} displays the results for the weighted copula model where $c_t$ is inverted Gumbel, $c_b$ is Student t, and the true underlying structure is Gaussian. The results for the second model where the true underlying structure is Galambos and the selected weighted copula model is Coles-Tawn as $c_t$ and Frank as $c_b$ are shown in Figure \ref{fig:partBstats1}. In both cases, we observe that the misspecified models capture the three dependence measures fairly well.

\section{Case study: ozone and temperature data} \label{section:casestudy}

\subsection{Data and background}\label{subsection:data}

\noindent The relationship between ozone concentration and temperature has been analysed previously in the literature. For instance, \citet{FinchPalmer2020} show that there is an increase of exceeding regulated thresholds for ozone when the temperature is high. More recently, \cite{GouldsbroughEtAll2022} study how extreme levels of ozone concentration are influenced by temperature in the UK by applying a temperature-dependent univariate extreme value model. They show that, with the increase in temperatures, the probability of exceeding a moderate regulated threshold of ozone concentration has increased over the last decade; this leads to this event no longer being considered extreme. The analysis of \citet{GouldsbroughEtAll2022} only considers the univariate distribution of ozone extremes conditional upon the value of temperature. Since both temperature and ozone concentration are measurements of random variables, we can apply our weighted copula model to learn about the relationship between these variables at all levels. Specifically, we study the dependence between temperature and ozone concentration at two UK sites: Blackpool (urban background) and Weybourne (rural background). Table \ref{tab:DAQI} shows the regulated threshold indexes for the levels of air pollution for Ozone in the UK.

\begin{table}[H]
    \caption{Daily Air Quality Index (DAQI) for ozone ($O_3$) concentrations in the UK.}
    \centering
    \begin{tabular}{ccccc}
        Levels & Low & Moderate & High & Very High\\
        \hline 
          $O_3 \,(\mu g\slash m^3)$ & $[0,100]$ & $[101,160]$ & $[161,240]$ & $>240$ \\ 
    \end{tabular}
    \label{tab:DAQI}
\end{table}

\begin{figure}[H]
    \centering
    \begin{subfigure}[b]{0.45\textwidth}
        \includegraphics[width=\textwidth]{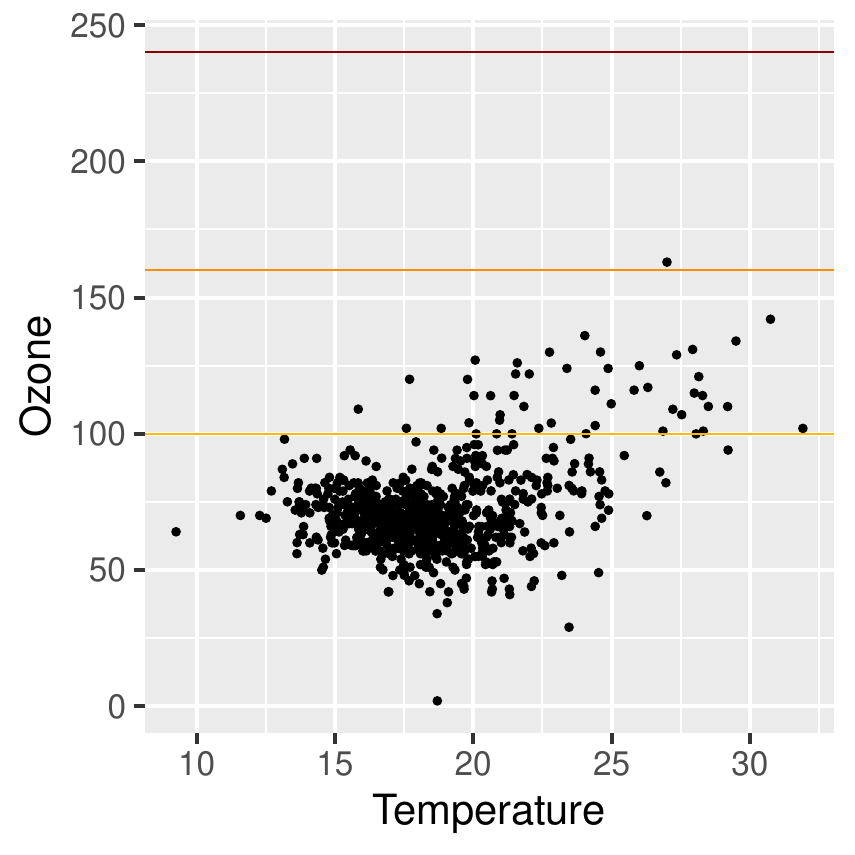}
        \caption{Daily maxima of temperature and ozone. The moderate, high and very high DAQI are represented by the yellow, orange and red lines, respectively.$\qquad\qquad\qquad$ \linebreak  $\qquad \qquad \qquad \qquad\qquad\qquad$}
        \label{subfig:blackpoolscatter}
    \end{subfigure}
    \hfill
    \begin{subfigure}[b]{0.45\textwidth}
        \includegraphics[width=\textwidth]{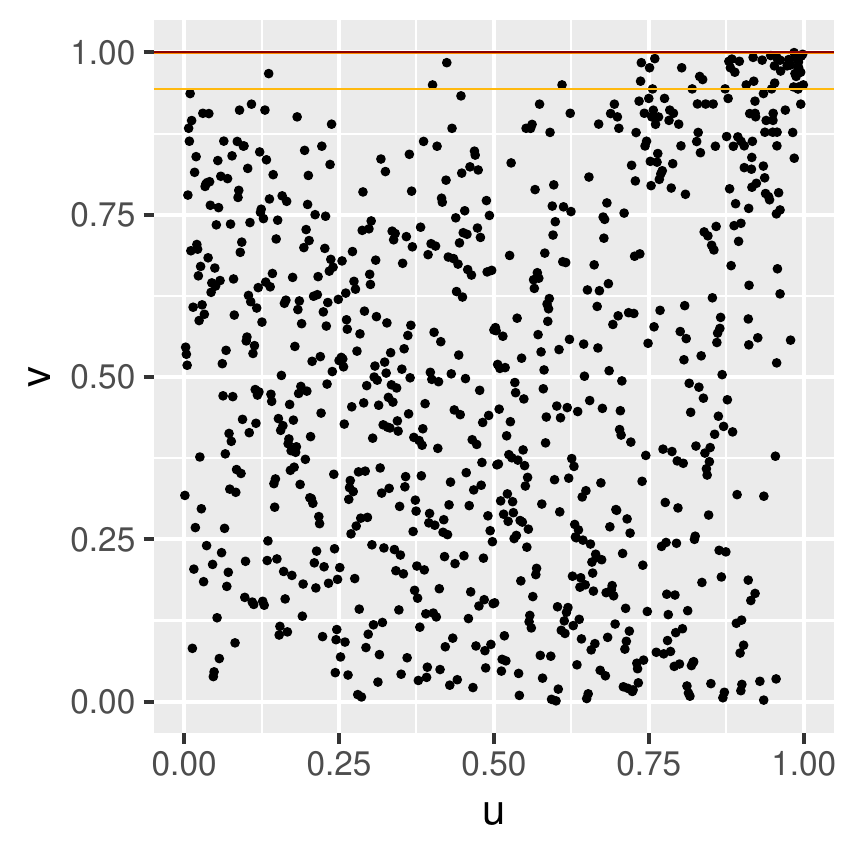}
        \caption{Daily maxima of temperature (u) and ozone (v) on uniform margins. The corresponding moderate, high and very high DAQI are represented by the yellow, orange and red lines, respectively.}
        \label{subfig:blackpooluniform}
    \end{subfigure}
    \caption{Summer data from 2011 to 2019 for Blackpool, UK.}
    \label{fig:blackpooldataset}
\end{figure}

\noindent We took the daily maxima from 8-hour running means ozone concentration available on the UK's Automatic Urban and Rural Network (AURN) (\url{https://uk-air.defra.gov.uk}) and obtain the corresponding daily maximum temperature data from the Centre for Environmental Analysis (CEDA) archive (\url{https://archive.ceda.ac.uk}). Since higher temperatures are expected during summer, and in order to overcome the non-stationarity often present in temperature data, we restrict our analysis to the summer months (June-August). Based on the available data, we consider the years from 2011 to 2019 for Blackpool and from 2010 to 2019 for Weybourne; this results in 827 and 892 observations, respectively. Figure \ref{subfig:blackpoolscatter} shows the scatterplot of the daily maxima of temperature and the daily maxima of ozone for the summers of 2011 to 2019 in Blackpool and the res\-pective regulated UK thresholds, while Figure \ref{subfig:blackpooluniform} shows the relationship between the variables when transformed to uniform margins using a semi-parametric approach with a GPD fit to the tail of both distributions. That is, we estimate the CDF of each marginal distribution via 
\begin{equation}
    F(x)=\begin{cases}
        \widetilde{F}(x), & x\leq r, \\
        1-\phi_r\left[1+\mfrac{\xi(x-r)}{\sigma}\right]_+^{-1\slash \xi}, & x>r,
    \end{cases}
\end{equation}
where $\widetilde{F}(x)$ is the empirical distribution function, $\phi_r$ is the probability of exceeding a selected high threshold $r,$ and $\xi$ and $\sigma$ are the GPD shape and scale parameters, res\-pectively. The corresponding analysis for Weybourne is presented in the Supplementary Material; the results show similar conclusions to the analysis for Blackpool. 

\subsection{Model fitting} \label{subsection:fit}

\begin{figure}[H]
    \centering
    \includegraphics[width=0.9\textwidth]{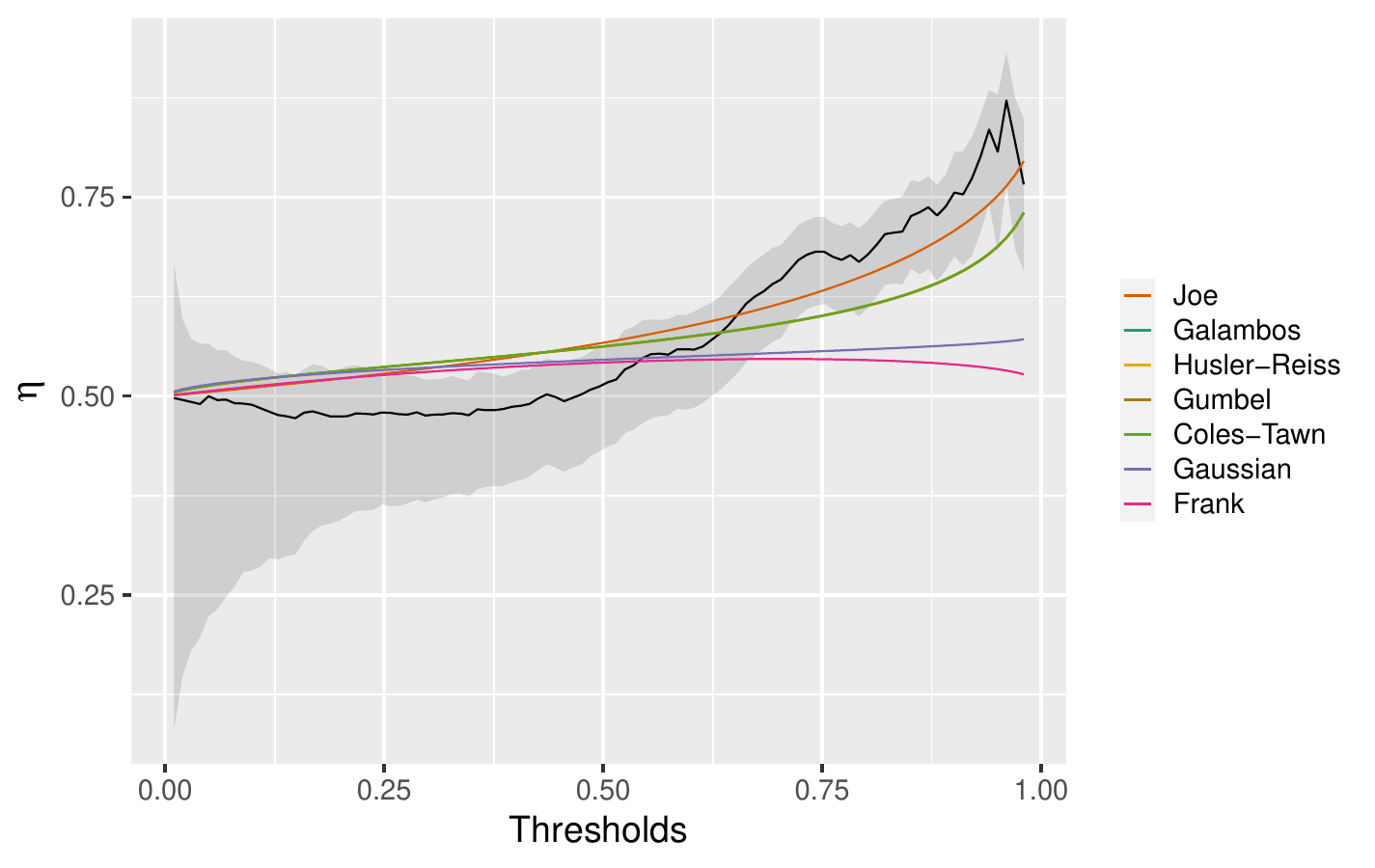}
        \caption{Empirical $\eta(r)$ (in black) and $\eta(r)$ for seven copulas (in colour) for $r\in(0,1).$ The 95$\%$ confidence bands were obtained by block bootstrapping. Note that the $\eta(r)$ for the Galambos, the H\"usler-Reiss, the Gumbel and the Coles-Tawn copulas overlap.}
    \label{fig:etasingle}
\end{figure}

\noindent We start by fitting a single copula model to the whole data set for comparison with the weighted copula model. Looking at Figure \ref{subfig:blackpooluniform}, the variables seem to exhibit positive correlation when they are both extreme, but negative dependence otherwise. We anti\-ci\-pate that the weighted copula model may be flexible enough to capture this, whereas a single copula is likely to be too rigid. Table \ref{tab:mlesingle} shows the MLEs obtained by fitting a range of copulas and the corresponding AIC values. From the copulas considered, the only ones capable of capturing negative dependence are the Gaussian and Frank, when their parameters are negative, and the Student t (which also exhibits lower and upper tail dependence). However, all parameter estimates are positive. In terms of AIC, the best fit is the Joe, followed by the Galambos, H\"usler-Reiss, Gumbel and Coles-Tawn copulas; these are all known to be asymptotically dependent copulas, which appears to agree with the dependence in the upper tail shown in Figure \ref{subfig:blackpooluniform}. As a further diagnostic, we compute the dependence measure $\eta(r)$ from equation \eqref{eq:etar} for $r\in (0,1)$ empirically, as well as for the five best models in terms of AIC, and for the Gaussian and Frank copulas; this is shown in Figure~\ref{fig:etasingle}. The confidence intervals in Figure \ref{fig:etasingle} were obtained via block bootstrapping the data with a block length of 14 days, to reflect temporal dependence in the extremes. It is evident that none of the copulas fit the model well in the whole support based on this measure. However, the Joe copula (in orange) appears to give the best fit in the tail, consistent with its AIC value being lowest.

\begin{table}[H]
\caption{MLEs for ten copulas and their AIC values. Lower AIC values are preferred.}
    \centering
    {\small\begin{tabular}{lccc}
         Copula & \multicolumn{2}{c}{Parameter} & AIC \\
         \hline
         Clayton & \multicolumn{2}{c}{$1.22\times 10^{-8}$} &\phantom{-1}2.0  \\ 
         Frank & \multicolumn{2}{c}{0.92}  & -15.8 \\ 
         Gumbel & \multicolumn{2}{c}{1.20}  & -97.4 \\ 
         Inverted Gumbel & \multicolumn{2}{c}{1.04} &\phantom{-1}0.1  \\ 
         Galambos & \multicolumn{2}{c}{0.46}  &-99.0 \\
         Gaussian & \multicolumn{2}{c}{0.19} & \phantom{1}-28.6 \\
         Joe & \multicolumn{2}{c}{1.41} & -143.6 \\
         Student t & 0.16 & 4.52 & \phantom{1}-52.8 \\
         H\"usler-Reiss & \multicolumn{2}{c}{0.82} & \phantom{1}-99.1 \\
          Coles-Tawn & 0.24 & 0.22 & \phantom{1}-95.9 \\
    \end{tabular}}
    \label{tab:mlesingle}
\end{table}

\noindent We next fit the weighted copula model to the whole data set taking the weighting function $\pi(u^*,v^*;\theta)=(u^*v^*)^\theta.$ We consider several copulas with different extremal dependence characteristics to fit both $c_b$ and $c_t;$ Table \ref{tab:mlemixed} shows the MLEs obtained by optimising the log-likelihood \eqref{eq:loglike} and their AIC values for some of the models considered. According to AIC, there is a preference for models with the Gaussian and Frank as candidates for $c_b$ and AD copulas, such as the Galambos, H\"usler-Reiss, Joe and Coles-Tawn copulas, as $c_t.$ In contrast to the single copula fits, the parameter estimates for the Gaussian and the Frank copulas are negative, which mirror the negative association visible in the body of Figure \ref{subfig:blackpooluniform}.

\begin{table}[H]
    \caption{MLEs for different weighted copula models and their AIC values when the weighting function used is $\pi(u^*,v^*;\theta)=(u^*v^*)^\theta.$ Lower AIC values are preferred.}
    \centering
    {\small\begin{tabular}{lllccccc}
         Model & $c_t$ & $c_b$ & \multicolumn{2}{c}{$\hat{\bm\alpha}$} & $\hat\beta$ & $\hat\theta$ &   AIC  \\
         \hline
         Model 1 & H\"usler-Reiss & Gaussian & \multicolumn{2}{c}{1.24} & -0.40 & 0.35 & -176.1 \\
         Model 2 & Galambos & Gaussian & \multicolumn{2}{c}{0.79} & -0.41 & 0.34 & -172.1 \\
         Model 3 & Coles-Tawn & Gaussian & 0.35 & 2.86 & -0.33 & 0.43 & -158.4 \\
         Model 4 & Coles-Tawn & Frank & 0.33 & 4.80 & -2.52 & 0.37 & -163.2 \\
         Model 5 & Joe & Frank & \multicolumn{2}{c}{1.61} & -4.11 & 0.18 & -184.9 \\
         Model 6 & Clayton & Gaussian & \multicolumn{2}{c}{12.10} & -0.20 & 2.10 & -129.9 \\
         Model 7 & Inverted Gumbel & Gaussian & \multicolumn{2}{c}{2.65} & -0.29 & 0.90 & -153.4 \\
         Model 8 & H\"usler-Reiss & Joe & \multicolumn{2}{c}{1.28} & 1.30 & 3.18 & -145.6 \\
         Model 9 & Student t & Galambos & 0.72 & 4.98 & 0.28 & 2.59 & -125.0 \\
         Model 10 & Gaussian & Clayton & \multicolumn{2}{c}{0.81} & $3.38\times 10^{-4}$ & 2.80 & -132.6 \\
         Model 11 & Gumbel & Joe & \multicolumn{2}{c}{1.52} & 1.18 & 0.91 & -145.1 \\
    \end{tabular}}
    \label{tab:mlemixed}
\end{table}

\noindent We next consider a different weighting function, $\pi(u^*,v^*;\theta)=\exp\{-\theta(1-u^*)(1-v^*)\},$ in the five models with the lowest AICs. The MLEs and the AIC values are shown in Table \ref{tab:mlemixed1}. In terms of AIC, these models are all better fits to the data, while the negative correlation is still captured by $c_b,$ and is now stronger. Because these models represent a better fit based on AIC, we focus on them for the rest of the analysis.

\begin{table}[H]
    \caption{MLEs for five weighted copula models and their AIC values when the weighting function used is $\pi(u^*,v^*;\theta)=\exp\{-\theta(1-u^*)(1-v^*)\}.$ Lower AIC values are preferred.}
    \centering
    {\small\begin{tabular}{lllccccc}
         Model & $c_t$ & $c_b$ & \multicolumn{2}{c}{$\hat{\bm\alpha}$} & $\hat\beta$ & $\hat \theta$ &  AIC  \\
         \hline
         Model 1 & H\"usler-Reiss & Gaussian & \multicolumn{2}{c}{1.33} & -0.74 & 3.32 & -240.1 \\
         Model 2 & Galambos & Gaussian & \multicolumn{2}{c}{0.90} & -0.72 & 3.55 & -237.2 \\
         Model 3 & Coles-Tawn & Gaussian & 0.85 & 0.79 & -0.74 & 3.25 & -234.8 \\
         Model 4 & Coles-Tawn & Frank & 0.869 & 1.02 & -4.51 & 4.33 & -235.7 \\
         Model 5 & Joe & Frank & \multicolumn{2}{c}{1.72} & -6.49 & 2.45 & -232.9 \\
    \end{tabular}}
    \label{tab:mlemixed1}
\end{table}

\subsection{Diagnostics} \label{subsection:diagnostics}

\noindent To check the adequacy of the model fits, we compare a variety of empirical dependence measures to their model-based counterparts. These include Kendall's $\tau,$ the dependence measures $\chi(r)$ and $\eta(r)$ for $r\in (0,1),$ and some probabilities of interest. Specifically, we look at the probability of ozone concentrations exceeding the so-called moderate threshold ({\it i.e.,} 100 $\mu g\slash m^3$) when the temperature is high or low, and the probability of $O_3$ exceeding this and the higher threshold of 160 $\mu g\slash m^3,$ knowing that the temperature is in a specific range. 

\begin{figure}[H]
    \centering
    \begin{subfigure}[b]{0.9\textwidth}
        \includegraphics[width=\textwidth]{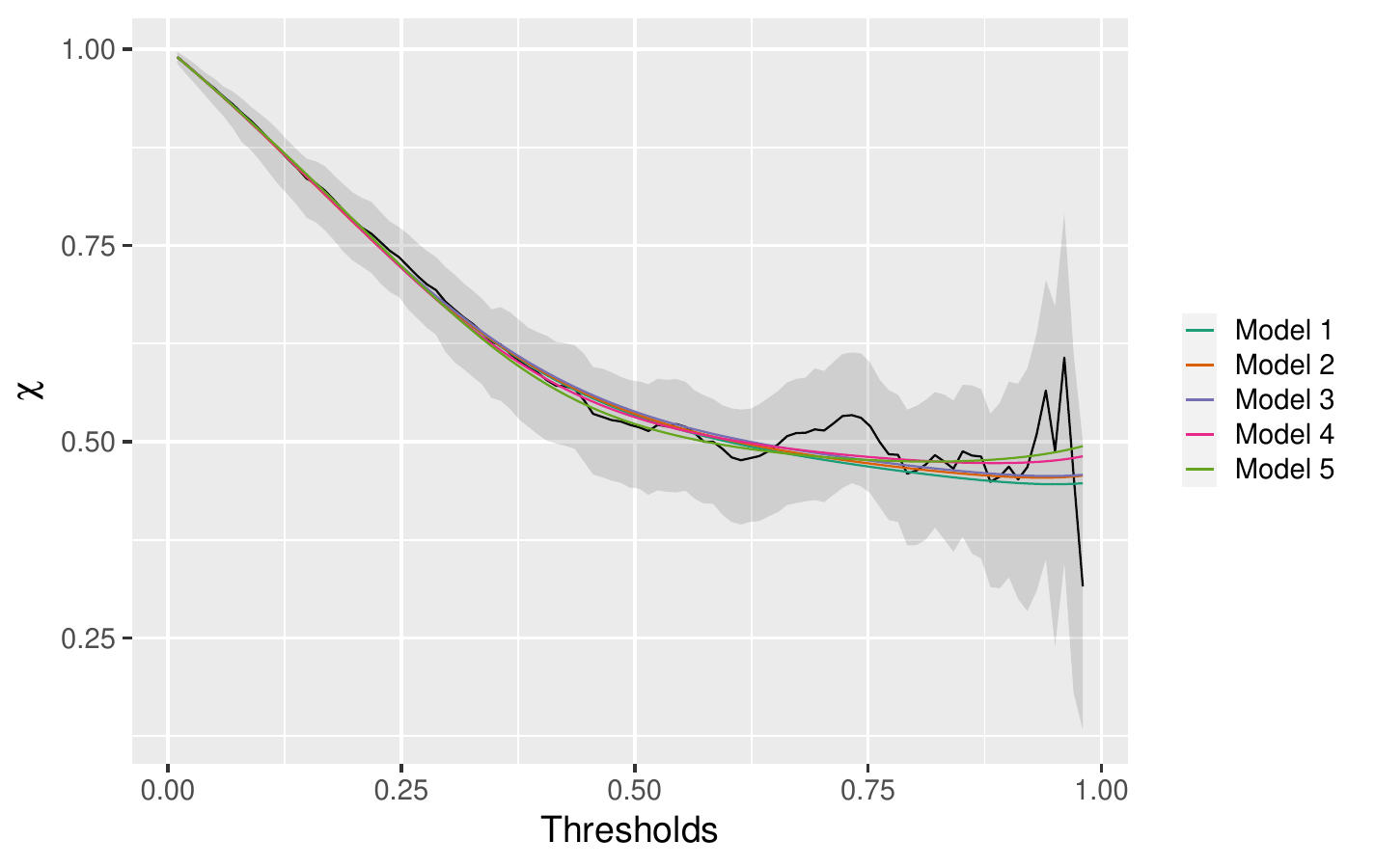}
        \caption{Empirical $\chi(r)$ (in black) and $\chi(r)$ for the five models (in colour) for $r\in(0,1).$ The 95$\%$ confidence bands were obtained by block bootstrapping.}
        \label{subfig:chimixed1}
    \end{subfigure}
    \begin{subfigure}[b]{0.9\textwidth}
        \includegraphics[width=\textwidth]{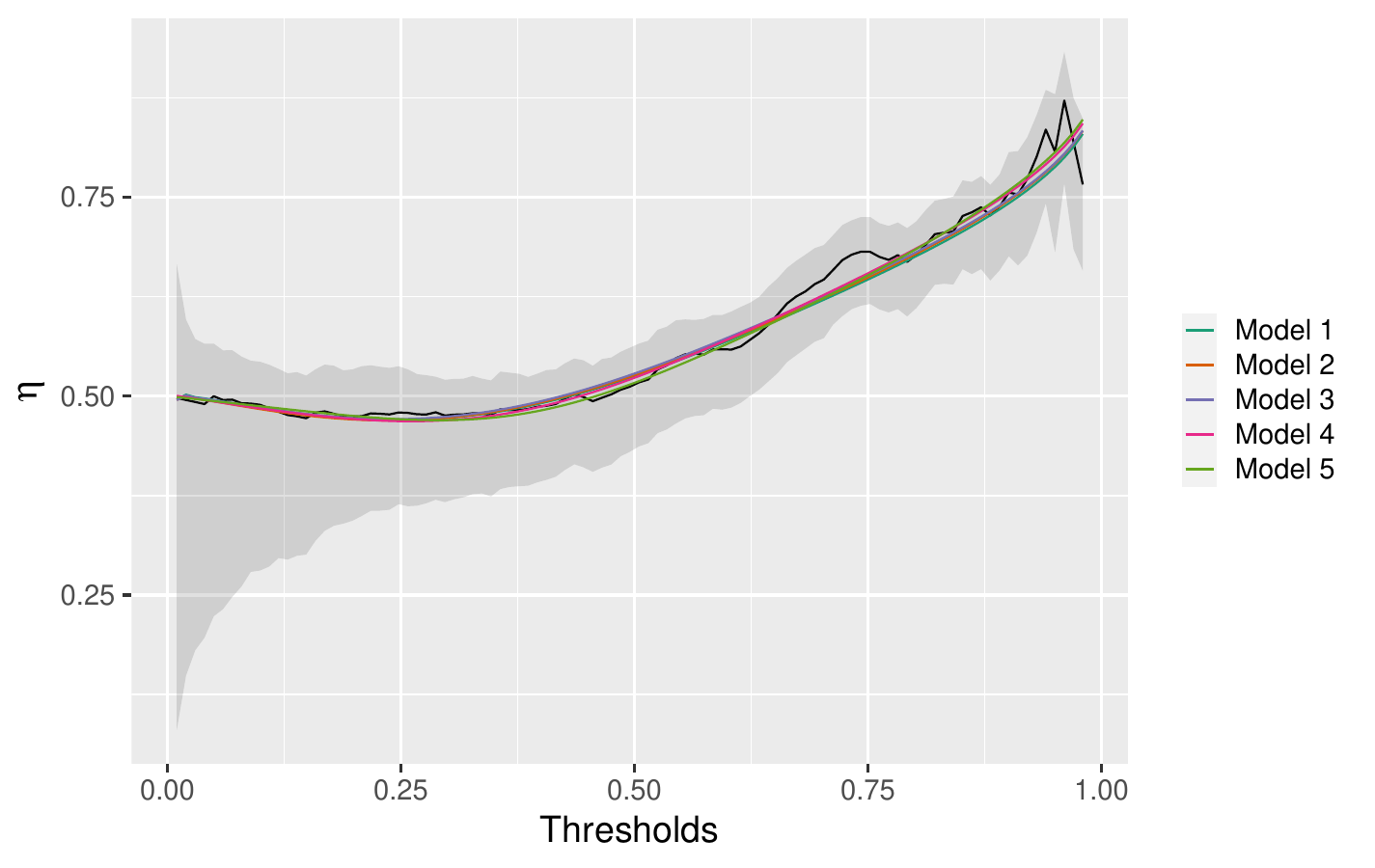}
        \caption{Empirical $\eta(r)$ (in black) and $\eta(r)$ for the five models (in colour) for $r\in(0,1).$ The 95$\%$ confidence bands were obtained by block bootstrapping.}
        \label{subfig:etamixed1}
    \end{subfigure}
    \caption{Dependence measures $\chi(r)$ and $\eta(r)$.}
    \label{fig:chietamixed1}
\end{figure}

\noindent Figure \ref{fig:chietamixed1} displays $\chi(r)$ and $\eta(r)$ for $r\in(0,1).$ A clear improvement from the single copula models shown in Figure \ref{fig:etasingle} can be seen as now all five models offer a reasonable fit throughout the whole support. In addition, model 5 (in light green) seems to provide slightly better $\chi(r)$ and $\eta(r)$ estimates at median values of $r$ and in the tail.

\noindent The average temperature in summer in Blackpool is between 17$^{\circ}$C and 20$^{\circ}$C and the observed 90th, 95th and 99th percentiles of the temperature are approximately 22$^{\circ}$C, 24$^{\circ}$C and 28$^{\circ}$C, respectively. Thus, we focus on probabilities based on these values of temperature; these are presented with Kendall's $\tau$ in Table \ref{tab:diag1}. We can see that the five models give very similar probabilities and  they are all inside the 95$\%$ confidence interval of the empirical values, except for $P[T\leq 16, O_3 \geq 100]$ and $P[O_{3}\geq 160 \mid 28\leq T\leq 29].$ The empirical probability and its 95$\%$ confidence interval of the latter are explained by the low number of observations present in the data set. When there are no observations in a certain region then this will be true of each bootstrap sample as well. \citet{GouldsbroughEtAll2022} obtained the mean probability of exceeding the high threshold 160 $\mu g\slash m^3$ at the 99th percentile of temperature for urban and rural backgrounds across the UK. These were $0.0002$ $\left([0,\,0.0004]\right)$ for an urban background and $0.006$ $\left([0.003,\,0.009]\right)$ for a rural background. We obtained higher probabilities of exceeding this threshold given that the temperature is close to the observed 99th percentile (we refer readers to the Supplementary Material for the results for Weybourne). This might be due to having only considered two sites within the UK, and potentially some of the characteristics of the relationship between temperature and ozone being better captured with the weighted copula model than with the univariate conditional model. 

\begin{table}[H]
    \caption{Diagnostics for the best five models based on their AIC values. The 95$\%$ confidence intervals for the empirical values were obtained by block bootstrapping.}
    \centering
    \setlength{\tabcolsep}{2pt}
    {
    \begin{tabular}{lccc}
         Model & Kendall's $\tau$ &  $P[T\leq 16, O_{3}\geq 100]$ & $P[T\geq 22, O_{3}\geq 100]$  \\ 
         \hline
        Empirical & 0.0821  & 0.0012 &  0.0363 \\ 
         \footnotesize{($95\%$ CI)} &\footnotesize{(0.0173\,,\,0.1867)} & \footnotesize{(0.0000\,,\,0.0011)}& \footnotesize{(0.0170\,,\,0.0601)}  \\ 
        Model 1 & 0.0690 & 0.0036 &  0.0332 \\ 
        Model 2 & 0.0663 & 0.0040 &  0.0336 \\ 
        Model 3 & 0.0770 & 0.0039 &  0.0338 \\ 
        Model 4 & 0.0779 & 0.0035 &  0.0348 \\ 
        Model 5 & 0.0718 & 0.0036 &  0.0353 \\ 
         \hline\hline
         Model & $P[T\geq 24, O_{3}\geq 100]$ & $P[O_{3}\geq 100 \mid 22\leq T\leq 23]$ & $P[O_{3}\geq 160 \mid 28\leq T\leq 29]$ \\
         \hline
         Empirical & 0.0302 & 0.1330 & 0.0000 \\ 
         \footnotesize{($95\%$ CI)}  & \footnotesize{(0.0147\,,\,0.0544)} & \footnotesize{(0.0227\,,\,0.1944)} & \footnotesize{(0.0000\,,\,0.0000)} \\  
        Model 1 & 0.0246 & 0.1441 & 0.0070 \\ 
        Model 2 & 0.0250 & 0.1412 & 0.0062 \\  
        Model 3 & 0.0251 & 0.1429 & 0.0061 \\ 
        Model 4 & 0.0262 & 0.1392 & 0.0055 \\ 
        Model 5 & 0.0267 & 0.1366 & 0.0050 \\ 
    \end{tabular}}
    \label{tab:diag1}
\end{table}

\noindent An advantage of this modelling approach in comparison to the conditional univariate modelling of \citet{GouldsbroughEtAll2022} is that we are able to extrapolate and consider probabilities of ozone exceeding certain thresholds at temperature values that have not been observed in the data set. In this way, we can consider probabilities such as $P[O_3 \geq 160 \mid 33 \leq T \leq 35],$ which we estimate to be 0.6944 for Model 1, for example.

\section{Conclusions and discussion} \label{section:conclusion}

\noindent In this paper, we introduced a dependence model that is able to capture both the body and tail of a bivariate data set. This is important when we aim to obtain an accurate representation of the data in both regions. The model has the advantage of not requiring a choice of thresholds above which we fit the copula tailored to the extreme observations. Moreover, it offers a smooth transition between the two copulas. Through simulation studies, we have shown that the model behaves as expected when only a single dependence structure is present, and that it is sufficiently flexible to capture misspecified dependence structures. We applied the weighted copula model to study the relationship between temperature and concentrations of air pollution in the UK and showed that this model performs substantially better than fitting a single copula model to the data. In fact, in this particular application, we were able to capture the negative dependence exhibited by the bulk and the positive association present in the upper tail, which was not possible through fitting a single copula.

\noindent A drawback of the weighted copula model is that it is computationally expensive due to the need for numerical integration and inversion. As shown in the simulation studies in Sections \ref{subsection:simulation1} and \ref{subsection:simulation2}, for a sample size of 1000, optimising the log-likelihood takes more than one hour to compute, although the run time also varies depending on the chosen copulas. Whilst in principle the weighted copula model could be extended to higher dimensions, doing so would exacerbate the computational issues. 

\noindent For the temperature and ozone data, we have $\chi(r)>0$ and $\eta(r) < 1,$ for the largest values of $r,$ which does not allow us to draw conclusions about the extremal dependence. This is a common situation in practice but results in complications if we wish to extrapolate for larger values than the ones observed. Incorporating a more flexible copula as the tail component of the proposed model is a possibility to overcome this issue. Such a copula could be the one proposed by \citet{HuserWadsworth2019}, which is able to capture both dependence classes with the transition between them occurring at an interior point of the parameter space. However, because it is computationally expensive on its own, when applied as the tail component in our model, the computational time required was not feasible.

\noindent It would be an advantage to have a copula model that could accommodate changes in the dependence structure due to covariates over the whole support of the distribution. Until now, we have been assuming stationarity, which is rarely the case in real world situations. Non-stationary multivariate extreme value methods naturally focus on capturing trends present in the extreme observations. However, data may be extreme in only one variable and thus studying the trends present in the body of the data is of importance as well. Incorporating covariates in the proposed model would also be an interesting avenue for future work. 

\noindent Finally, some theoretical aspects of the weighted copula model remain open for further work. For instance, it would be interesting to investigate bounds on differences between $c_b$ and/or $c_t$ with the copula $c$ of $c^*,$ or whether we could identify the family of the resulting copulas  in specific cases such as when both $c_t$ or $c_b$ are from the same family. Further theoretical exploration of extremal dependence properties of the weighted copula model would also be valuable as only particular cases were considered.

\bigskip

\noindent {\bf Declarations of Interest:} None.

\section*{Acknowledgments}
\noindent This paper is based on work completed while L\'idia Andr\'e was part of the EPSRC funded STOR-i centre for doctoral training (EP/S022252/1). We are grateful to the three referees for constructive comments that have improved this article. We are also grateful to Jonathan Tawn for advice on calculations in the Supplementary Material.

\bibliography{references}{}
\bibliographystyle{apalike}

\begin{appendices}
\section{Copula densities}\label{section:appendix}

\noindent In this appendix we give the copula distribution function $C$ and density function $c$ for all copulas used in the paper.

\subsection{Gaussian copula}

\noindent The Gaussian copula with correlation parameter $\rho\in(-1,1)$ is given by
\begin{equation*}
    C(u,v;\rho)=\Phi_2\left(\Phi_1^{-1}(u),\Phi_1^{-1}(v);\rho\right), \quad u,\,v\in(0,1),
\end{equation*}
where $\Phi_2(\cdot,\cdot;\rho)$ is the bivariate standard normal distribution function with correlation $\rho$ and $\Phi_1^{-1}(\cdot)$ is the inverse of the univariate standard normal distribution function. The Gaussian copula density can be written as
\begin{equation*}
    c(u,v;\rho)=\frac{1}{\sqrt{1-\rho^2}}\exp\left\{-\frac{\rho^2x^2+\rho^2y^2-2\rho xy}{2(1-\rho^2)}\right\}, \quad u,\,v\in(0,1),
\end{equation*}
where $x=\Phi_1^{-1}(u)$ and $y=\Phi_1^{-1}(v).$

\subsection{Student t copula}

\noindent The Student t copula with correlation parameter $\rho\in(-1,1)$ and $\nu>0$ degrees of freedom is given by
\begin{equation*}
    C(u,v;\rho,\nu)=T_{2,\nu}\left(T_\nu^{-1}(u),T_\nu^{-1}(v);\rho\right), \quad u,\,v\in(0,1),
\end{equation*}
where $T_{2,\nu}(\cdot,\cdot;\rho)$ is the bivariate t distribution function with correlation parameter $\rho$ and $T_\nu^{-1}(\cdot)$ is the inverse of the univariate t distribution function. The Student t copula density can be written as
\begin{equation*}
    c(u,v;\rho,\nu)=\frac{1}{\sqrt{1-\rho^2}}\frac{\Gamma\left(\frac{\nu+2}{2}\right)\Gamma\left(\frac{\nu}{2}\right)}{\Gamma\left(\frac{\nu+1}{2}\right)^2}\frac{\left[\left(1+\frac{x^2}{\nu}\right)\left(1+\frac{y^2}{\nu}\right)\right]^{(\nu+1)\slash 2}}{\left[1+\frac{\left(x^2+y^2-2\rho s r\right)}{\nu(1-\rho)^2}\right]^{(\nu+2)\slash 2}}, \quad u,\,v\in(0,1),
\end{equation*}
where $x=T^{-1}(u)$ and $y=T^{-1}(v).$

\subsection{Frank copula}

\noindent The Frank copula with parameter $\alpha\in\mathbb{R}\setminus\{0\}$ is given by
\begin{equation*}
    C(u,v;\alpha)=-\frac{1}{\alpha}\log\left(1-\frac{\left(1-e^{-\alpha u}\right)\left(1-e^{-\alpha v}\right)}{1-e^{-\alpha}}\right), \quad u,\,v\in(0,1),
\end{equation*}
and its density can be written as
\begin{equation*}
    c(u,v;\alpha)=\frac{\alpha(1-e^{-\alpha})e^{-\alpha(u+v)}}{\left[1-e^{-\alpha}-(1-e^{-\alpha u})(1-e^{-\alpha v})\right]^2}, \quad u,\,v\in(0,1).
\end{equation*}

\subsection{Clayton copula}

\noindent The Clayton copula with parameter $\alpha\in\mathbb{R}^{+}$ is given by
\begin{equation*}
    C(u,v;\alpha)=\left(u^{-\alpha}+v^{-\alpha}-1\right)^{-1\slash \alpha}, \quad u,\,v\in(0,1),
\end{equation*}
and its density can be written as
\begin{equation*}
    c(u,v;\alpha)=\frac{(\alpha+1)(uv)^\alpha}{\left(u^\alpha+v^\alpha-(uv)^\alpha\right)^{1\slash\alpha +2}}, \quad u,\,v\in(0,1).
\end{equation*}

\subsection{Joe copula}

\noindent The Joe copula with parameter $\alpha>1$ is given by
\begin{equation*}
    C(u,v;\alpha)=1-\left[\left(1-u\right)^{\alpha}+\left(1-v\right)^{\alpha}-\left(1-u\right)^{\alpha}\left(1-v\right)^{\alpha}\right]^{1\slash \alpha}, \quad u,\,v\in(0,1),
\end{equation*}
and its density can be written as
\begin{equation*}
    c(u,v;\alpha)=\left(x^\alpha+y^\alpha-(xy)^\alpha\right)^{1\slash \alpha -2}(xy)^{\alpha-1}\left(\alpha-1+x^\alpha+y^\alpha-(xy)^\alpha\right), \quad u,\,v\in(0,1),
\end{equation*}
where $x=1-u$ and $y=1-v.$

\subsection{Gumbel copula}

\noindent The Gumbel copula with parameter $\alpha>1$ is given by
\begin{equation*}
    C(u,v;\alpha)=\exp\left\{-\left(x^{\alpha}+y^{\alpha}\right)^{1\slash\alpha}\right\}, \quad u,\,v\in(0,1),
\end{equation*}
where $x=-\log(u)$ and $y=-\log(v).$ The Gumbel copula density can be written as 
\begin{equation*}
    c(u,v;\alpha)=\frac{C(u,v;\alpha)}{uv}(xy)^{\alpha-1}\left(x^\alpha+y^\alpha\right)^{1\slash \alpha -2}\left[\left(x^\alpha+y^\alpha\right)^{1\slash \alpha} +\alpha-1\right], \quad u,\,v\in(0,1).
\end{equation*}
The Inverted Gumbel copula density is obtained if we substitute $u$ and $v$ by $(1-u)$ and $(1-v),$ respectively.

\subsection{H\"usler-Reiss copula}

\noindent The H\"usler-Reiss copula with parameter $\alpha\in \mathbb{R}^+$ is given by
\begin{equation*}
    C(u,v;\alpha)=\exp\left\{-x\Phi\left(\frac{1}{\alpha}+\frac{\alpha}{2}\log\left(\frac{x}{y}\right)\right)-y\Phi\left(\frac{1}{\alpha}+\frac{\alpha}{2}\log\bigg(\frac{y}{x}\bigg)\right)\right\}, \quad u,\,v\in(0,1),
\end{equation*}
where $x=-\log(u)$ and $y=-\log(v).$ The H\"usler-Reiss copula density can be written as 
\begin{align*}
    c(u,v;\alpha)=\frac{C(u,v;\alpha)}{uv}&\left[\Phi\left(\frac{1}{\alpha}+\frac{\alpha}{2}\log\left(\frac{x}{y}\right)\right)\Phi\left(\frac{1}{\alpha}+\frac{\alpha}{2}\log\bigg(\frac{y}{x}\bigg)\right)\right.\nonumber \\& \left.+\frac{\alpha}{2y}\phi\left(\frac{1}{\alpha}+\frac{\alpha}{2}\log\left(\frac{x}{y}\right)\right)\right], \quad u,\,v\in(0,1). 
\end{align*}

\subsection{Galambos copula}

\noindent The Galambos copula with parameter $\alpha\in \mathbb{R}^+$ is given by
\begin{equation*}
    C(u,v;\alpha)=\exp\left\{-x-y+\left(x^{-\alpha}+y^{-\alpha}\right)^{-1\slash \alpha}\right\}, \quad u,\,v\in(0,1),
\end{equation*}
where $x=-\log(u)$ and $y=-\log(v).$ For $u,\,v\in(0,1),$ the Galambos copula density can be written as 
\begin{align*}
    c(u,v;\alpha)=\frac{C(u,v;\alpha)}{uv}&\left[1-(x^{-\alpha}+y^{-\alpha})^{-1-1\slash \alpha}(x^{-\alpha-1}+y^{-\alpha-1})\right.\nonumber \\& \left.+(x^{-\alpha}+y^{-\alpha})^{-2-1\slash\alpha}(xy)^{-\alpha-1}\left(1+\alpha+(x^{-\alpha}+y^{-\alpha})^{-1\slash\alpha}\right)\right]. 
\end{align*}

\subsection{Coles-Tawn copula}

\noindent The Coles-Tawn copula with parameters $\alpha,\beta\in \mathbb{R}^+$ is given by
\begin{equation*}
    C(u,v;\alpha,\beta)=\exp\left\{-x\left(1-\mbox{Be}(q;\alpha+1,\beta)\right)-y\mbox{Be}(q;\alpha,\beta+1)\right\}, \quad u,\,v\in(0,1),
\end{equation*} 
where $x=-\log(u),$ $y=-\log(v),$ $q=\displaystyle{\frac{\alpha x}{\alpha y+\beta x}}$ and $\mbox{Be}(q;a,b)$ represents the Beta distribution function with shape parameters $a>0$ and $b>0.$ The Coles-Tawn copula density can be written as
\begin{align*}
    c(u,v;\alpha,\beta)=\frac{C(u,v;\alpha,\beta)}{uvx^2y^2}&\left[x^2y^2\left(1-\mbox{Be}\left(q;\alpha+1,\beta\right)\right)\mbox{Be}\left(q;\alpha,\beta+1\right)\phantom{\frac{1}{2}}\right. \nonumber\\&\left.+\frac{\alpha\beta\Gamma(\alpha+\beta+1)}{\Gamma(\alpha)\Gamma(\beta)}\frac{q^{\alpha-1}(1-q)^{\beta-1}}{(\alpha\slash x + \beta\slash y)^3}\right], \quad u,\,v\in(0,1).
\end{align*}
\clearpage
\end{appendices}

\title{Supplementary Material for \emph{Joint modelling of the body and tail of bivariate data}}
\author{L. M. Andr\'e$^{1}$, J. L. Wadsworth$^{2}$, A. O'Hagan$^{3,4}$\\
\small $^{1}$ STOR-i Centre for Doctoral Training, Lancaster University, UK \\
\small $^{2}$ Department of Mathematics and Statistics, Lancaster University, UK \\
\small $^{3}$ School of Mathematics and Statistics, University College Dublin, Ireland \\
\small $^{4}$ Insight Centre for Data Analytics, University College Dublin, Ireland }
\date{August 25, 2023}

\maketitle
\pagenumbering{arabic}

\setcounter{section}{0}
\setcounter{figure}{0}
\setcounter{table}{0}

\section{Extremal dependence properties}

\noindent The extremal dependence measures $\chi$ and $\eta$ of the weighted copula model presented in Section 2 of the main text were derived for the case where $c_b$ is a Frank copula and $c_t$ a Gumbel copula, with two different weighting functions, and are presented in this Section. From equation (2) of Section 1.3 of the main text, we have
\begin{align*}
    \chi&=\lim_{r\rightarrow 1}\chi(r)=\lim_{r\rightarrow 1}\frac{P[U^*>r,V^*>r]}{P[U^*>r]}\\
    &=\lim_{r\rightarrow 1}\frac{(1\slash K)\int_{r}^{1}\int_{r}^{1}f_{c_t}(u^*,v^*;\bm\alpha,\theta)\text{d}v^*\,\text{d}u^*+(1\slash K)\int_{r}^{1}\int_{r}^{1}f_{c_b}(u^*,v^*;\bm\beta,\theta)\text{d}v^*\,\text{d}u^*}{(1\slash K)\int_{r}^{1}\int_{0}^{1}f_{c_t}(u^*,v^*;\bm\alpha,\theta)\text{d}v^*\,\text{d}u^*+(1\slash K)\int_{r}^{1}\int_{0}^{1}f_{c_b}(u^*,v^*;\bm\beta,\theta)\text{d}v^*\,\text{d}u^*}\\
    &=\lim_{r\rightarrow 1}\frac{\int_{r}^{1}\int_{r}^{1}f_{c_t}(u^*,v^*;\bm\alpha,\theta)\text{d}v^*\,\text{d}u^*+\int_{r}^{1}\int_{r}^{1}f_{c_b}(u^*,v^*;\bm\beta,\theta)\text{d}v^*\,\text{d}u^*}{\int_{r}^{1}\int_{0}^{1}f_{c_t}(u^*,v^*;\bm\alpha,\theta)\text{d}v^*\,\text{d}u^*+\int_{r}^{1}\int_{0}^{1}f_{c_b}(u^*,v^*;\bm\beta,\theta)\text{d}v^*\,\text{d}u^*},
\end{align*}
where $f_{c_t}=K_tf_t$ and $f_{c_b}=K_bf_b$ with $K_t,$ $K_b,$ $f_t,$ $f_b$ and $K$ as defined in Section 2.2 of the main text. 

\subsection{Case 2: $c_b$ is a Frank copula, $c_t$ is a Gumbel copula and $\pi(u^*,v^*;\theta)=(uv)^\theta$}
\noindent Assuming $\pi(u^*,v^*;\theta)=(u^*v^*)^\theta,$ we have
\begin{align*}
f_{c_b}(u^*,v^*;\beta,\theta)&=[1-(u^*v^*)^\theta]\frac{\beta(1-\exp\{-\beta\})\exp\{-\beta(u^*+v^*)\}}{\left[1-\exp\{-\beta\}-(1-\exp\{-\beta u^*\})(1-\exp\{-\beta v^*\})\right]^2}
\end{align*}
and
\begin{align*}
f_{c_t}(u^*,v^*;&\alpha,\theta)=(u^*v^*)^\theta\frac{C_t(u^*,v^*;\alpha)}{u^*v^*}(xy)^{\alpha-1}\left(x^\alpha+y^\alpha\right)^{1\slash \alpha -2}\left[\left(x^\alpha+y^\alpha\right)^{1\slash \alpha} +\alpha-1\right]\\
=&(u^*v^*)^{\theta-1}C_t(u^*,v^*;\alpha)(xy)^{\alpha-1}\left(x^\alpha+y^\alpha\right)^{1\slash \alpha -2}\left[\left(x^\alpha+y^\alpha\right)^{1\slash \alpha} +\alpha-1\right],
\end{align*}
with $x=-\log(u^*),$ $y=-\log(v^*)$ and $C_t(u^*,v^*;\alpha)=\exp\left\{-\left(x^{\alpha}+y^{\alpha}\right)^{1\slash\alpha}\right\}.$

\subsubsection{Effect of the body copula $c_b$}
\noindent Since the interest is on the limit when $u^*$ and $v^*$ are very near (1,1) and $f_{c_b}(u^*,v^*;\beta,\theta)$ is defined at (1,1), a Taylor approximation of order 1 can be used about (1,1) with point $(1-s,1-t)$ for $\int_{r}^{1}\int_{r}^{1}f_{c_b}(u^*,v^*)\text{d}v^*\,\text{d}u^*,$ where $s,t\rightarrow 0.$ Therefore, for some norm $\|\cdot\|$ near 0, we have
\begin{equation*}
    f_{c_b}(1-s,1-t;\beta,\theta)= f_{c_b}(1,1)-s\frac{\partial f_{c_b}}{\partial s}(1,1)-t\frac{\partial f_{c_b}}{\partial t}(1,1)+\mathcal{O}\left(\|(s,t)\|^2\right),
    \end{equation*}
where
\begin{align*}
    \frac{\partial f_{c_b}}{\partial s}=&\frac{2\beta^2[1-(st)^\theta](1-\exp\{-\beta\})(1-\exp\{-\beta t\})\exp\{-\beta(2s+t)\}}{[1-\exp\{-\beta\}-(1-\exp\{-\beta s\})(1-\exp\{-\beta t\})]^3}\\
    &-\frac{\beta\theta s^{\theta-1}t^\theta(1-\exp\{-\beta\})\exp\{-\beta(s+t)\}}{[1-\exp\{-\beta\}-(1-\exp\{-\beta s\})(1-\exp\{-\beta t\})]^2}\\
    &-\frac{\beta^2[1-(st)^\theta](1-\exp\{-\beta\})\exp\{-\beta(s+t)\}}{[1-\exp\{-\beta\}-(1-\exp\{-\beta s\})(1-\exp\{-\beta t\})]^2}.
\end{align*}
At the point (1,1), $f_{c_b}(1,1)=0$ and $$\displaystyle{\frac{\partial f_{c_b}}{\partial s}(1,1)=\frac{\partial f_{c_b}}{\partial t}(1,1)=-\beta\theta\left(1-\exp\{-\beta\}\right)^{-1}}.$$ So, 
\begin{equation*}
   f_{c_b}(1-s,1-t;\beta,\theta)=\beta\theta\left(1-\exp\{-\beta\}\right)^{-1}(s+t)+\mathcal{O}\left(\|(s,t)\|^2\right). 
\end{equation*}
Taking $s=1-u^*$ and $t^*=1-v^*,$ we have
\begin{align*}
\int_{r}^{1}\int_{r}^{1}f_{c_b}&(u^*,v^*)\text{d}v^*\,\text{d}u^*\\
=&\int_{0}^{1-r}\int_{0}^{1-r}\beta\theta\left(1-\exp\{-\beta\}\right)^{-1}(s+t)\text{d}t\,\text{d}s+\mathcal{O}\left((1-r)^4\right)\\
=&\beta\theta\left(1-\exp\{-\beta\}\right)^{-1}\int_{0}^{1-r}\int_{0}^{1-r}(s+t)\text{d}t\,\text{d}s+\mathcal{O}\left((1-r)^4\right)\\
=&\beta\theta\left(1-\exp\{-\beta\}\right)^{-1}(1-r)^3+\mathcal{O}\left((1-r)^4\right).
\end{align*}

\noindent Similarly, for $\int_{r}^{1}\int_{0}^{1}f_{c_b}(u^*,v^*)\text{d}v^*\,\text{d}u^*,$ a Taylor approximation of order 1 can be used about $(1,v^*)$ with point $(u^*,v^*).$ Thus, we have 
\begin{equation*}
    f_{c_b}(u^*,v^*;\beta,\theta)= f_{c_b}(1,v^*)+(u^*-1)\frac{\partial f_{c_b}}{\partial u^*}(1,v^*)+\mathcal{O}\left((u^*-1)^2\right),
\end{equation*}
where
\begin{align*}
    f_{c_b}(1,v^*)=& \frac{(1-(v^*)^\theta)\beta \exp\{-\beta(1-v^*)\}}{1-\exp\{-\beta\}}=A_{v^*,\beta,\theta} 
\end{align*}
and
\begin{align*}
\frac{\partial f_{c_b}}{\partial u^*}(1,v^*)=&\frac{2\beta^2(1-(v^*)^\theta)(1-\exp\{-\beta v^*\})\exp\{-2\beta(1-v^*)\}}{(1-\exp\{-\beta\})^2}\\
&-\frac{\beta\theta (v^*)^\theta \exp\{-\beta(1-v^*)\}}{1-\exp\{-\beta\}}-\frac{\beta^2(1-(v^*)^\theta)\exp\{-\beta(1-v^*)\}}{1-\exp\{-\beta\}}\\=&B_{v^*,\beta,\theta}.
\end{align*}
So, $f_{c_b}(u^*,v^*)= A_{v^*,\beta,\theta}+B_{v^*,\beta,\theta}(u^*-1)+\mathcal{O}\left((u^*-1)^2\right),$ and we obtain
\begin{align*}
\int_{r}^{1}\int_{0}^{1}&f_{c_b}(u^*,v^*)\text{d}v^*\,\text{d}u^*\\
=&\int_{r}^{1}\int_{0}^{1}[A_{v^*,\beta,\theta}+B_{v^*,\beta,\theta}(u^*-1)]\text{d}v^*\,\text{d}u^*+\mathcal{O}\left((1-r)^3\right) \\
=&\int_{0}^{1}A_{v^*,\beta,\theta}\int_{r}^{1}\text{d}u^*\,\text{d}v^*+\int_{0}^{1}B_{v^*,\beta,\theta}\int_{r}^{1}(u^*-1)\text{d}u^*\,\text{d}v^* +\mathcal{O}\left((1-r)^3\right)\\
=&(1-r)\underbrace{\int_{0}^{1}A_{v^*,\beta,\theta}\text{d}v^*}_{C_{\beta,\theta}} - \frac{1}{2}(1-r)^2\underbrace{\int_{0}^{1}B_{v^*,\beta,\theta}\text{d}v^*}_{D_{\beta,\theta}} +\mathcal{O}\left((u^*)^2\right)\\
=&C_{\beta,\theta}(1-r)-\frac{D_{\beta,\theta}}{2}(1-r)^2+\mathcal{O}\left((1-r)^3\right)
\end{align*}

\subsubsection{Effect of the tail copula $c_t$}

\noindent Contrarily to $f_{c_b}(\cdot),$ $f_{c_t}(u^*,v^*;\alpha,\theta)$ is not finite at (1,1). For this reason, it is not possible to use a Taylor approximation about (1,1). Instead, we use asymptotics near this point. Specifically, we now write $u^*$ and $v^*$ in terms of $s$ and $t,$ where $s,t>0$ and $u^*=1-s+o(s)$ and $v^*= 1-t+o(t)$ as $s,t\rightarrow 0.$ This describes the behaviour of $u^*$ and $v^*$ as they tend to 1. Thus, for the first term of $f_{c_t},$ we have 
\begin{align*}
    (u^*v^*)^{\theta-1} =& (1-s)^{\theta-1}(1-t)^{\theta-1} +o(s)+o(t)\\
    =& [1-(\theta-1)s][1-(\theta-1)t]+o(s)+o(t),
\end{align*}
as $s,t\rightarrow 0.$ 

\noindent Let us first consider the case when $x=-\log(u^*)>y=-\log(v^*).$ For $(u^*,v^*)\rightarrow (1,1),$ i.e., $s\rightarrow0$ and $t\rightarrow 0,$ with $t\slash s \rightarrow c$ for $c\in(0,1),$ the copula density term follows asymptotically 
\begin{equation*}
    c_t(u^*,v^*;\alpha)\sim (\alpha-1)x^{-\alpha}y^{\alpha-1}\left[1+\left(\frac{y}{x}\right)^\alpha\right]^{1\slash \alpha -2}.
\end{equation*}
Analogously, when $x<y,$ i.e., $s\rightarrow0$ and $t\rightarrow 0,$ with $t\slash s \rightarrow c$ for $c\in(1,\infty),$
\begin{equation*}
    c_t(u^*,v^*;\alpha)\sim (\alpha-1)y^{-\alpha}x^{\alpha-1}\left[1+\left(\frac{x}{y}\right)^\alpha\right]^{1\slash \alpha -2}.
\end{equation*}
 
\noindent Moreover, $x=s+o(s)$ and $y=t+o(t)$ as $s,t\rightarrow 0.$ So, considering the symmetry between cases $x>y$ and $x<y,$ and recalling $u^*=1-s+o(s)$ and $v^*= 1-t+o(t),$
\begin{equation*}
    \int_{r}^{1}\int_{r}^{1}f_{c_t}(u^*,v^*)\text{d}v^*\,\text{d}u^*= P[1-S>r,1-T>r]=2P[S<1-r,T<S]. 
\end{equation*}

\noindent So, we have
\begin{align*}
    P[S&<1-r,T<S]=\int_{0}^{1-r}\int_{0}^{s}f_{c_t}^*(s,t;\alpha,\theta)\text{d}t\,\text{d}s \\ 
    &=\int_{0}^{1-r}\int_{0}^{s} [1-(\theta-1)s][1-(\theta-1)t] (\alpha-1)s^{-\alpha}t^{\alpha-1}\\
    &\phantom{\hspace{55mm}}\times\left[1+\left(\frac{t}{s}\right)^\alpha\right]^{1\slash \alpha -2} \text{d}t\,\text{d}s+o\left((1-r)^2\right)\\
    &=(\alpha-1)\int_{0}^{1-r}[1-(\theta-1)s]s^{-\alpha}\\
    &\phantom{\hspace{25mm}}\underbrace{\int_{0}^{s} [1-(\theta-1)t]t^{\alpha-1}\left[1+\left(\frac{t}{s}\right)^\alpha\right]^{1\slash \alpha -2} \text{d}t}_{A(s)}\,\text{d}s+o\left((1-r)^2\right)
\end{align*}
as $r\rightarrow 1.$ Evaluating $A(s)$ by parts, we get
\begin{align*}
    \int_{0}^{s} [1-(\theta-1)&t]t^{\alpha-1}\left[1+\left(\frac{t}{s}\right)^\alpha\right]^{1\slash \alpha -2} \text{d}t  \\
    &=\frac{2^{1\slash \alpha -1}s^\alpha}{1-\alpha}-\frac{2^{1\slash \alpha -1}(\theta-1)s^{\alpha+1}}{1-\alpha}-\frac{s^\alpha}{1-\alpha}-\frac{(1-\theta)s^{\alpha+1}}{1-\alpha}C_{\alpha},
\end{align*}
with $C_{\alpha}=\displaystyle{\int_0^1 (1+q^\alpha)^{1\slash \alpha -1}}\text{d}q.$ And, by substituting $A(s)$ in the outer integral, we obtain
\begin{align*}
    P[S<1-r,T<s]=&(1-2^{1\slash \alpha -1})(1-r)+\left[(2^{1\slash \alpha}-1-C_{\alpha})(\theta-1)\slash 2\right](1-r)^2\\
    &+o\left((1-r)^2\right), \quad \text{as } r\rightarrow 1.
\end{align*}
Then, as $r\rightarrow 1,$
\begin{align*}
    \int_{r}^{1}&\int_{r}^{1}f_{c_t}(u^*,v^*)\text{d}v^*\,\text{d}u^*\\=&\,2(1-2^{1\slash \alpha -1})(1-r)+2\left[(2^{1\slash \alpha}-1-C_{\alpha})(\theta-1)\slash 2\right](1-r)^2+o\left((1-r)^2\right)\\
    =&(2-2^{1\slash \alpha})(1-r)+(2^{1\slash \alpha}-1-C_{\alpha})(\theta-1)(1-r)^2+o\left((1-r)^2\right),
\end{align*}

\noindent Since for $\int_{r}^{1}\int_{0}^{1}f_{c_t}(u^*,v^*)\text{d}v^*\,\text{d}u^*$ we need to integrate over the support for $v^*,$ it is not possible to approximate $f_{c_t}(\cdot)$ as above. Instead, we take the change of variable $y=xz,$ with $z=y\slash x \in \mathbb{R^+},$ so we have $u^*=\exp\{-x\}$ and $v^*=\exp\{-xz\}.$ Thus, we obtain

\begin{align*}
\int_{r}^{1}&\int_{0}^{1}f_{c_t}(u^*,v^*;\alpha,\theta)\text{d}v^*\,\text{d}u^* \\
=&\int_{r}^{1}\int_{0}^{1}(u^*v^*)^{\theta-1}C_t(u^*,v^*;\alpha)(xy)^{\alpha-1}\left(x^\alpha+y^\alpha\right)^{1\slash \alpha -2}\\
&\phantom{\hspace{50mm}}\times\left[\left(x^\alpha+y^\alpha\right)^{1\slash \alpha} +\alpha-1\right]\text{d}v^*\,\text{d}u^* \\
=&\int_{0}^{-\log(r)}\int_{0}^{\infty}\exp\left\{-x\left[\theta(1+z)+(1+z^\alpha)^{1\slash \alpha}\right]\right\}z^{\alpha-1}\left(1+z^\alpha\right)^{1\slash \alpha -2}\\
&\phantom{\hspace{50mm}}\times\left[x\left(1+z^\alpha\right)^{1\slash \alpha} +\alpha-1\right]\text{d}z\,\text{d}x\\
=&\int_{0}^{-\log(r)}\int_{0}^{\infty}x\underbrace{z^{\alpha-1}\left(1+z^\alpha\right)^{2\slash \alpha -2}}_{g(z)}\exp\left\{-x\underbrace{\left[\theta(1+z)+(1+z^\alpha)^{1\slash \alpha}\right]}_{h(z)}\right\}\text{d}z\,\text{d}x\\
&+(\alpha-1)\int_{0}^{-\log(r)}\int_{0}^{\infty}\underbrace{z^{\alpha-1}\left(1+z^\alpha\right)^{1\slash \alpha -2}}_{f(z)}\\
& \phantom{\hspace{56mm}}\times\exp\left\{-x\underbrace{\left[\theta(1+z)+(1+z^\alpha)^{1\slash \alpha}\right]}_{h(z)}\right\}\text{d}z\,\text{d}x\\
=&\int_{0}^{\infty}g(z)\underbrace{\int_{0}^{-\log(r)}x \exp\{-xh(z)\}\text{d}x}_{B(z,r)}\,\text{d}z\phantom{\hspace{60mm}}\\
&+(\alpha-1)\int_{0}^{\infty}f(z)\underbrace{\int_{0}^{-\log(r)}\exp\{-xh(z)\}\text{d}x}_{C(z,r)}\,\text{d}z.
\end{align*} 

\noindent Evaluating $B(z,r)$ by parts, we get
\begin{align*}
    \int_{0}^{-\log(r)}x&\exp\{-xh(z)\}\text{d}x\\
    =&\left[-\frac{x}{h(z)}\exp\{-xh(z)\}\right]^{x=-\log(r)}_{x=0}-\left[\frac{1}{h^2(z)}\exp\{-xh(z)\}\right]^{x=-\log(r)}_{x=0} \\
    =&\frac{\log(r)}{h(z)}r^{h(z)}-\frac{1}{h^2(z)}r^{h(z)}+\frac{1}{h^2(z)}.
\end{align*}

\noindent Analogously, by evaluating $C(z,r)$, we have
\begin{align*}
    \int_{0}^{-\log(r)}\exp\{-xh(z)\}\text{d}x=&\left[-\frac{1}{h(z)}\exp\{-h(z)x\}\right]^{x=-\log(r)}_{x=0}\\
    =&-\frac{1}{h(z)}r^{h(z)}+\frac{1}{h(z)}.
\end{align*}

\noindent Substituting $B(z,r)$ and $C(z,r)$ in the outer integral, we obtain
\begin{align*}
\int_{r}^{1}\int_{0}^{1}f_{c_t}(u^*,v^*;\alpha,\theta)\text{d}v^*\,\text{d}u^* =&\log(r)\int_{0}^{\infty}\frac{g(z)}{h(z)}r^{h(z)}\text{d}z+\int_{0}^{\infty}\frac{g(z)}{h^2(z)}\left(1-r^{h(z)}\right)\text{d}z\\
&+(\alpha-1)\int_{0}^{\infty}\frac{f(z)}{h(z)}\left(1-r^{h(z)}\right)\text{d}z.
\end{align*}

\noindent Evaluating $\displaystyle \int_{0}^{\infty}\frac{f(z)}{h(z)}\left(1-r^{h(z)}\right)\text{d}z$ by parts, we have
\begin{align*}
	\int_{0}^{\infty}\frac{f(z)}{h(z)}&\left(1-r^{h(z)}\right)\text{d}z\\
 =&\left[\frac{1}{1-\alpha} \left(1+z^\alpha\right)^{1\slash\alpha-1}\frac{1-r^{h(z)}}{h(z)}\right]_0^{\infty} \\
	&-\underbrace{\int_{0}^{\infty}\frac{1}{1-\alpha} \left(1+z^\alpha\right)^{1\slash\alpha-1}\left(\frac{h'(z)\left(r^{h(z)}-1\right)}{h^2(z)}-\frac{\log(r)h'(z)r^{h(z)}}{h(z)}\right)\text{d}z}_{D(r)} \\
	=&\frac{1}{1-\alpha}\lim_{z\rightarrow \infty}\left(1+z^\alpha\right)^{1\slash\alpha-1}\frac{1-r^{\theta(1+z)+(1+z^\alpha)^{1\slash \alpha}}}{\theta(1+z)+(1+z^\alpha)^{1\slash \alpha}}\\
	&-\frac{1}{1-\alpha}\frac{1-r^{\theta+1}}{\theta+1}-D(r) \\
	=&\frac{1}{1-\alpha}\lim_{z\rightarrow \infty}z^{1-\alpha}\frac{r^{z(\theta+1)}}{z(1+\theta)}+\frac{1-r^{\theta+1}}{(\alpha-1)(\theta+1)}-D(r)\\
	=&\frac{1}{1-\alpha}\lim_{z\rightarrow \infty}z^{-\alpha}\frac{r^{z(\theta+1)}}{1+\theta}+\frac{1-r^{\theta+1}}{(\alpha-1)(\theta+1)}-D(r) \\	
    =&\frac{1-r^{\theta+1}}{(\alpha-1)(\theta+1)}-D(r).
\end{align*}
Noting that $h'(z)=\theta+z^{\alpha-1}(1+z^{\alpha})^{1\slash
\alpha-1},$ and recalling that $g(z)=z^{\alpha-1}(1+z^{\alpha})^{2\slash
\alpha-2},$ $D(r)$ can be simplified as below

\begin{align*}
    D(r)=&\int_{0}^{\infty}\frac{1}{1-\alpha} \left(1+z^\alpha\right)^{1\slash\alpha-1}\left(\frac{h'(z)\left(r^{h(z)}-1\right)}{h^2(z)}-\frac{\log(r)h'(z)r^{h(z)}}{h(z)}\right)\text{d}z\\
    =&\int_{0}^{\infty}\frac{1}{1-\alpha} \left(1+z^\alpha\right)^{1\slash\alpha-1}\left[\theta+z^{\alpha-1}(1+z^{\alpha})^{1\slash
\alpha-1}\right]\frac{\left(r^{h(z)}-1\right)}{h^2(z)}\text{d}z\\
	&-\log(r)\int_{0}^{\infty}\frac{1}{1-\alpha} \left(1+z^\alpha\right)^{1\slash\alpha-1}\left[\theta+z^{\alpha-1}(1+z^{\alpha})^{1\slash
\alpha-1}\right]\frac{r^{h(z)}}{h(z)}\text{d}z\\
 	=&-\theta\int_{0}^{\infty}\frac{1}{1-\alpha} \left(1+z^\alpha\right)^{1\slash\alpha-1}\frac{\left(1-r^{h(z)}\right)}{h^2(z)}\text{d}z\\
  &-\int_{0}^{\infty}\frac{1}{1-\alpha} z^{\alpha-1}\left(1+z^\alpha\right)^{2\slash\alpha-2}\frac{\left(1-r^{h(z)}\right)}{h^2(z)}\text{d}z\\
	&-\theta\log(r)\int_{0}^{\infty}\frac{1}{1-\alpha} \left(1+z^\alpha\right)^{1\slash\alpha-1}\frac{r^{h(z)}}{h(z)}\text{d}z\\
 &-\log(r)\int_{0}^{\infty}\frac{1}{1-\alpha} z^{\alpha-1}\left(1+z^\alpha\right)^{2\slash\alpha-2}\frac{r^{h(z)}}{h(z)}\text{d}z\\
	=&\frac{\theta}{\alpha-1}\int_{0}^{\infty}\left(1+z^\alpha\right)^{1\slash\alpha-1}\frac{\left(1-r^{h(z)}\right)}{h^2(z)}\text{d}z+\frac{1}{\alpha-1} \int_{0}^{\infty}\frac{g(z)}{h^2(z)}\left(1-r^{h(z)}\right)\text{d}z\\
	&+\frac{\theta\log(r)}{\alpha-1}\int_{0}^{\infty}\left(1+z^\alpha\right)^{1\slash\alpha-1}\frac{r^{h(z)}}{h(z)}\text{d}z+\frac{\log(r)}{\alpha-1}\int_{0}^{\infty}\frac{g(z)}{h(z)}r^{h(z)}\text{d}z.
\end{align*}

\noindent Thus, we have
\begin{align*}
\int_{r}^{1}\int_{0}^{1}&f_{c_t}(u^*,v^*;\alpha,\theta)\text{d}v^*\,\text{d}u^* =\log(r)\int_{0}^{\infty}\frac{g(z)}{h(z)}r^{h(z)}\text{d}z+\int_{0}^{\infty}\frac{g(z)}{h^2(z)}\left(1-r^{h(z)}\right)\text{d}z\\
&+(\alpha-1)\frac{1-r^{\theta+1}}{(\alpha-1)(\theta+1)}-(\alpha-1)\frac{\theta}{\alpha-1}\int_{0}^{\infty}\left(1+z^\alpha\right)^{1\slash\alpha-1}\frac{\left(1-r^{h(z)}\right)}{h^2(z)}\text{d}z\\
&-(\alpha-1)\frac{1}{\alpha-1} \int_{0}^{\infty}\frac{g(z)}{h^2(z)}\left(1-r^{h(z)}\right)\text{d}z\\
&-(\alpha-1)\frac{\theta\log(r)}{\alpha-1}\int_{0}^{\infty}\left(1+z^\alpha\right)^{1\slash\alpha-1}\frac{r^{h(z)}}{h(z)}\text{d}z\\
&-(\alpha-1)\frac{\log(r)}{\alpha-1}\int_{0}^{\infty}\frac{g(z)}{h(z)}r^{h(z)}\text{d}z\\
=&\frac{1-r^{\theta+1}}{\theta+1}-\theta\int_{0}^{\infty}\left(1+z^\alpha\right)^{1\slash\alpha-1}\frac{\left(1-r^{h(z)}\right)}{h^2(z)}\text{d}z\\
&-\theta\log(r)\int_{0}^{\infty}\left(1+z^\alpha\right)^{1\slash\alpha-1}\frac{r^{h(z)}}{h(z)}\text{d}z \\
=&1-r-\frac{\theta}{2}(1-r)^2+o\left((1-r)^2\right),
\end{align*}
where $-\theta\int_{0}^{\infty}\left(1+z^\alpha\right)^{1\slash\alpha-1}\frac{\left(1-r^{h(z)}\right)}{h^2(z)}\text{d}z-\theta\log(r)\int_{0}^{\infty}\left(1+z^\alpha\right)^{1\slash\alpha-1}\frac{r^{h(z)}}{h(z)}\text{d}z=o((1-r)^2)$ as $r\rightarrow 1.$ Additionally, $r^{\theta+1}=1-(\theta+1)(1-r)+\left[\theta(\theta+1)\slash 2\right](1-r)^2+o\left((1-r)^2\right)$ as $r\rightarrow 1$ by the Binomial expansion.

\subsubsection{Extremal dependence $\chi$ for this case}

\noindent Let 
\begin{align*}
    c_1&=2-2^{1\slash \alpha} =\chi_{Gumbel}, && c_5=-\theta\slash 2+o\left((1-r)^2\right),\\
    c_2&=(2^{1\slash \alpha}-1-C_{\alpha})(\theta-1), && c_6=C_{\beta,\theta}=\beta\left(1-\exp\{-\beta\}\right)^{-1} \\
    c_3&=\beta\theta\left(1-\exp\{-\beta\}\right)^{-1}, && \phantom{c_7=}\times\int_{0}^{1}(1-(v^*)^\theta)e^{-\beta(1-v^*)}\text{d}v^*, \\
    c_4&=1, &&  c_7=-D_{\beta,\theta}\slash 2.
\end{align*}
We then have
\begin{align}
    \chi=&\lim_{r\rightarrow 1}\frac{c_1(1-r)+c_2(1-r)^2+c_3(1-r)^3+o\left((1-r)^3\right)}{c_4(1-r)+c_5(1-r)^2+c_6(1-r)+c_7(1-r)^2+o\left((1-r)^2\right)} \nonumber \\
    =&\lim_{r\rightarrow 1} \left(\frac{c_1}{c_4+c_6}+\left[\frac{c_2-c_1(c_5+c_7)}{(c_4+c_6)^2}\right](1-r)+\mathcal{O}\left((1-r)^2\right)\right) \nonumber \\
    =&\frac{c_1}{c_4+c_6}=\frac{2-2^{1\slash \alpha}}{1+\beta\left(1-\exp\{-\beta\}\right)^{-1}\int_{0}^{1}(1-(v^*)^\theta)e^{-\beta(1-v^*)}\text{d}v^*} \label{eq:chi}
\end{align}

\begin{figure}[H]
    \centering
    \includegraphics[width=0.7\textwidth]{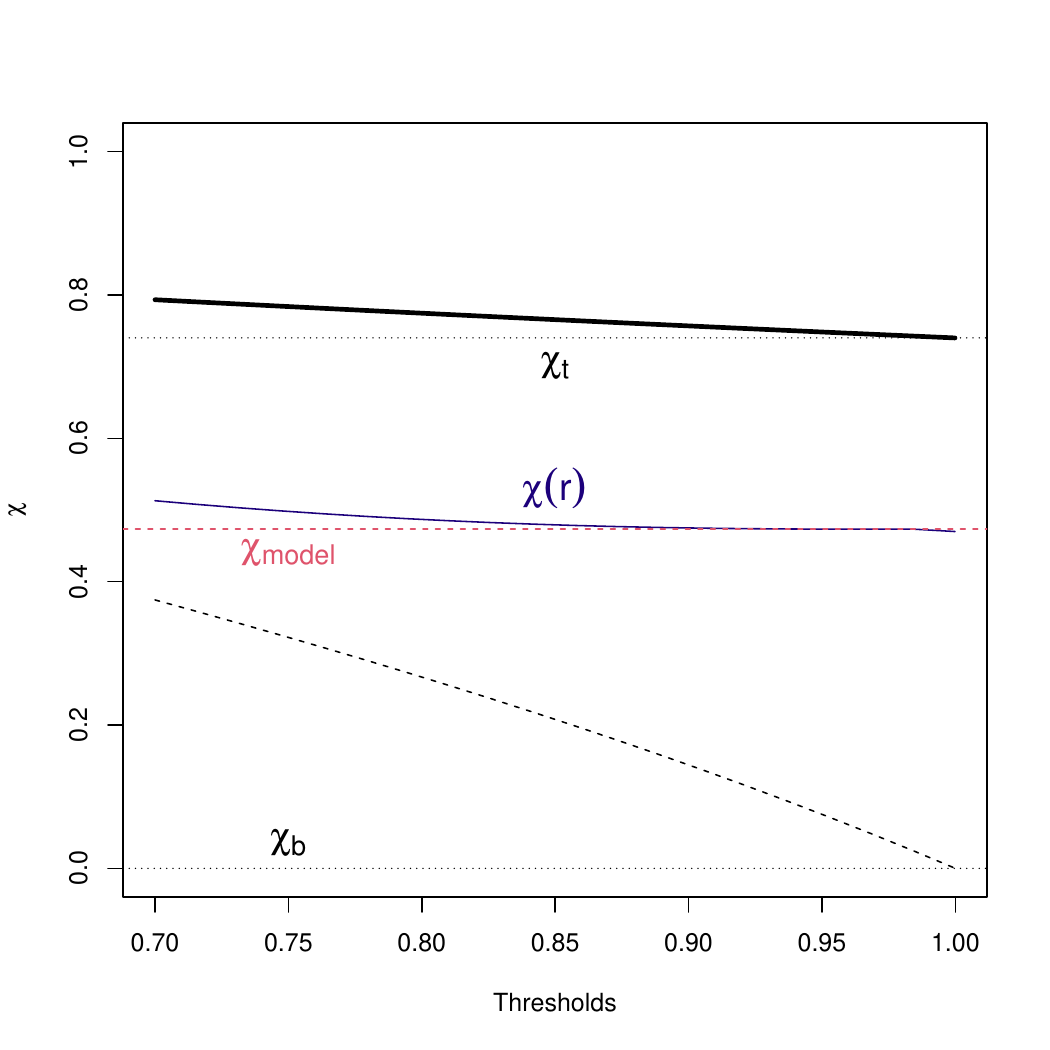}
    \caption{The blue line represents $\chi(r)$ for $r\in[0.7,1)$ with weighting function $\pi(u^*,v^*;\theta)=(u^*v^*)^{\theta}$ and $\theta=1.84444.$ The thick black lines represent the single copula models - Frank (dashed) and Gumbel (solid). The theoretical values for the Frank and Gumbel copulas based on Table 2 of Section 2.3 from the main text are represented by the horizontal dashed lines, and the value derived for the model is represented by the pink dashed line.} 
    \label{fig:chi1}
\end{figure}

\noindent For the vector of parameters $\bm \gamma=(3,1,1.844444),$ $c_1\approx 0.740079,$ $c_5=1$ and $c_7\approx 0.5630892.$ Thus, from equation \eqref{eq:chi}, we have $\chi\approx 0.473472.$ Moreover, from the numerical investigation, $\chi(r)\approx0.4699556$ with $r=0.9998779.$ Figure \ref{fig:chi1} shows this comparison.

\begin{figure}[H]
    \centering
    \includegraphics[width=0.7\textwidth]{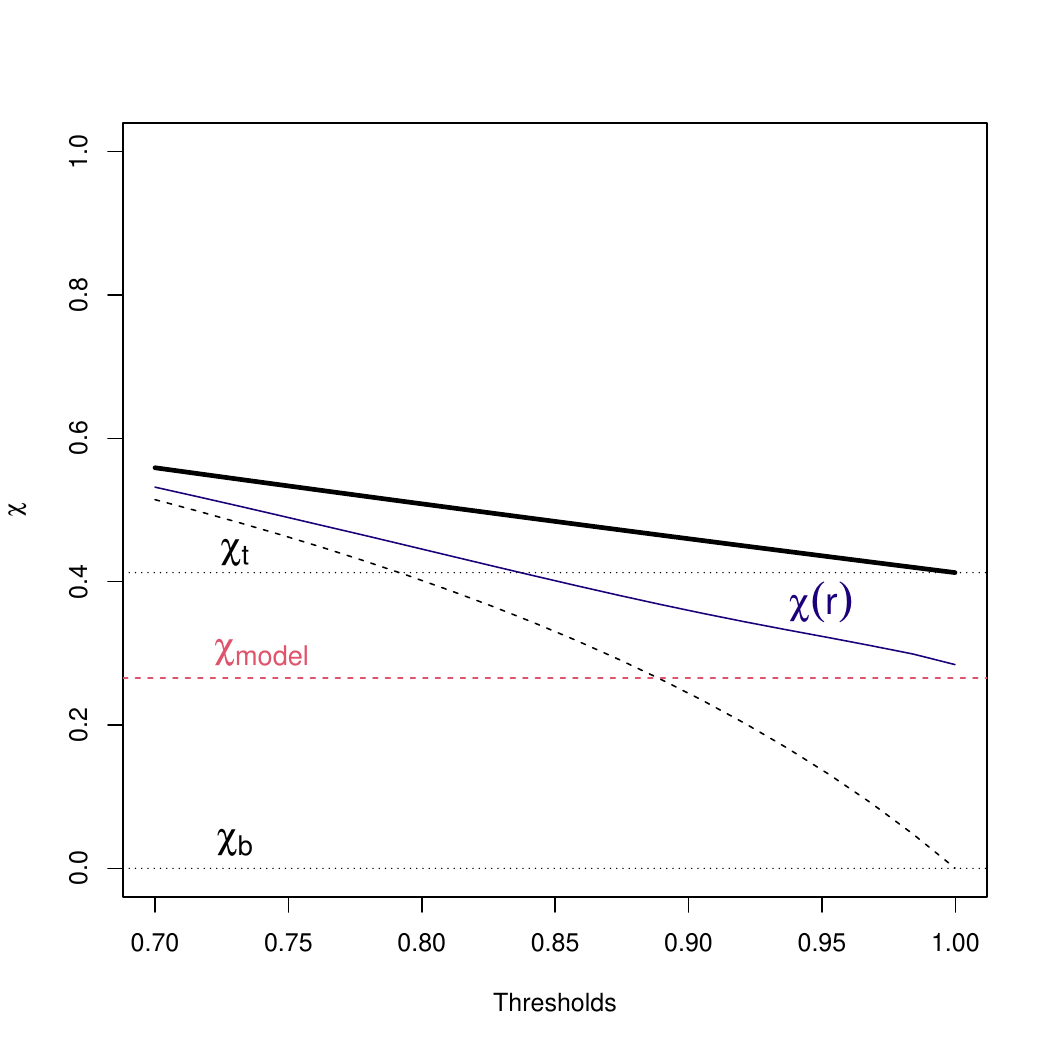} \caption{The blue line represents $\chi(r)$ for $r\in[0.7,1)$ with weighting function $\pi(u^*,v^*;\theta)=(u^*v^*)^{\theta}$ and $\theta=3.488889.$ The thick black lines represent the single copula models - Frank (dashed) and Gumbel (solid). The theoretical values for the Frank and Gumbel copulas based on Table 2 of Section 2.3 from the main text are represented by the horizontal dashed lines, and the value derived for the model is represented by the pink dashed line.} 
    \label{fig:chi2}
\end{figure}

\noindent For the vector of parameters $\bm \gamma=(1.5,3,3.488889),$ $c_1\approx 0.4125989,$ $c_5=1$ and $c_7\approx 0.5555462.$ Thus, from equation \eqref{eq:chi}, we have $\chi\approx 0.2652438.$ Moreover, from the numerical investigation, $\chi(r)\approx0.2842924$ with $r=0.9998779.$ Figure \ref{fig:chi2} shows this comparison.

\subsubsection{Extremal dependence $\eta$ for this case}

\noindent As $\chi>0,$ we should expect $\eta=1.$ Following equation (4) of Section 1.3 from the main text, we have
\begin{align*}
    \eta=&\lim_{r\rightarrow 1}\frac{\log\left(P[U^*>r]\right)}{\log\left(P[U^*>r,V^*>r]\right)} \\
    =&\lim_{r\rightarrow 1}\frac{\log\left[c_4(1-r)+c_5(1-r)^2+c_6(1-r)+c_7(1-r)^2+o\left((1-r)^2\right)\right]}{\log\left[c_1(1-r)+c_2(1-r)^2+c_3(1-r)^3+o\left((1-r)^3\right)\right]} \\
    \overset{\left(\frac{\infty}{\infty}\right)}{=}&\lim_{r\rightarrow 1}\frac{-c_4-c_6-2(c_5+c_7)(1-r)+o(1-r)}{-c_1-2c_2(1-r)-3c_3(1-r)^2+o\left((1-r)^2\right)}\\
    &\phantom{\hspace{10mm}}\times\frac{c_1+c_2(1-r)+c_3(1-r)^2+o((1-r)^2)}{c_4+c_6+(c_5+c_7)(1-r)+o\left(1-r\right)}\\
    =&\frac{c_4+c_6}{c_1}\frac{c_1}{c_4+c_6}=1,
\end{align*}
by L'H\^opital's Rule.

\subsection{Case 2.1: $c_b$ is a Frank copula, $c_t$ is a Gumbel copula and $\pi(u^*,v^*;\theta)=\exp\{-\theta(1-u^*)(1-v^*)\}$}
\noindent Let us now assume a different weighting function $\pi(u^*,v^*;\theta)=\exp\{-\theta(1-u^*)(1-v^*)\}.$ We have
\begin{align*}
f_{c_b}(u^*,v^*;\beta,\theta)=&[1-\exp\{-\theta(1-u^*)(1-v^*)\}]\\
&\times\frac{\beta(1-\exp\{-\beta\})\exp\{-\beta(u^*+v^*)\}}{\left[1-\exp\{-\beta\}-(1-\exp\{-\beta u^*\})(1-\exp\{-\beta v^*\})\right]^2}
\end{align*}
and
\begin{align*}
f_{c_t}(u^*,v^*;\alpha,\theta)=&\exp\{-\theta(1-u^*)(1-v^*)\}\frac{C_t(u^*,v^*;\alpha)}{u^*v^*}(xy)^{\alpha-1}\left(x^\alpha+y^\alpha\right)^{1\slash \alpha -2}\\
&\times\left[\left(x^\alpha+y^\alpha\right)^{1\slash \alpha} +\alpha-1\right],
\end{align*}
with $x=-\log(u^*),$ $y=-\log(v^*)$ and $C_t(u^*,v^*;\alpha)=\exp\left\{-\left(x^{\alpha}+y^{\alpha}\right)^{1\slash\alpha}\right\}.$

\subsubsection{Effect of the body copula $c_b$}

\noindent As the above case, a Taylor approximation of order 1 can be used about (1,1) with point $(1-s,1-t)$ for $\int_{r}^{1}\int_{r}^{1}f_{c_b}(u^*,v^*)\text{d}v^*\,\text{d}u^*,$ where $s,t\rightarrow 0.$ Therefore, for some norm $\|\cdot\|$ near 0, we have
\begin{equation*}
    f_{c_b}(1-s,1-t;\beta,\theta)= f_{c_b}(1,1)-s\frac{\partial f_{c_b}}{\partial s}(1,1)-t\frac{\partial f_{c_b}}{\partial t}(1,1)+\mathcal{O}\left(\|(s,t)\|^2\right),
    \end{equation*}
where
\begin{align*}
    \frac{\partial f_{c_b}}{\partial s}=&-\exp\{-\theta(1-s)(1-t)\}\\
    &\times\frac{2\beta^2(1-\exp\{-\beta\})(1-\exp\{-\beta t\})\exp\{-\beta(2s+t)\}}{[1-\exp\{-\beta\}-(1-\exp\{-\beta s\})(1-\exp\{-\beta t\})]^3}\\
    &-\exp\{-\theta(1-s)(1-t)\}\\
    &\times\frac{\beta(1-\exp\{-\beta\})[\theta(1-t)-\beta]\exp\{-\beta(s+t)\}}{[1-\exp\{-\beta\}-(1-\exp\{-\beta s\})(1-\exp\{-\beta t\})]^2}.
\end{align*}
At the point (1,1), $f_{c_b}(1,1)=0$ and $$\displaystyle{\frac{\partial f_{c_b}}{\partial s}(1,1)=\frac{\partial f_{c_b}}{\partial t}(1,1)=-\beta^2\left(1-\exp\{-\beta\}\right)^{-1}}.$$ So, 
\begin{equation*}
   f_{c_b}(1-s,1-t;\beta,\theta)=\beta^2\left(1-\exp\{-\beta\}\right)^{-1}(s+t)+\mathcal{O}\left(\|(s,t)\|^2\right), 
\end{equation*}
and we obtain
\begin{align*}
\int_{r}^{1}\int_{r}^{1}f_{c_b}&(u^*,v^*)\text{d}v^*\,\text{d}u^*\\
=&\int_{r}^{1}\int_{r}^{1}\beta^2\left(1-\exp\{-\beta\}\right)^{-1}(s+t)\text{d}t\,\text{d}s +\mathcal{O}\left((1-r)^4\right)\\
=&\beta^2\left(1-\exp\{-\beta\}\right)^{-1}(1-r)^3+\mathcal{O}\left((1-r)^4\right).
\end{align*}

\noindent Similarly, for $\int_{r}^{1}\int_{0}^{1}f_{c_b}(u^*,v^*)\text{d}v^*\,\text{d}u^*,$ a Taylor approximation of order 1 can be used about $(1,v^*)$ with point $(u^*,v^*).$ Thus, we have 
\begin{equation*}
    f_{c_b}(u^*,v^*;\beta,\theta)= f_{c_b}(1,v^*)+(u^*-1)\frac{\partial f_{c_b}}{\partial u^*}(1,v^*)+\mathcal{O}\left((u^*-1)^2\right),
\end{equation*}
where $f_{c_b}(1,v^*)=0$ and
\begin{align*}
\frac{\partial f_{c_b}}{\partial u^*}(1,v^*)=&-\frac{2\beta^2(1-\exp\{-\beta\})\exp\{-2\beta(1-v^*)\}}{(1-\exp\{-\beta\})^2}\\
    &-\frac{\beta\theta(1-v^*)\exp\{-\beta(1-v^*)\}}{1-\exp\{-\beta\}}\\
    &+\frac{\beta^2\exp\{-\beta(1-v^*)\}}{1-\exp\{-\beta\}}=A_{v^*,\beta,\theta}.
\end{align*}
So, $f_{c_b}(u^*,v^*)= A_{v^*,\beta,\theta}+\mathcal{O}\left((u^*-1)^2\right),$ and we obtain
\begin{align*}
\int_{r}^{1}\int_{0}^{1}f_{c_b}(u^*,v^*)\text{d}v^*\,\text{d}u^*=&\int_{r}^{1}\int_{0}^{1}A_{v^*,\beta,\theta}(u^*-1)\text{d}v^*\,\text{d}u^*+\mathcal{O}\left((1-r)^3\right) \\
=&\int_{0}^{1}A_{v^*,\beta,\theta}\int_{r}^{1}(u^*-1)\text{d}u^*\,\text{d}v^*+\mathcal{O}\left((1-r)^3\right) \\
=& - \frac{1}{2}(1-r)^2\underbrace{\int_{0}^{1}A_{v^*,\beta,\theta}\text{d}v^*}_{B_{\beta,\theta}}+\mathcal{O}\left((1-r)^3\right) \\
=&-\frac{B_{\beta,\theta}}{2}(1-r)^2+\mathcal{O}\left((1-r)^3\right)
\end{align*}

\subsubsection{Effect of the tail copula $c_t$}

\noindent Let us again write $u^*$ and $v^*$ in terms of $s$ and $t,$ where $s,t> 0$ and $u^*=1-s+o(s)$ and $v^*=1-t+o(t)$ as $s,t\rightarrow 0.$ As before, this describes the behaviour of $u^*$ and $v^*$ as they tend to 1. For the weighting function term of $f_{c_t}$, we have 
\begin{align*}
    \exp\{-\theta(1-u^*)(1-v^*)\}= \exp\{-\theta st\}+o(s)+o(t),
\end{align*}
as $s,t\rightarrow 0.$

\noindent Similarly to the previous case, we consider $x=-\log(u^*)>y=-\log(v^*).$ For $(u^*,v^*)\rightarrow (1,1),$ i.e, $s\rightarrow 0$ and $t\rightarrow 0,$ with $t\slash s\rightarrow c$ for $c\in(0,1),$ the copula density term follows asymptotically 
\begin{equation*}
    c_t(u^*,v^*;\alpha)\sim (\alpha-1)x^{-\alpha}y^{\alpha-1}\left[1+\left(\frac{y}{x}\right)^\alpha\right]^{1\slash \alpha -2}.
\end{equation*}
And, when $x<y,$ i.e, $s\rightarrow 0$ and $t\rightarrow 0,$ with $t\slash s\rightarrow c$ for $c\in(1,\infty),$
\begin{equation*}
    c_t(u^*,v^*;\alpha)\sim (\alpha-1)y^{-\alpha}x^{\alpha-1}\left[1+\left(\frac{x}{y}\right)^\alpha\right]^{1\slash \alpha -2}.
\end{equation*}

\noindent Finally, $x=s+o(s)$ and $y=t+o(t)$ as $s,t\rightarrow0.$ Thus, considering the symmetry between cases $x>y$ and $x<y,$ and recalling $u^*=1-s+o(s)$ and $v^*=1-t+o(t),$
\begin{equation*}
    \int_{r}^{1}\int_{r}^{1}f_{c_t}(u^*,v^*)\text{d}v^*\,\text{d}u^*= P[1-S>r,1-T>r]=2P[S<1-r,T<s].
\end{equation*}

\noindent So, we have
\begin{align*}
    &P[S<1-r,T<s]=\int_{0}^{1-r}\int_{0}^{s}f_{c_t}^*(s,t;\alpha,\theta)\text{d}t\,\text{d}s \\
    &=(\alpha-1)\int_{0}^{1-r}\int_{0}^{s} \exp\{-\theta st\}s^{-\alpha}t^{\alpha-1}\left[1+\left(\frac{t}{s}\right)^\alpha\right]^{1\slash \alpha -2} \text{d}t\,\text{d}s +o\left((1-r)^2\right) \\
    &=(\alpha-1)\int_{0}^{1-r}s^{-\alpha}\underbrace{\int_{0}^{s} \exp\{-\theta st\}t^{\alpha-1}\left[1+\left(\frac{t}{s}\right)^\alpha\right]^{1\slash \alpha -2} \text{d}t}_{A(s)}\,\text{d}s +o\left((1-r)^2\right)
\end{align*}
as $r\rightarrow 1.$ Evaluating $A(s)$ by parts, we get
\begin{align*}
    \int_{0}^{s} \exp\{-\theta st\}t^{\alpha-1}\left[1+\left(\frac{t}{s}\right)^\alpha\right]^{1\slash \alpha -2} \text{d}t =&\frac{2^{1\slash \alpha -1}\exp\{-\theta s^2\}s^\alpha}{1-\alpha}-\frac{s^\alpha}{1-\alpha}\\
    &+\frac{\theta s^{\alpha+2}}{1-\alpha}C_{\alpha}-\frac{\theta^2s^{\alpha+4}}{1-\alpha} C^*_{\alpha},
\end{align*}
with $C_{\alpha}=\displaystyle{\int_0^1 (1+q^\alpha)^{1\slash \alpha -1}}\text{d}q$ and $C^*_{\alpha}=\displaystyle{\int_0^1 q(1+q^\alpha)^{1\slash \alpha -1}}\text{d}q.$ By substituting $A(s)$ in the outer integral, we obtain
\begin{align*}
    P[S<1-r,T<s]=&-2^{1\slash \alpha -1}\int_{0}^{1-r}e^{-\theta s^2}\text{d}s+\int_0^{1-r}\text{d}s \\
    &-\theta C_{\alpha}\int_0^{1-r}s^2\text{d}s+\theta^2 C^*_{\alpha}\int_0^{1-r}s^4\text{d}s+o\left((1-r)^2\right) \\
    =&-2^{1\slash \alpha -1}\int_{0}^{1-r}(1-\theta s^2)\text{d}s+(1-r)+o\left((1-r)^2\right)\\
    =&(1-2^{1\slash \alpha -1})(1-r)+o\left((1-r)^2\right),
\end{align*}
as $r\rightarrow 1$ and where $\exp\{-\theta s^2\}= 1-\theta s^2+\mathcal{O}\left((1-r)^4\right)$ as $s\rightarrow 0.$ Thus,
\begin{align*}
   \int_{r}^{1}\int_{r}^{1}f_{c_t}(u^*,v^*)\text{d}v^*\,\text{d}u^*= & \,2(1-2^{1\slash \alpha -1})(1-r)+o\left((1-r)^2\right) \\
    =&(2-2^{1\slash \alpha})(1-r)+o\left((1-r)^2\right),
\end{align*}
as $r\rightarrow 1.$

\noindent As before, for $\int_{r}^{1}\int_{0}^{1}f_{c_t}(u^*,v^*)\text{d}v^*\,\text{d}u^*,$ we take the change of variable $y=xz,$ with $z=y\slash x \in \mathbb{R^+},$ so we have $u^*=\exp\{-x\}$ and $v^*=\exp\{-xz\}.$ Thus, we obtain
\begin{align*}
\int_{r}^{1}&\int_{0}^{1}f_{c_t}(u^*,v^*)\text{d}v^*\,\text{d}u^* \\
=&\int_{r}^{1}\int_{0}^{1}\exp\{-\theta(1-u^*)(1-v^*)\}\frac{C_t(u^*,v^*;\alpha)}{u^*v^*}(xy)^{\alpha-1}\left(x^\alpha+y^\alpha\right)^{1\slash \alpha -2}\\
&\times\left[\left(x^\alpha+y^\alpha\right)^{1\slash \alpha} +\alpha-1\right]\text{d}v^*\,\text{d}u^* \\
=&\int_{0}^{-\log(r)}\hspace{-3mm}\int_{0}^{\infty}\hspace{-2mm}\exp\{-\theta(1-\exp\{-x\}\!-\!\exp\{-xz\}\!+\!\exp\{-x-xz\})\!-\!x(1+z^\alpha)^{1/\alpha}\}\\
&\times z^{\alpha-1}\left(1+z^\alpha\right)^{1\slash \alpha -2}\left[x\left(1+z^\alpha\right)^{1\slash \alpha} +\alpha-1\right]\text{d}z\,\text{d}x 
\end{align*} 

\noindent We have $\exp\{-x\}=\displaystyle{1-x+\frac{x^2}{2}+\mathcal{O}\left(x^3\right)},$ $\exp\{-xz\}=\displaystyle{1-xz+\frac{x^2z^2}{2}+\mathcal{O}\left(x^3\right)}$ and $\exp\{-x(1+z)\}=\displaystyle{1-x(1+z)+\frac{x^2(1+z)^2}{2}+\mathcal{O}\left(x^3\right)}$ as $x\rightarrow 0.$ So, the exponential term 
\begin{align*}
    \exp&\{-\theta(1-\exp\{-x\}-\exp\{-xz\}+\exp\{-x(1+z)\})-x(1+z^\alpha)^{1/\alpha}\} \\
    =&\exp\left\{-\theta\left[1-\left(1-x+\frac{x^2}{2}\right)-\left(1-xz+\frac{x^2z^2}{2}\right)\right.\right. \\
     & \left.\left.+\left(1-x(1+z)+\frac{x^2(1+z)^2}{2}\right)\right]-x(1+z^\alpha)^{1/\alpha}\right\}+\mathcal{O}\left(x^3\right)\\
     =&\exp\left\{-\theta x^2z-x(1+z^\alpha)^{1/\alpha}\right\}+\mathcal{O}\left(x^3\right) =\exp\{-x(1+z^\alpha)^{1/\alpha}\} +\mathcal{O}\left(x^2\right)
\end{align*}
as $x\rightarrow 0.$ 

\noindent  So, we have 
\begin{align*}
\int_{r}^{1}\int_{0}^{1}&f_{c_t}(u^*,v^*)\text{d}v^*\,\text{d}u^*\\
=&\int_{0}^{-\log(r)}\int_{0}^{\infty}\exp\{-x(1+z^\alpha)^{1/\alpha}\}z^{\alpha-1}\left(1+z^\alpha\right)^{1\slash \alpha -2}\\
&\times\left[x\left(1+z^\alpha\right)^{1\slash \alpha}+\alpha-1\right]\text{d}z\,\text{d}x\\
=&\int_{0}^{-\log(r)}\int_{x}^{\infty} \exp\{-w\}\left(\frac{x}{w}\right)^\alpha (w+\alpha-1)\frac{1}{x}\text{d}w\,\text{d}x \\
=&\int_{0}^{-\log(r)}\int_{0}^{w}x^{\alpha-1} \exp\{-w\} w^{-\alpha} (w+\alpha-1)\text{d}x\,\text{d}w \\
&+ \int_{-\log(r)}^{\infty}\int_{0}^{-\log(r)}x^{\alpha-1} \exp\{-w\} w^{-\alpha} (w+\alpha-1)\text{d}x\,\text{d}w \\
=&\int_{0}^{-\log(r)} \exp\{-w\} w^{-\alpha} (w+\alpha-1)\left[\frac{x^\alpha}{\alpha}\right]^{w}_{0}\text{d}w \\
&+ \int_{-\log(r)}^{\infty} \exp\{-w\} w^{-\alpha} (w+\alpha-1)\left[\frac{x^\alpha}{\alpha}\right]^{-\log(r)}_{0}\text{d}w\\
&\frac{1}{\alpha}\int_{0}^{-\log(r)} \exp\{-w\}(w+\alpha-1)\text{d}w \\
&+\frac{[-\log(r)]^\alpha}{\alpha}\int_{-\log(r)}^{\infty} \exp\{-w\} w^{-\alpha} (w+\alpha-1)\text{d}w
\end{align*} 

\begin{align*}
\phantom{\int_{r}^{1}\int_{0}^{1}}=&\frac{1}{\alpha}\int_{0}^{-\log(r)} w\exp\{-w\}\text{d}w + \frac{\alpha-1}{\alpha}\int_{0}^{-\log(r)} \exp\{-w\}\text{d}w\\
&+\frac{[-\log(r)]^\alpha}{\alpha}\int_{1}^{\infty} r^t [-\log(r)t]^{-\alpha} (-\log(r)t+\alpha-1)(-\log(r)\text{d}t \\
=& \frac{1}{\alpha}\left(r\log(r)-r+1\right) + \frac{\alpha-1}{\alpha}\left(-r+1\right)\\
&+\frac{-\log(r)[-\log(r)]^{-\alpha}[-\log(r)]^\alpha}{\alpha}\int_{1}^{\infty} r^t t^{-\alpha} (-\log(r)t+\alpha-1)\text{d}t \\
=&\frac{r\log(r)}{\alpha}+1-r -\frac{\log(r)}{\alpha}\int_{1}^{\infty} r^t t^{-\alpha} (-\log(r)t+\alpha-1)\text{d}t,
\end{align*} 
where $w=x(1+z^\alpha)^{1\slash \alpha}$ and $t=\frac{w}{-\log(r)}.$ As $r\rightarrow 1,$ we have
\begin{align*}
    \int_{r}^{1}\int_{0}^{1}&f_{c_t}(u^*,v^*)\text{d}v^*\,\text{d}u^*\\
    =&\frac{r\log(r)}{\alpha}+1-r-\frac{\log(r)}{\alpha}\int_{1}^{\infty} t^{-\alpha} (\alpha-1)\text{d}t\\
    =&\frac{r\log(r)}{\alpha}+1-r-\frac{\log(r)}{\alpha}\\
    =& \left(1-\frac{\log(r)}{\alpha}\right)(1-r)=\left(1-\frac{-(1-r)+\mathcal{O}\left((1-r)^2\right)}{\alpha}\right)(1-r) \\
    =& (1-r)+\frac{1}{\alpha}(1-r)^2+\mathcal{O}\left((1-r)^3\right).
\end{align*}

\subsubsection{Extremal dependence $\chi$ for this case}

\noindent Let 
\begin{align*}
    c_1&=2-2^{1\slash \alpha} =\chi_{Gumbel}, && c_3=1\slash\alpha,\\
    c_2&=1, && c_4=-B_{\beta,\theta}\slash 2. 
\end{align*}
we then have
\begin{align}
    \chi=&\lim_{r\rightarrow 1}\frac{c_1(1-r)+o\left((1-r)^2\right)}{c_2(1-r)+c_3(1-r)^2+c_4(1-r)^2+o\left((1-r)^2\right)} \nonumber \\
     =&\lim_{r\rightarrow 1}\left(\frac{c_1}{c_2}-\frac{c_3+c_4}{c_2^2}(1-r)+\mathcal{O}\left((1-r)^2\right)\right) =\frac{c_1}{c_2}=2-2^{1\slash\alpha}  \label{eq:chi1}
\end{align}

\begin{figure}[H]
    \centering
    \includegraphics[width=0.7\textwidth]{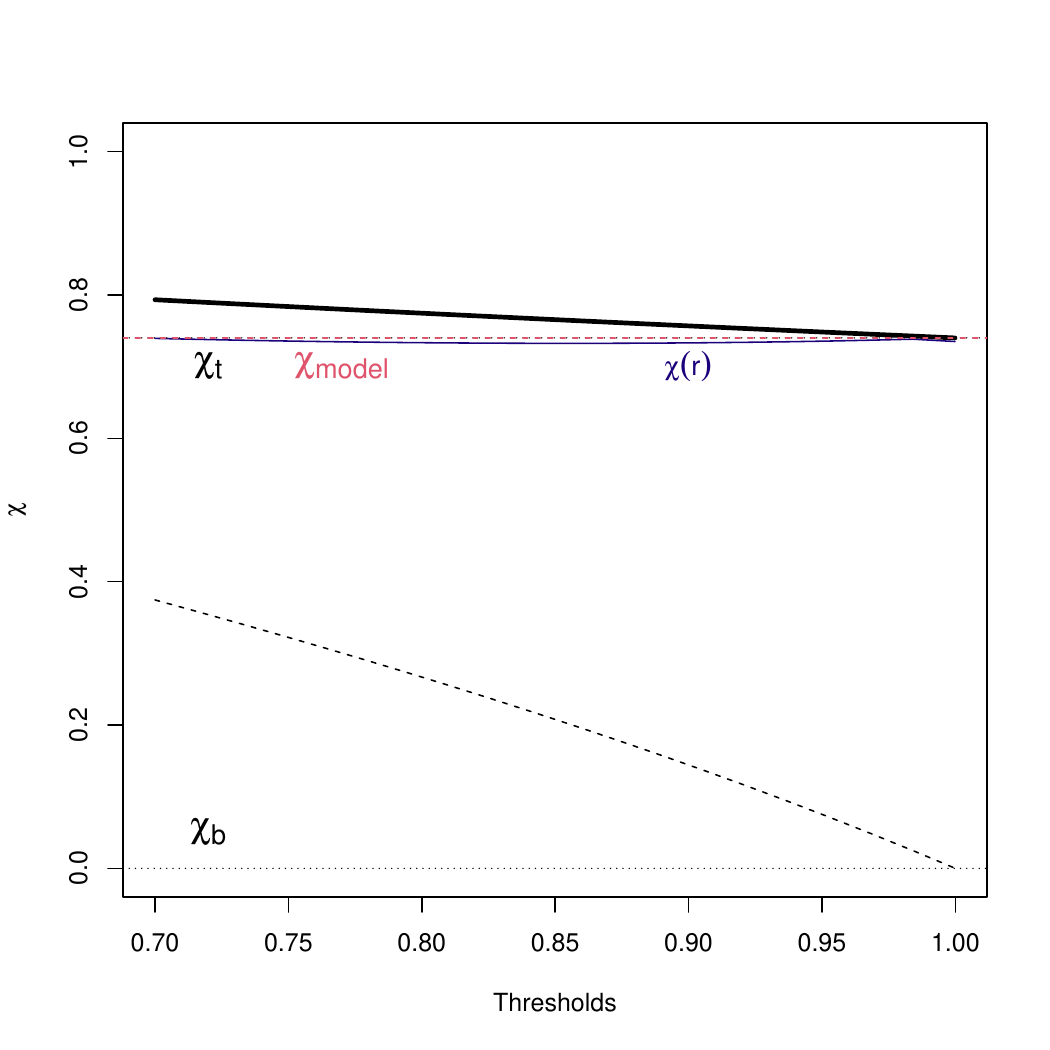}
    \caption{The blue line represents $\chi(r)$ for $r\in[0.7,1)$ with weighting function $\pi(u^*,v^*;\theta)=\exp\{-\theta(1-u^*)(1-v^*)\}$ and $\theta=1.84444.$ The thick black lines represent the single copula models - Frank (dashed) and Gumbel (solid). The theoretical values for the Frank and Gumbel copulas based on Table 2 of Section 2.3 from the main text are represented by the horizontal dashed lines, and the value derived for the model is represented by the pink dashed line. Note that the theoretical value for the Gumbel copula, $\chi_t,$ is the same as the one derived for the model, $\chi_{\text{Model}}.$} 
    \label{fig:chi1a}
\end{figure}

\noindent For the vector of parameters $\bm \gamma=(3,1,1.844444),$ $c_1\approx 0.740079$ and $c_5=1.$ Thus, from equation \eqref{eq:chi1}, we have $\chi\approx 0.740079.$ Moreover, from the numerical investigation, $\chi(r)\approx 0.7350891$ with $r=0.9998779.$ Figure \ref{fig:chi1a} shows this comparison.

\begin{figure}[H]
    \centering
    \includegraphics[width=0.7\textwidth]{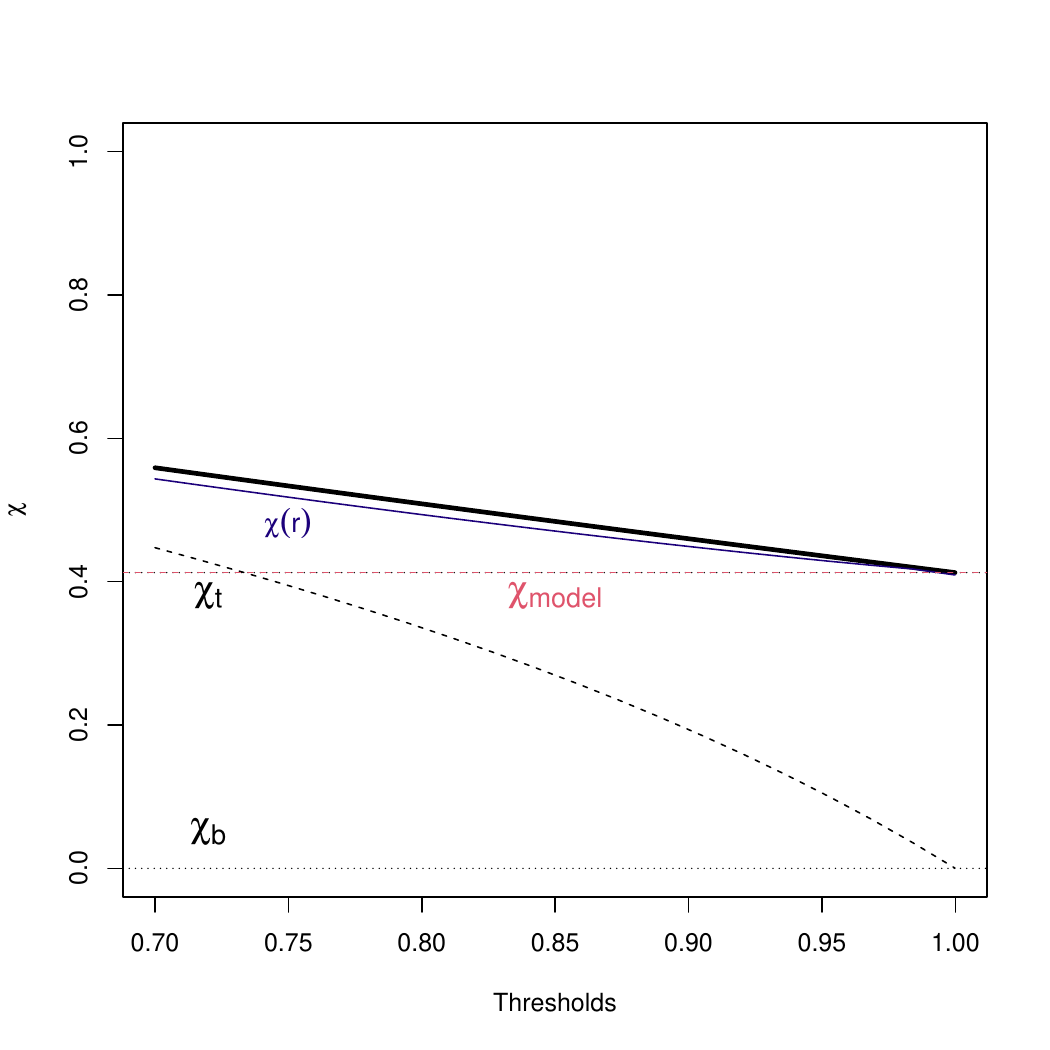} \caption{The blue line represents $\chi(r)$ for $r\in[0.7,1)$ with weighting function $\pi(u^*,v^*;\theta)=\exp\{-\theta(1-u^*)(1-v^*)\}$ and $\theta=3.488889.$ The thick black lines represent the single copula models - Frank (dashed) and Gumbel (solid). The theoretical values for the Frank and Gumbel copulas based on Table 2 of Section 2.3 from the main text are represented by the horizontal dashed lines, and the value derived for the model is represented by the pink dashed line. Note that the theoretical value for the Gumbel copula, $\chi_t,$ is the same as the one derived for the model, $\chi_{\text{Model}}.$} 
    \label{fig:chi2a}
\end{figure}

\noindent For the vector of parameters $\bm \gamma=(1.5,2,3.488889),$ $c_1\approx 0.4125989$ and $c_5=1.$ Thus, from equation \eqref{eq:chi}, we have $\chi\approx 0.4125989.$ Moreover, from the numerical investigation, $\chi(r)\approx 0.4093587$ with $r=0.9998779.$ Figure \ref{fig:chi2a} shows this comparison.

\subsubsection{Extremal dependence $\eta$ for this case}

\noindent As $\chi>0,$ we should expect $\eta=1.$ Following equation (4) of Section 1.3 from the main text, we have

\begin{align*}
    \eta=&\lim_{r\rightarrow 1}\frac{\log\left(P[U^*>r]\right)}{\log\left(P[U^*>r,V^*>r]\right)} \\
    =&\lim_{r\rightarrow 1}\frac{\log\left[c_2(1-r)+c_3(1-r)^2+c_4(1-r)^2+o\left((1-r)^2\right)\right]}{\log\left[c_1(1-r)+o\left((1-r)^2\right)\right]} \\
    \overset{\left(\frac{\infty}{\infty}\right)}{=}&\lim_{r\rightarrow 1}\frac{-c_2-2(c_3+c_4)(1-r)+o\left(1-r\right)}{-c_1+o\left(1-r\right)}\frac{c_1+o\left((1-r)^2\right)}{c_2+(c_3+c_4)(1-r)+o\left((1-r)^2\right)}\\
    =&\frac{c_2}{c_1}\frac{c_1}{c_2}=1
\end{align*}
by L'H\^opital's Rule.

\section{Extremal dependence properties: numerical investigation}

\begin{figure}[p]
    \vspace{-2cm}
    \centering
    \begin{subfigure}[b]{0.75\textwidth}
        \includegraphics[width=\textwidth]{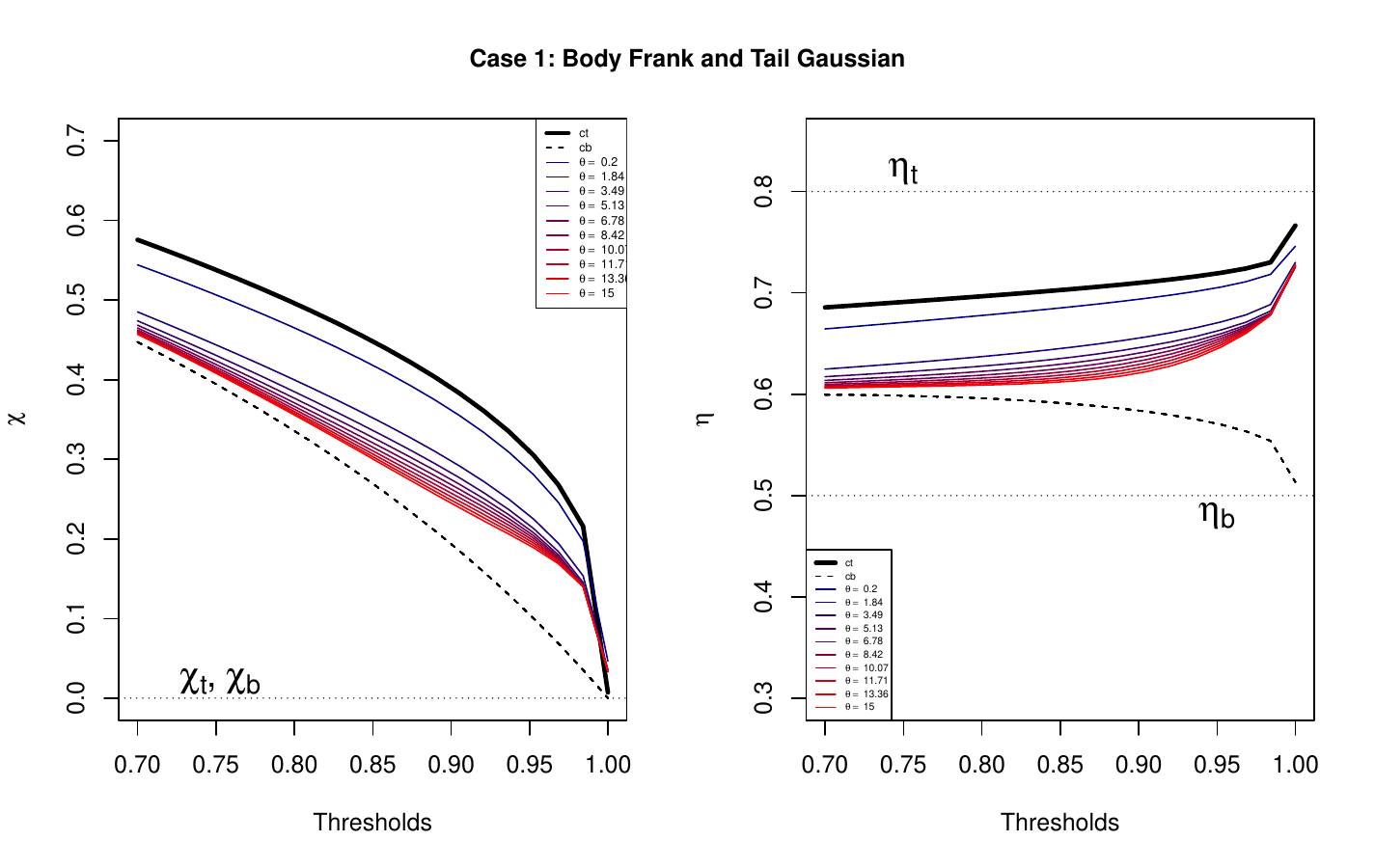}
    \end{subfigure}
    \begin{subfigure}[b]{0.75\textwidth}
        \includegraphics[width=\textwidth]{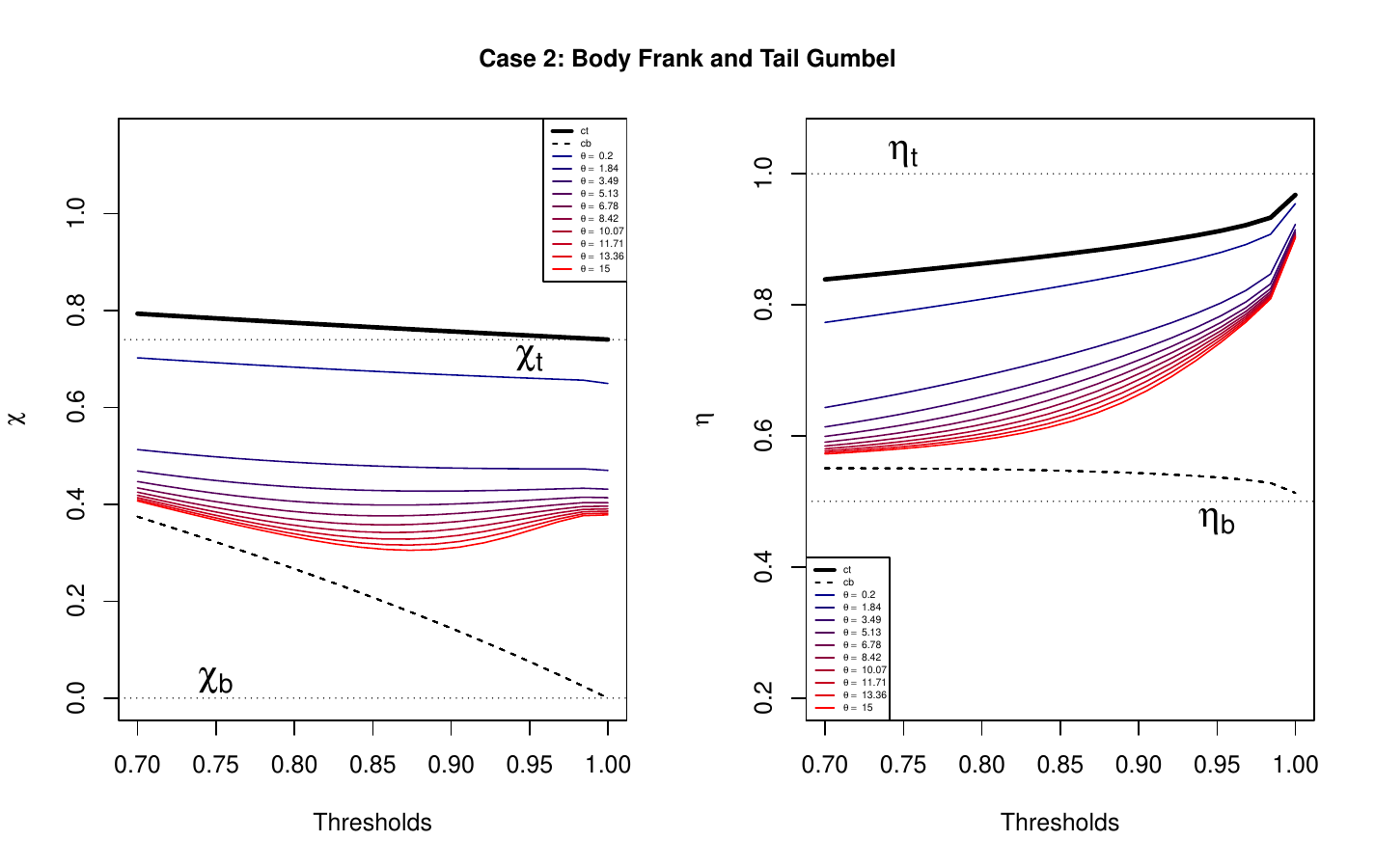}
    \end{subfigure}
    \begin{subfigure}[b]{0.75\textwidth}
        \includegraphics[width=\textwidth]{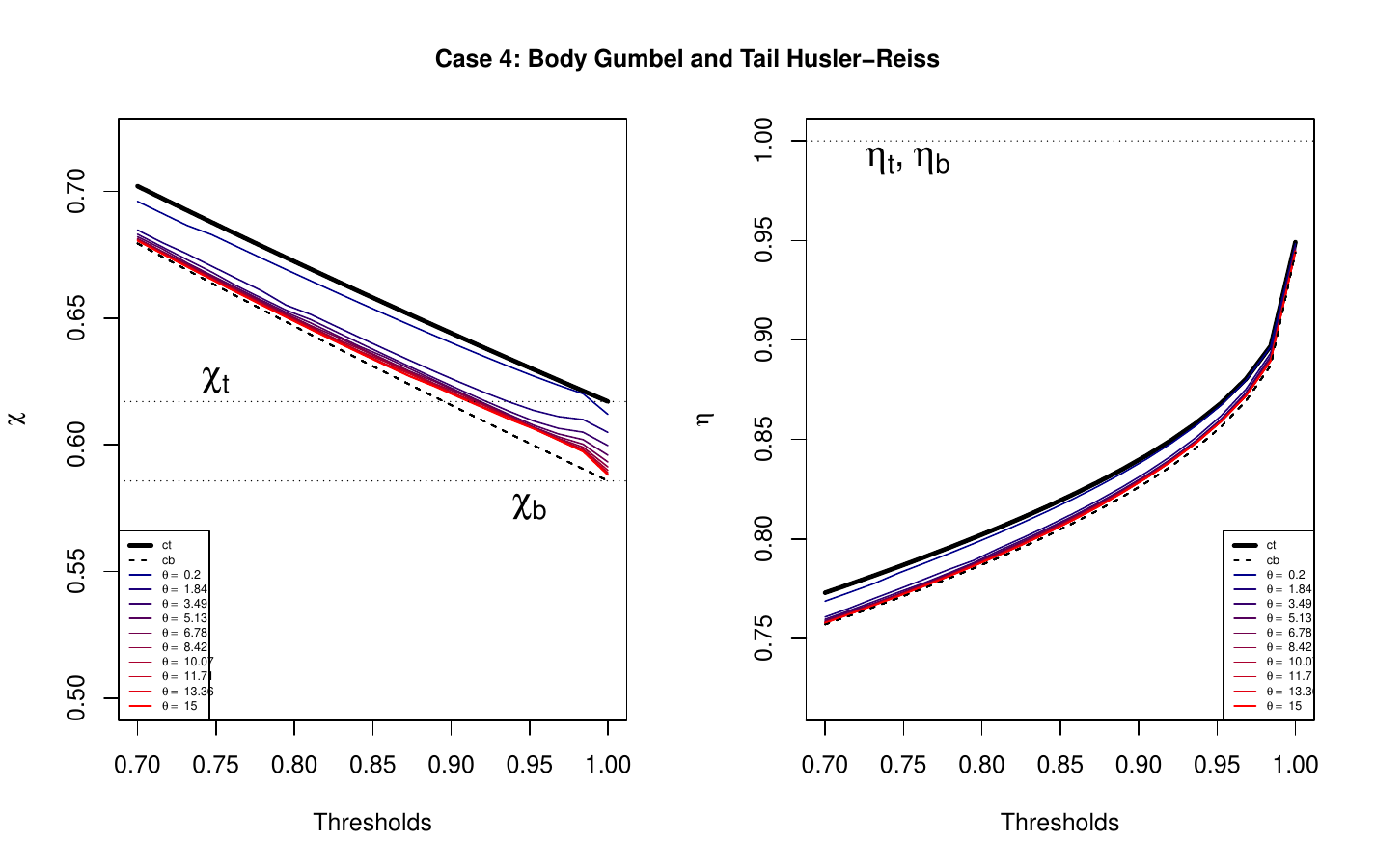}
    \end{subfigure}
    \caption{$\chi(r)$ and $\eta(r)$ for $r\in[0.7,1)$ with weighting function $\pi(u^*,v^*;\theta)=(u^*v^*)^{\theta}.$}
        \label{fig:depprop1}
\end{figure}

\begin{figure}[p]
\vspace{-2cm}
    \centering
    \begin{subfigure}[b]{0.75\textwidth}
        \includegraphics[width=\textwidth]{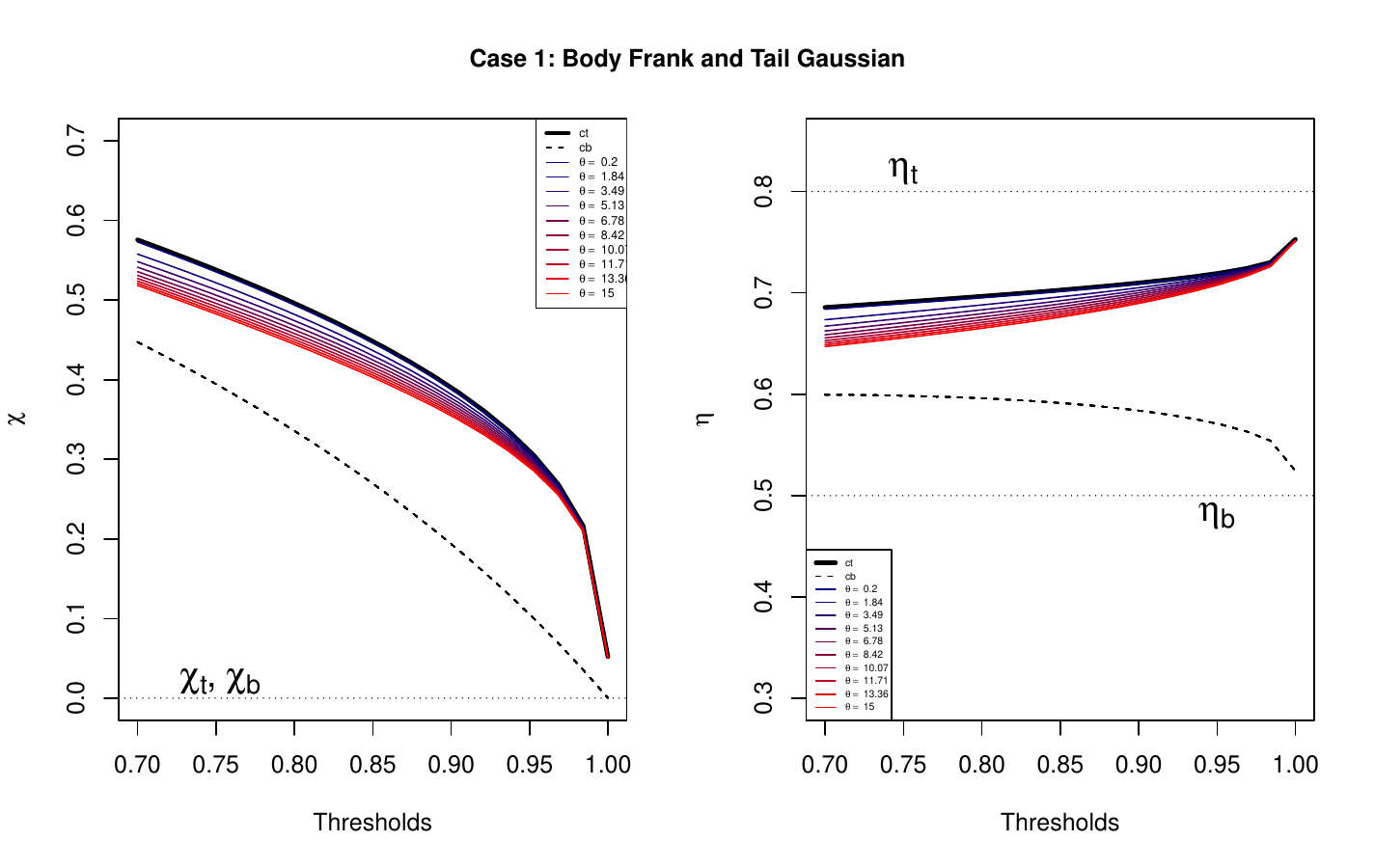}
    \end{subfigure}
    \begin{subfigure}[b]{0.75\textwidth}
        \includegraphics[width=\textwidth]{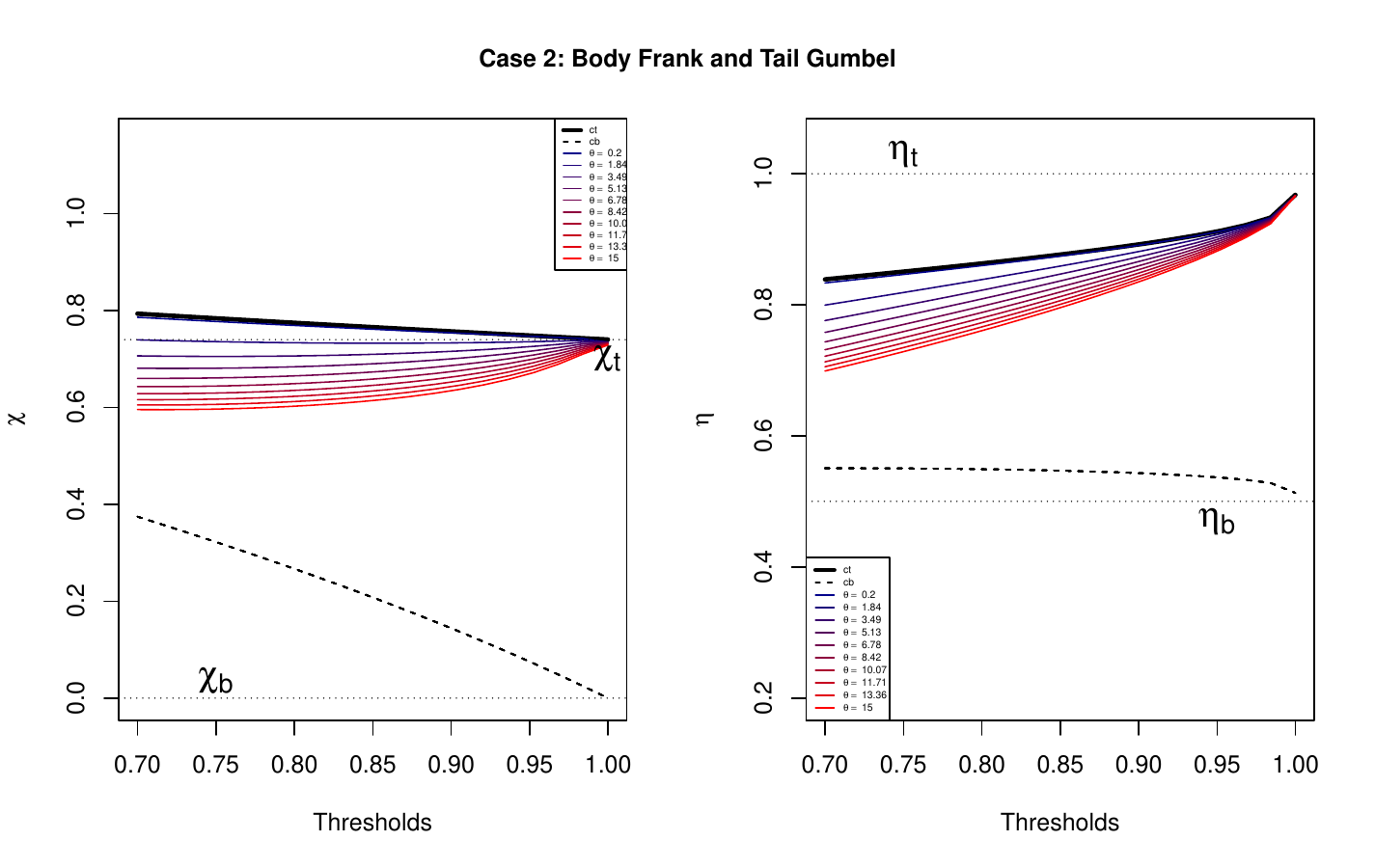}
    \end{subfigure}
    \begin{subfigure}[b]{0.75\textwidth}
        \includegraphics[width=\textwidth]{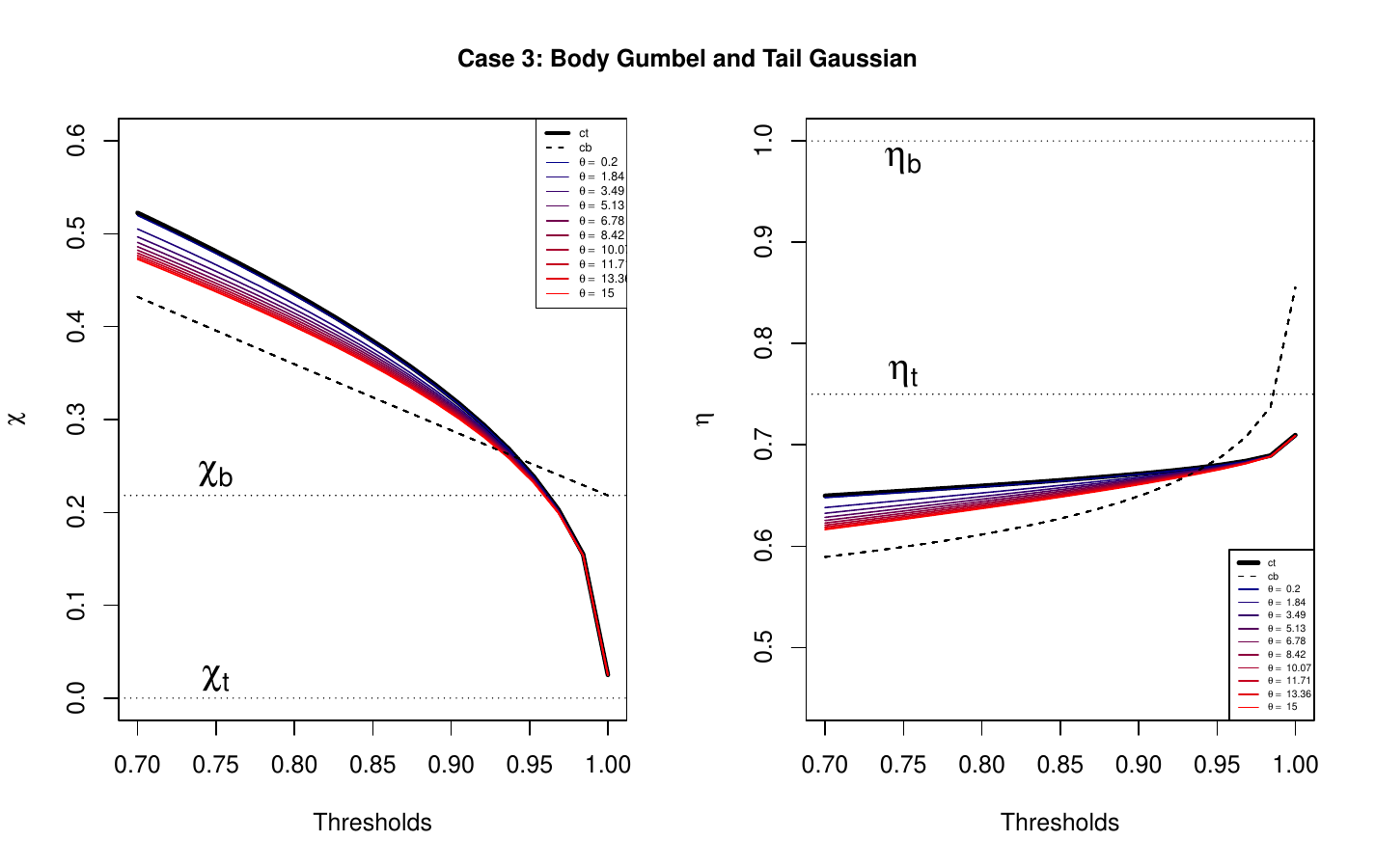}
    \end{subfigure}
    \vspace{-2mm}
    \caption{$\chi(r)$ and $\eta(r)$ with weighting function $\pi(u^*,v^*;\theta)=\exp\{-\theta(1-u^*)(1-v^*)\},$  for $r\in[0.7,1).$}
        \label{fig:depprop2}
\end{figure}

\noindent Figures \ref{fig:depprop1} and \ref{fig:depprop2} show the results of the numerical study presented in Section 2.3 of the main text for the remaining three models considered.

\clearpage
\section{Ozone and temperature analysis for Weybourne, UK}

\begin{figure}[H]
    \centering
    \begin{subfigure}[b]{0.45\textwidth}
        \includegraphics[width=\textwidth]{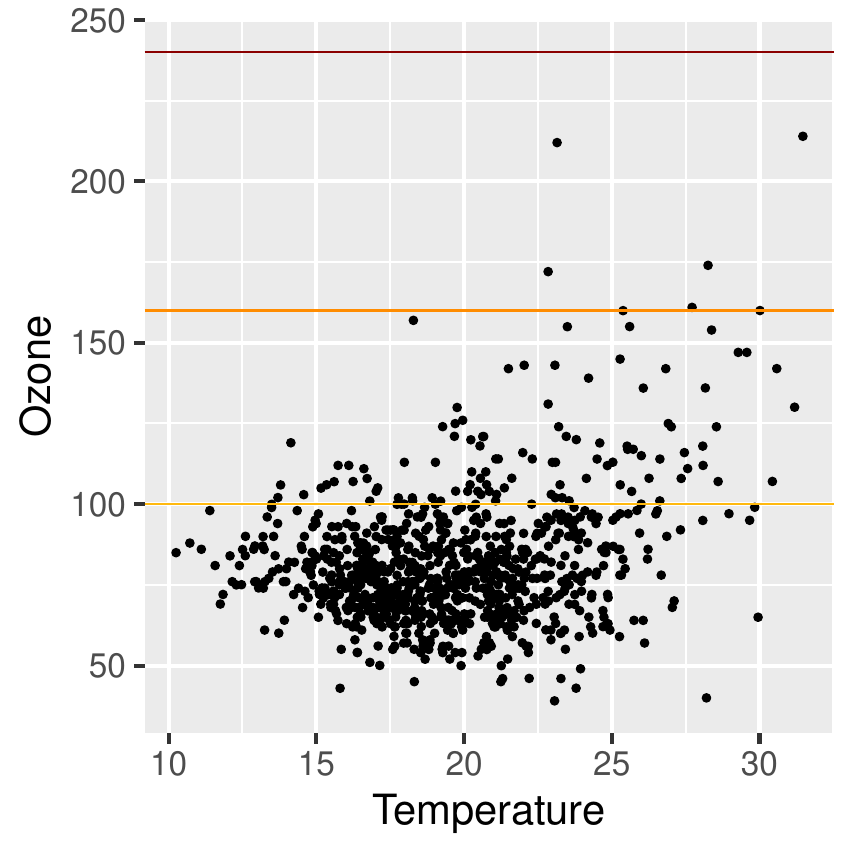}
        \caption{Daily maxima of temperature and ozone. The moderate, high and very high DAQI are represented by the yellow, orange and red lines, respectively.$\qquad\qquad\qquad$ \linebreak  $\qquad \qquad \qquad \qquad\qquad\qquad$}
        \label{subfig:weybournescatter}
    \end{subfigure}
    \hfill
    \begin{subfigure}[b]{0.45\textwidth}
        \includegraphics[width=\textwidth]{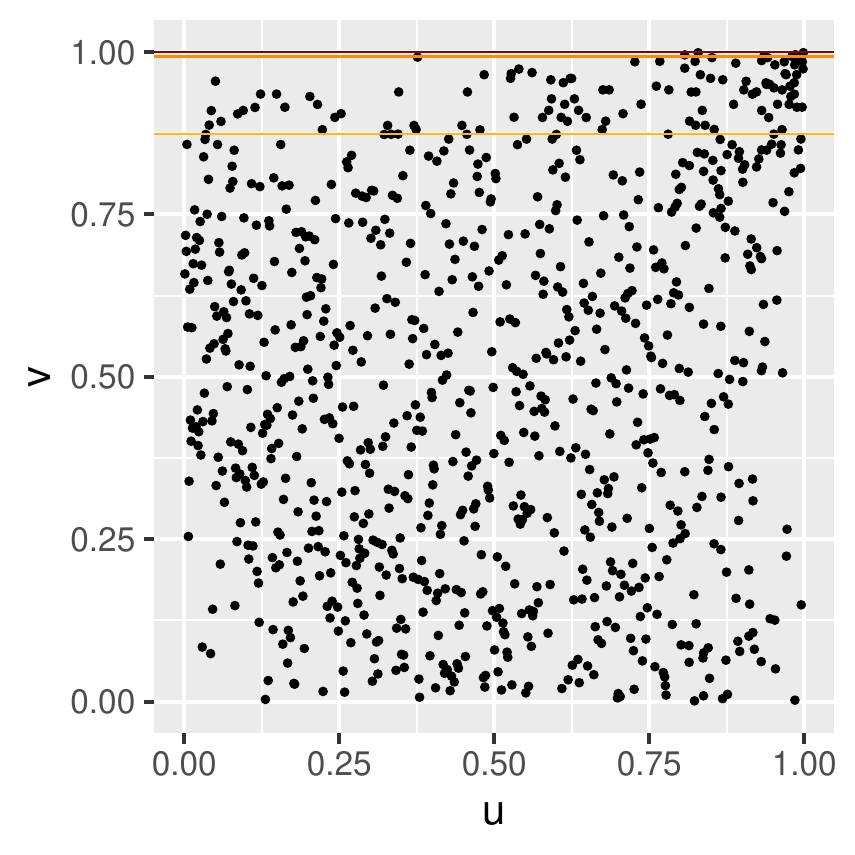}
        \caption{Daily maxima of temperature (u) and ozone (v) on uniform margins. The corresponding moderate, high and very high DAQI are represented by the yellow, orange and red lines, respectively.}
        \label{subfig:weybourneuniform}
    \end{subfigure}
    \caption{Summer data from 2010 to 2019 for Weybourne, UK.}
    \label{fig:weybournedataset}
\end{figure}

\noindent Following the same structure as the case study in Section 4 in the main paper, the analysis for the summers of 2010 to 2019 of Weybourne, UK, is presented here. Figures \ref{subfig:weybournescatter} and \ref{subfig:weybourneuniform} show the scatterplots of the daily maxima of temperature and the daily maxima of ozone on the original scale and on uniform margins, respectively.

\subsection{Model fitting}

\begin{figure}[H]
    \centering
    \includegraphics[width=0.9\textwidth]{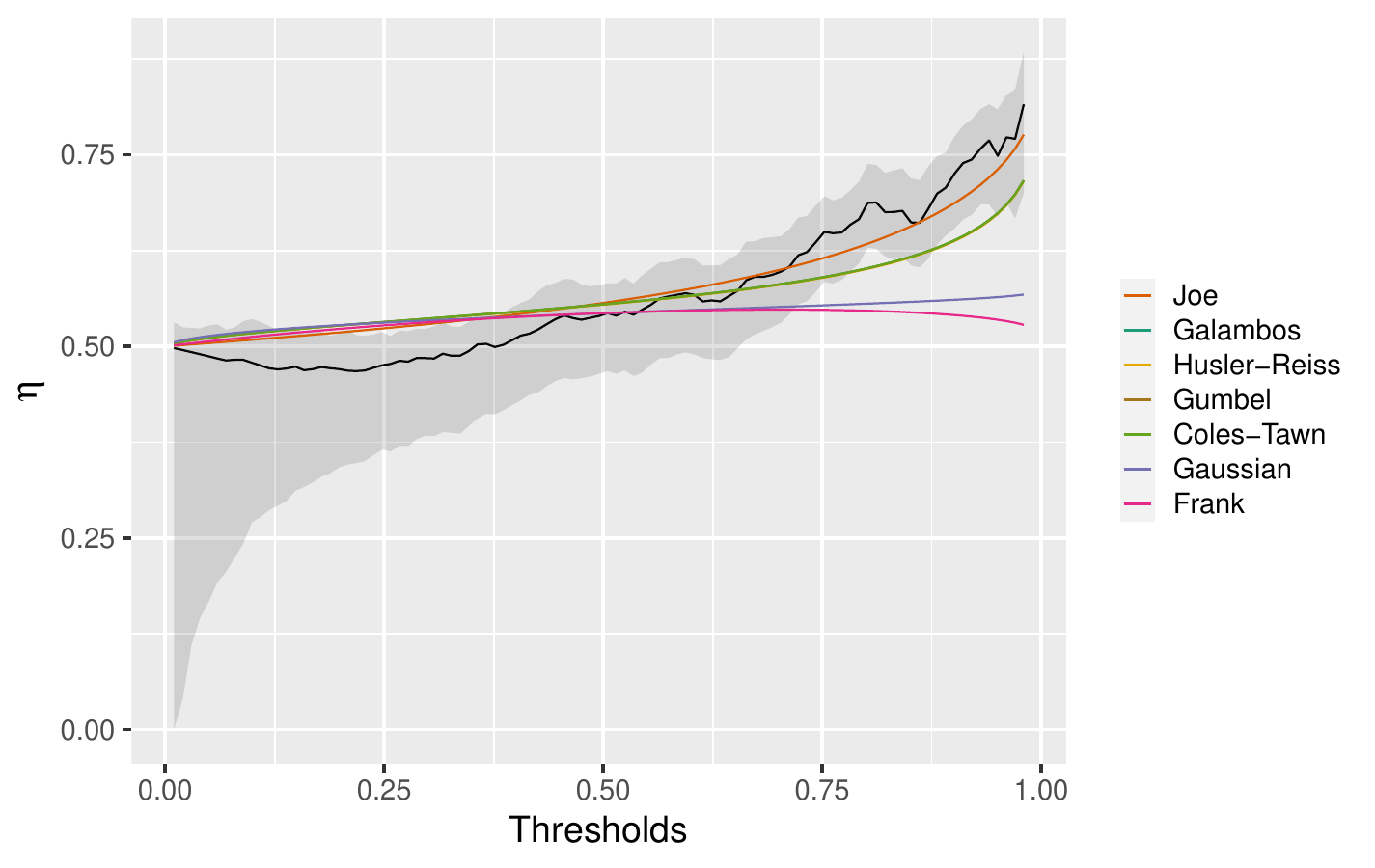}
        \caption{Empirical $\eta(r)$ (in black) and $\eta(r)$ for seven copulas (in colour) for $r\in(0,1).$ The 95$\%$ confidence bands were obtained by block bootstrapping. Note that the $\eta(r)$ for the Galambos, the H\"usler-Reiss, the Gumbel and the Coles-Tawn copulas overlap.}
    \label{fig:etasinglewey}
\end{figure}

\noindent Table \ref{tab:mlesinglewey} shows the MLEs obtained by fitting a range of single copulas and the corres\-ponding AIC values, whereas Figure \ref{fig:etasinglewey} illustrates the comparison between the empirical extremal dependence measure $\eta(r)$ for $r\in (0,1)$ and the model-derived ones.


\begin{table}[H]
\caption{MLEs for ten copulas and their AIC values. Lower AIC values are preferred.}
    \centering
    \begin{tabular}{lccc}
         Copula & \multicolumn{2}{c}{Parameter} & AIC \\
         \hline
         Clayton & \multicolumn{2}{c}{$7.21\times 10^{-9}$} & \phantom{-1}2.0  \\ 
         Frank & \multicolumn{2}{c}{0.94}  & -19.2 \\ 
         Gumbel & \multicolumn{2}{c}{1.18}  & -81.7 \\ 
         Inverted Gumbel & \multicolumn{2}{c}{1.03}  & \phantom{-1}0.9  \\ 
         Galambos & \multicolumn{2}{c}{0.43}  & -82.9 \\
         Gaussian & \multicolumn{2}{c}{0.18} & \phantom{1}-27.6 \\
         Joe & \multicolumn{2}{c}{1.34} & -113.8 \\
         Student t & 0.17 & 8.95 & \phantom{1}-34.9 \\
         H\"usler-Reiss & \multicolumn{2}{c}{0.82} & \phantom{1}-99.1 \\
         Coles-Tawn & 0.16 & 0.24 & \phantom{1}-80.4 \\
    \end{tabular}
    \label{tab:mlesinglewey}
\end{table}

\noindent Table \ref{tab:mlemixedwey} shows the MLEs when fitting a range of weighted copula models with $\pi(u^*,v^*;\theta)=(u^*v^*)^\theta$ and their AIC values. Table \ref{tab:mlemixed1wey} shows the MLEs of the five best models according to AIC when the weighting function is $\pi(u^*,v^*;\theta)=\exp\{-\theta(1-u^*)(1-v^*)\}.$

\begin{table}[H]
    \centering
    \caption{MLEs for different weighted copula models and their AIC values when the weighting function used is $\pi(u^*,v^*;\theta)=(u^*v^*)^\theta.$ Lower AIC values are preferred.}
    \renewcommand*{\arraystretch}{1.1}
    {\small\begin{tabular}{lllccccc}
         Model & $c_t$ & $c_b$ & \multicolumn{2}{c}{$\hat{\bm\alpha}$} & $\hat\beta$ & $\hat\theta$ &   AIC  \\
         \hline
         Model 1 & H\"usler-Reiss & Gaussian & \multicolumn{2}{c}{1.08} & -0.23 & 0.34 & -124.2 \\
         Model 2 & Galambos & Gaussian & \multicolumn{2}{c}{0.66} & -0.23 & 0.33 & -121.9 \\
         Model 3 & Coles-Tawn & Gaussian & 0.29 & 1.10 & -0.22 & 0.34 & -122.5 \\
         Model 4 & Coles-Tawn & Frank & 0.30 & 1.22 & -1.59 & 0.32 & -123.8 \\
         Model 5 & Joe & Frank & \multicolumn{2}{c}{1.46} & -1.95 & 0.16 & -126.7 \\
         Model 6 & Clayton & Gaussian &  \multicolumn{2}{c}{14.99} & -0.05 & 4.33 & \phantom{1}-92.8  \\
         Model 7 & Inverted Gumbel & Gaussian &  \multicolumn{2}{c}{2.33} & -0.15 & 0.96 & -105.4 \\
         Model 8 & H\"usler-Reiss & Joe &  \multicolumn{2}{c}{1.19} & 1.26 & 4.93 & -112.2 \\
         Model 9 & Student t & Galambos & 0.69 & 4.82 & 0.27 & 2.71 & \phantom{1}-98.0 \\
         Model 10 & Gaussian & Clayton & \multicolumn{2}{c}{0.75} & $1.16\times 10^{-5}$ & 2.45 & \phantom{1}-99.1 \\
         Model 11 & Gumbel & Joe & \multicolumn{2}{c}{1.47} & 1.26 & 4.27 & -111.8 \\
    \end{tabular}}
    \label{tab:mlemixedwey}
\end{table}

\begin{table}[H]
    \centering
    \caption{MLEs for five weighted copula models and their AIC values when the weighting function used is $\pi(u^*,v^*;\theta)=\exp\{-\theta(1-u^*)(1-v^*)\}.$ Lower AIC values are preferred.}
    \renewcommand*{\arraystretch}{1.1}
    {\small\begin{tabular}{lllccccc}
         Model & $c_t$ & $c_b$ & \multicolumn{2}{c}{$\hat{\bm\alpha}$} & $\hat\beta$ & $\hat \theta$ &  AIC  \\
         \hline
         Model 1 & H\"usler-Reiss & Gaussian & \multicolumn{2}{c}{1.12} & -0.52 & 3.21 & -158.5 \\
         Model 2 & Galambos & Gaussian & \multicolumn{2}{c}{0.72} & -0.51 & 3.48 & -159.2 \\
         Model 3 & Coles-Tawn & Gaussian & 0.46 & 0.82 & -0.48 & 4.13 & -158.1 \\
         Model 4 & Coles-Tawn & Frank & 0.48 & 0.74 & -3.05 & 3.61 & -150.0 \\
         Model 5 & Joe & Frank & \multicolumn{2}{c}{1.52} & -2.63 & 2.85 & -147.1 \\
    \end{tabular}}
    \label{tab:mlemixed1wey}
\end{table}

\subsection{Diagnostics} \label{subsubsection:diagnostics}

\begin{figure}[H]
    \centering
    \begin{subfigure}[b]{0.9\textwidth}
        \includegraphics[width=\textwidth]{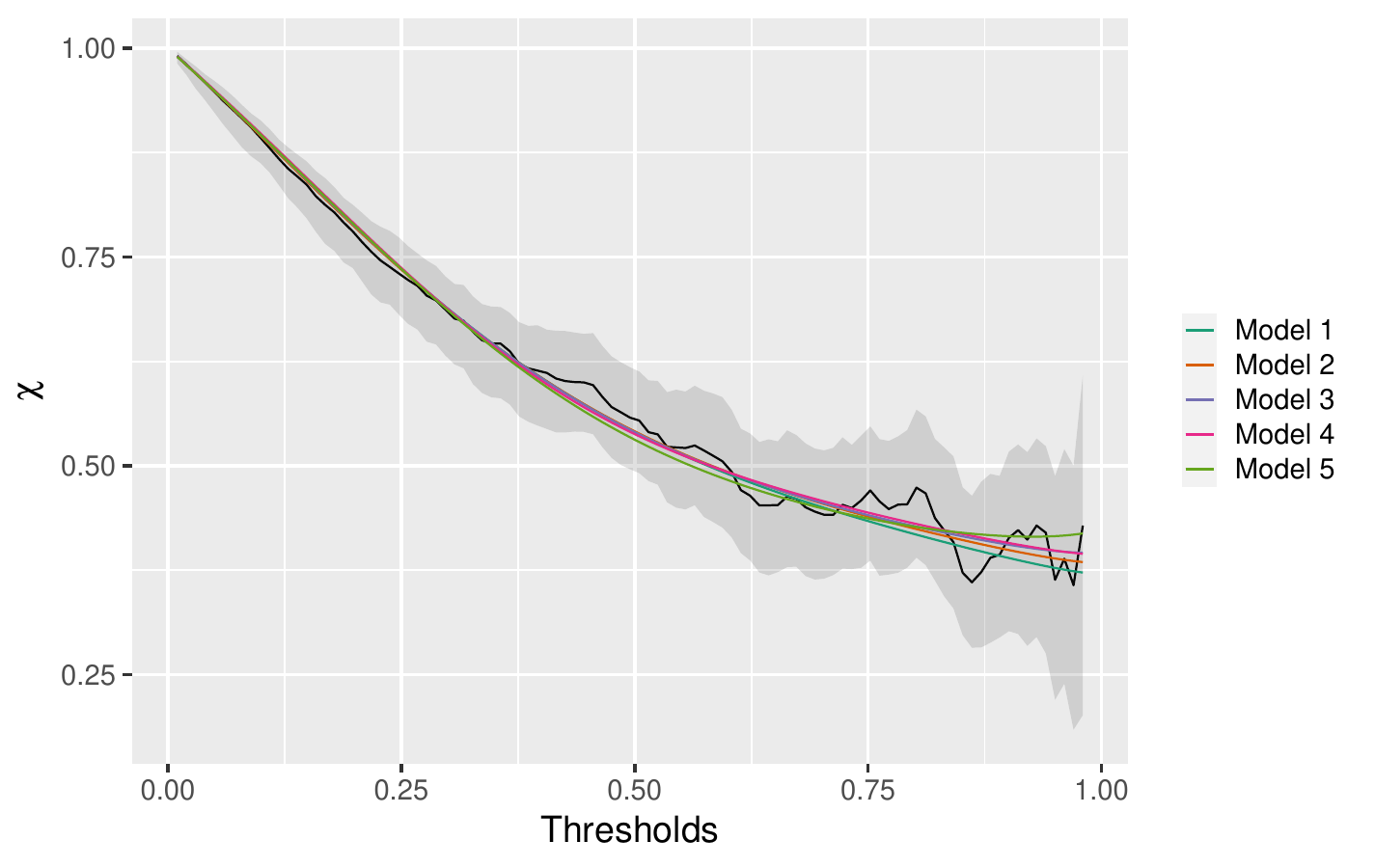}
        \caption{Empirical $\chi(r)$ (in black) and $\chi(r)$ for the five models (in colour) for $r\in(0,1).$ The 95$\%$ confidence bands were obtained by block bootstrapping.}
        \label{subfig:chimixed1}
    \end{subfigure}
    \begin{subfigure}[b]{0.9\textwidth}
        \includegraphics[width=\textwidth]{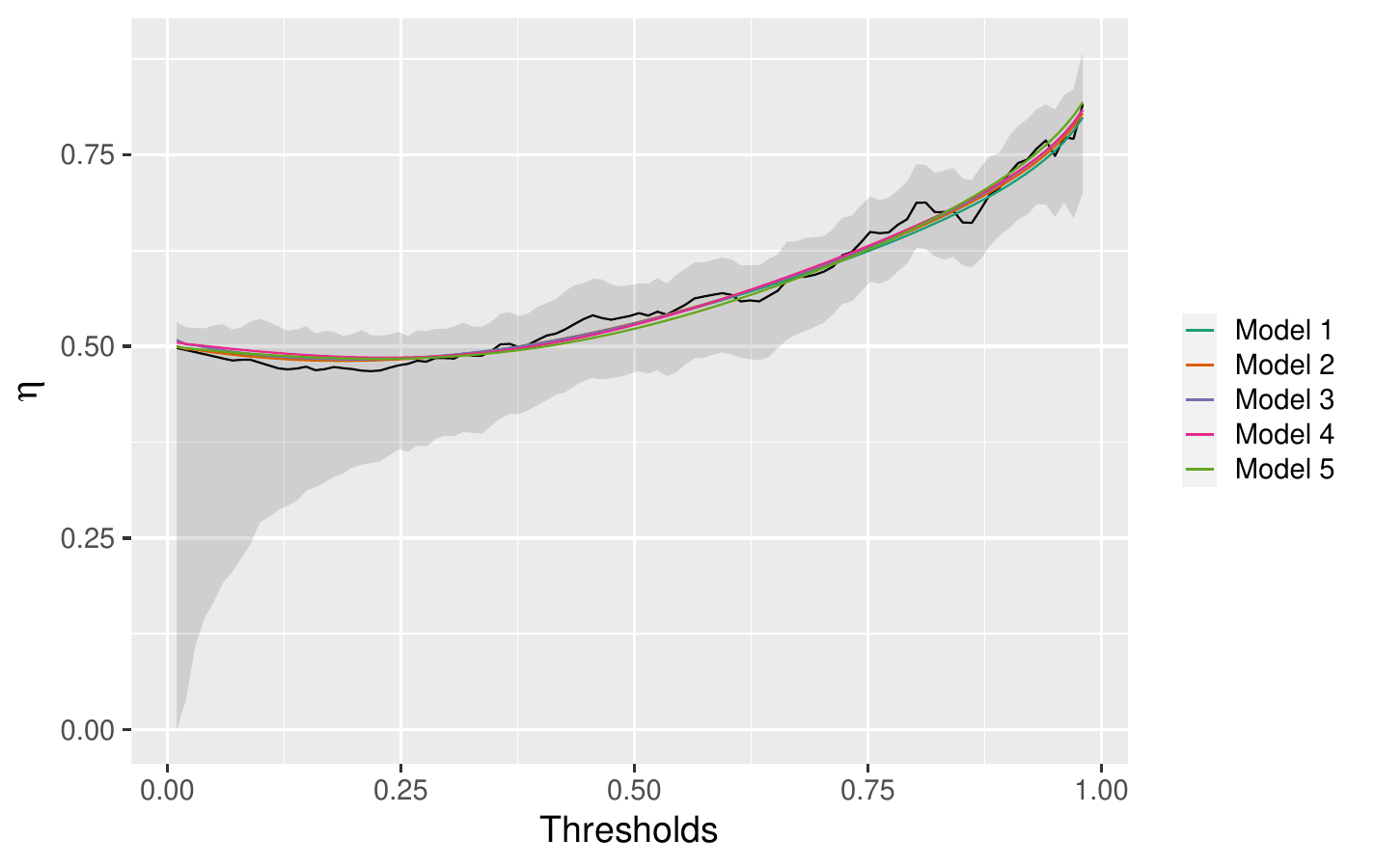}
        \caption{Empirical $\eta(r)$ (in black) and $\eta(r)$ for the five models (in colour) for $r\in(0,1).$ The 95$\%$ confidence bands were obtained by block bootstrapping.}
        \label{subfig:etamixed1}
    \end{subfigure}
    \caption{Dependence measures $\chi(r)$ and $\eta(r)$.}
    \label{fig:chietamixed1wey}
\end{figure}

\noindent Figure \ref{fig:chietamixed1wey} displays $\chi(r)$ and $\eta(r)$ for $r\in(0,1)$ for the five models considered. A clear improvement from the single copula models shown in Figure \ref{fig:etasinglewey} can be seen as now all five models offer a reasonable fit throughout the whole support of the data. In summer, the average temperature in Weybourne is between 18$^{\circ}$C and 22$^{\circ}$C and the observed 90th, 95th and 99th percentiles of the temperature are around 24$^{\circ}$C, 26$^{\circ}$C and 29$^{\circ}$C, respectively. Table \ref{tab:diag1wey} shows Kendall's $\tau$ and some probabilities of interest.

\begin{table}[H]
    \centering
    \caption{Diagnostics for the best five models according to their AIC values. The 95$\%$ confidence intervals for the empirical values were obtained by block bootstrapping. The empirical probability $P[O_{3}\geq 160 \mid 29\leq T\leq 30]$ and its 95$\%$ confidence interval are explained by the low number of observations present in the data set.}
    \renewcommand*{\arraystretch}{1.1}\setlength{\tabcolsep}{2pt} 
    {
    \begin{tabular}{lccc}
         Model & Kendall's $\tau$ &  $P[T\leq 15, O_{3}\geq 100]$ & $P[T\geq 24, O_{3}\geq 100]$  \\ 
         \hline
        Empirical & 0.0966  & 0.0045 & 0.0460 \\ 
         \footnotesize{($95\%$ CI)} &\footnotesize{(0.0555\,,\,0.1934)} & \footnotesize{(0.0000\,,\,0.0050)}& \footnotesize{(0.0338\,,\,0.0667)}\\ 
        Model 1 & 0.0881 & 0.0072 & 0.0491 \\ 
        Model 2 & 0.0900 & 0.0076 & 0.0502 \\ 
        Model 3 & 0.0853 & 0.0084 & 0.0509 \\ 
        Model 4 & 0.0944 & 0.0069 & 0.0512 \\ 
        Model 5 & 0.0882 & 0.0068 & 0.0517 \\ 
         \hline\hline
         Model & $P[T\geq 26, O_{3}\geq 100]$ & $P[O_{3}\geq 100 \mid 24\leq T\leq 25]$ &$P[O_{3}\geq 160 \mid 29\leq T\leq 30]$ \\ 
         \hline
         Empirical & 0.0291  & 0.1520 & 0.0000 \\ 
         \footnotesize{($95\%$ CI)} & \footnotesize{(0.0189\,,\,0.0438)} & \footnotesize{(0.0488\,,\,0.2800)} & \footnotesize{(0.0000\,,\,0.0000)} \\ 
        Model 1 & 0.0283 & 0.2557 & 0.1912 \\ 
        Model 2 & 0.0287 & 0.2617 & 0.1982 \\ 
        Model 3 & 0.0300 & 0.2516 & 0.1894 \\ 
        Model 4 & 0.0298 & 0.2573 & 0.1921 \\ 
        Model 5 & 0.0297 & 0.2646 & 0.2176 \\  
    \end{tabular}}
    \label{tab:diag1wey}
\end{table}


\end{document}